# Subleading Corrections to Hadronic Cross-Sections at High Energies

James D Cockburn

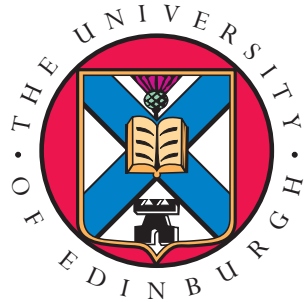

Doctor of Philosophy
The University of Edinburgh
2017

# Lay Summary

The startup of the Large Hadron Collider at CERN has marked the beginning of a new age of particle physics. We will have access to data about the most energetic particle collisions we have ever been able to produce. There are many hopes about what this data will reveal and a main one is that it will shed some light on how we can go beyond the Standard Model, which to date is our best theory of particle physics. The Standard Model has been incredibly successful but yet we know that it must be incomplete because it fails to explain, for example, the force of gravity. There are a number of potential ways we could extend the Standard Model and all of them predict either the existence of new particles that we have not yet seen or some modifications to the properties of the particles we already know about – in many cases, both. In the latter case, these modifications can be very small and hard to detect and thus it is of paramount importance that we as a theory community provide accurate and precise predictions for how particles should behave within the bounds of the Standard Model.

In this thesis, we argue that the usual method of how these predictions are made is not satisfactory when describing the collisions at a high energy. Instead, we should think about how the calculation is done in a different way which guarantees that these high energy considerations are correctly taken into account. This is done in a fairly straightforward way for most of the potential collisions at the Large Hadron Collider, but requires more effort and thought when it is being expanded to include more possibilities. It is this latter point that constitutes new work on the part of the author.

We show also that the formalism is simple enough that a common approximation in Higgs boson processes need not be employed. This original result means we can therefore make predictions for these processes in a fundamentally new way.

# Abstract

The Large Hadron Collider (LHC) has provided, and will continue to provide, data for collisions at the highest energies ever seen in a particle accelerator. A strong knowledge of the properties of amplitudes for Quantum Chromodynamics in the High Energy Limit is therefore important to interpret this data. We study this limit in the context of the High Energy Jets (HEJ) formalism. This formalism resums terms in the perturbative expansion of the cross-section that behave like $\alpha_s^n \log\left(\frac{s}{-t}\right)^{n-1}$, which are enhanced in this limit. Understanding this region is particularly important in certain key analyses at the LHC: for example, Higgs-boson-plus-dijet analyses where cuts are applied to pick out events with a large $m_{jj}$ and in many searches for new physics.

In this thesis, we discuss two directions in which HEJ's accuracy has been improved. Firstly, we look at adding descriptions of partonic subprocesses which are formally sub-leading in the jet cross-section but Leading Logarithmic (LL) in the particular subprocess itself. This required the derivation of new effective vertices that describe the emission of a quark/anti-quark pair in a way that is consistent with the resummation procedure. The inclusion of such processes reduces HEJ's dependence on fixed-order calculations and marks an important step towards full Next-to-Leading Logarithmic (NLL) accuracy in the inclusive dijet cross-section.

The second extension was to improve our description of events involving the emission of a Higgs boson along with jets. Specifically, we derive new effective vertices which keep the full dependence on the quark mass that appears in the loops that naturally arise in such amplitudes. The formalism is also simple enough to allow for any number of extra final state jets in the process. Therefore, HEJ is unique in its ability to provide predictions for high-multiplicity Higgs-plus-jets processes with full finite quark mass effects. Such a calculation is far beyond the reach of any fixed order approach.

# Declaration

I declare that this thesis was composed by myself, that the work contained herein is my own except where explicitly stated otherwise in the text, and that this work has not been submitted for any other degree or professional qualification except as specified.

A brief overview of the work performed in chapter 3 has been published in the proceedings of DIS 2016 [23]. It is anticipated that two publications will be produced within the next year corresponding to the results of chapters 3 and 4.

*(James D Cockburn, 2017)*

# Acknowledgements

Thanks first and foremost must go to my wonderful parents who have never once discouraged me from following any path in life other than the one I wanted to go down. Without their constant love and support, I could never have got to where I am today. This thesis is specifically dedicated to them and I hope they will find comfort in the fact that all my pondering has turned into something useful!

I also owe a great deal to the rest of the 'PPT crew' that started their PhD journeys at the same time as me. Andries, Gustav, Susi, James, Ava and slightly-late-to-the-party Rafa – thank you. You made the transition into PhD life very smooth, efficient and dare I say, a jolly good laugh. I've learnt so much from all of you over these past three-and-a-half years and I couldn't have finished it without you.

The largest thanks I have to offer must go to my supervisor, Dr Jenni Smillie. I could not have asked for a more patient, willing, enthusiastic, fun and intelligent person to guide me. I have often suspected that I would not have had anywhere near as good a time doing my PhD if it were not for her. She has never made me feel inadequate or that I did not belong, and having the visible belief of one's supervisor is a huge part of successful postgraduate study.

Lastly, a big thankyou to the rest of the High Energy Jets collaboration of past and present - Jeppe, Jack, Helen, Andreas and Tuomas. Some combination of you were always there to answer any questions I ever had and your constant drive inspired me to persevere even when the going was tough.

# Contents

# List of Figures

# List of Tables

# Chapter 1

# Background and Theoretical Basis

In this section, we present an overview of Particle Physics and Quantum Field Theory (QFT) upon which the work contained in this thesis depends. We begin by discussing the Standard Model (SM) of Particle Physics and how modern experiments at the Large Hadron Collider (LHC) test it. We then focus on Quantum Chromodynamics (QCD), which is the part of the SM that describes the interactions of quarks and gluons, the fundamental particles that come together to form protons, neutrons and a whole range of other composite particles [48]. Finally, we delve into the Spinor Helicity formalism [33] to introduce notation and derive results that will allow for a simple and clear method of performing the work presented in the later chapters of this thesis.

## 1.1 The LHC and the Standard Model

The SM, shown in figure 1.1, is currently our best theoretical model for understanding the fundamental particles of nature and their interactions. It is also one of the most rigorously tested models in the history of physics and to date it has stood up to all the tests thrown at it [32]. For instance, the measurement of the anomalous magnetic dipole moment of the electron agrees with that predicted by the model to ten significant figures [8]. Despite these successes, however, we know that the SM cannot be a *complete* model of our universe. For instance, it does not include gravity in its formulation. On top of that, cosmological measurements such as that of the velocities of spinning galaxies and the expansion rate of the universe imply that there is a whole other sector we are currently ignorant of: that of Dark Matter and Dark Energy [16]. Indeed, some estimates suggest that approximately 96% of the universe consists of these mysterious quantities [44]. Clearly, we need to find a way forward in order to develop a theory 'Beyond the Standard Model' that can describe them. A large number of proposed theories predict that non-SM particles interact with SM particles, but so weakly that very precise measurements of known SM processes must be made before we could observe the effects. Theorists and experimentalists must continue to come together in order to face this challenge; theorists by improving how calculations are conducted and experimentalists by reducing experimental uncertainties and collecting ever more data.

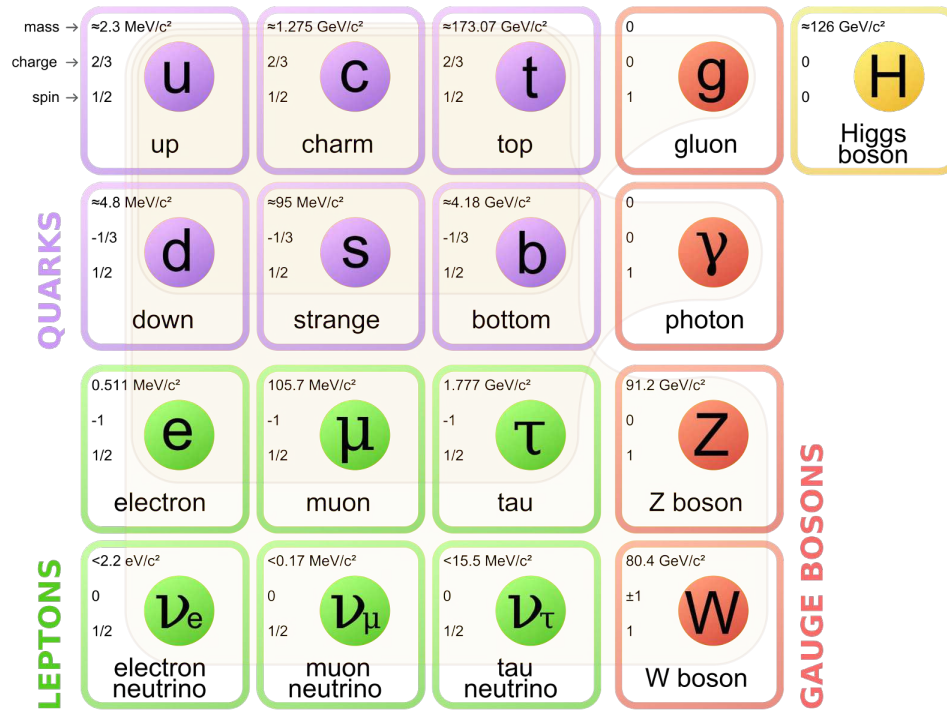

**Figure 1.1** *The Standard Model of Particle Physics. Image from the Wikipedia page on the Standard Model.*

The LHC has given physicists the opportunity to probe the SM at the highest energies yet, ultimately being able to reach a centre-of-mass energy of 14 TeV. This is approximately 7 times higher than what was achievable at its spiritual predecessor, the Tevatron at Fermilab [43]. The headline story for the LHC, of course, was its discovery of the Higgs boson in 2012, the physical remnant of 'Electroweak Symmetry Breaking' (EWSB) [11, 41]. In effect, the masses of the fundamental particles are generated via interactions with the Higgs field, and so it is often said that the discovery of the Higgs boson is the discovery of the 'origin of mass'. At this stage, we do not know whether the Higgs boson that has been discovered is exactly the one predicted by the Standard Model. Much more research now needs to be done in order to determine its fundamental properties, such as the strength of its coupling to the massive gauge bosons, before we can know one way or the other. This presents the very exciting theoretical challenge of providing accurate predictions for Higgs processes, and Chapter 4 of this thesis will explore this issue further. The theory that we will mostly concern ourselves with, however, is not that of EW physics but rather QCD. The following sections will provide an overview of the theory in order to discuss some important aspects and also to provide a basic introduction to QFT, which underpins the entirety of the SM.

## 1.2 Quantum Chromodynamics

QCD is described by a non-Abelian $SU(3)$ gauge group within the SM and provides the theoretical underpinning for the *strong interaction*, the interaction between *quarks* and *gluons*. This then confines our study to the purple area and the very top of the red area shown in figure 1.1. This may seem restrictive, but because the LHC collides protons which consist of quarks and gluons, it is clearly QCD that is the underlying theory we need to use to understand the collisions that happen there.

### 1.2.1 Lagrangian Formalism and Dirac Fermions

QFTs have their foundations in Lagrangian mechanics and so we will here review the background of the formalism. The fundamental quantity is the *action*, the time integral of the Lagrangian $L$:

$$S = \int L \, dt. \tag{1.1}$$

In a local field theory, we can write the Lagrangian as a integral over all space of the *Lagrangian Density*, $\mathscr{L}$, which can depend on a set of fields and the derivatives of those fields, $\mathscr{L} = \mathscr{L}(\phi_i, \partial_\mu \phi_i)$. From this point onwards, we will refer to the

'Lagrangian density' as just the 'Lagrangian'. We therefore have

$$S = \int \mathscr{L}(\phi_i, \partial_\mu \phi_i) \, dt \, d^3x = \int \mathscr{L}(\phi_i, \partial_\mu \phi_i) \, d^4x, \qquad (1.2)$$

where we note that we work in units such that $c = \hbar = 1$. The *principle of stationary action* states that as a system evolves from one configuration to another, it does so in a way that minimises the action; $\delta S = 0$. By considering a small change on the right-hand side of the previous equation, we arrive at the famous *Euler-Lagrange* equations of motion,

$$\partial_\mu \left( \frac{\partial \mathscr{L}}{\partial(\partial_\mu \phi_i)} \right) - \frac{\partial \mathscr{L}}{\partial \phi_i} = 0. \qquad (1.3)$$

The field content of QCD consists of quarks and gluons, so they must be included in the Lagrangian in such a way that the appropriate equations of motion are recovered from it. We will begin with quarks. They are fermionic particles and so (when not interacting or, more simply, 'free') will obey the Dirac equation

$$(i\slashed{\partial} - m)\psi(x) = 0, \qquad (1.4)$$

where $\slashed{\partial} \equiv \gamma^\mu \partial_\mu$ and $m$ is the mass of the quark. Suppressed here is the fact that this is actually a matrix equation in spinor space; $\psi$ is a four-component spinor and the $\gamma^\mu$ are a set of four-by-four matrices with the anti-commutation relation $\{\gamma^\mu, \gamma^\nu\} = 2\eta^{\mu\nu}$. An explicit form for the matrices $\gamma^\mu$ will be presented later in the chapter since it is not necessary to have one at this point. We introduce a conjugate field $\bar{\psi}(x) = \psi^\dagger(x)\gamma^0$ and write a Lagrangian

$$\mathscr{L} = \bar{\psi}(x)(i\slashed{\partial} - m)\psi(x), \qquad (1.5)$$

where now we can evaluate the Euler-Lagrange equation with respect to the field $\bar{\phi}$ to yield the Dirac equation for $\psi$, as we wanted. The reason $\bar{\psi}$ is used as a conjugate field as opposed to the more obvious $\psi^\dagger$ comes from the requirement that the Lagrangian is invariant under a Lorentz transformation; in other words, $\mathscr{L}$ is a Lorentz scalar. We could also, of course, take the variation of this Lagrangian with respect to $\psi$ and doing so we get

$$\bar{\psi}(x)(i\slashed{\partial} + m) = 0, \qquad (1.6)$$

which is the Hermitian conjugate of the previous equation. The Dirac equation has plane wave solutions of the form

$$\psi(x) = u(\vec{p})\, e^{-ip\cdot x} + v(\vec{p})\, e^{ip\cdot x}, \tag{1.7}$$

where $p$ is the four-momentum and $\vec{p}$ the three-momentum. In practice it is the $u$ and $v$ spinors that are used in calculations and they are interpreted as describing fermions and anti-fermions respectively.

This prediction of the existence of anti-particles is one of the great successes of the Dirac equation. To understand precisely why the equation's solutions imply their existence, imagine that we are in a frame such that $\vec{p} = 0$. Since $p \cdot x = Et - \vec{p} \cdot \vec{x}$ and $E = \sqrt{\vec{p}^2 + m^2}$, then the second term in equation 1.7 will have an exponential $e^{imt}$ – a state with negative energy[1]. Because Quantum Mechanics requires a complete set of basis states, we cannot simply throw the solution away as 'unphysical'. The solutions have to be interpreted as either positive energy fermions moving backwards in time or negative energy anti-fermions moving forwards in time. We will see later how the former interpretation lends itself rather neatly to a graphical representation of scattering involving fermionic particles. In any case, substituting this form for $\psi(x)$ into the Dirac equation, we find the following conditions:

$$(\slashed{p} - m)u(\vec{p}) = 0, \tag{1.8a}$$
$$(\slashed{p} + m)v(\vec{p}) = 0. \tag{1.8b}$$

### 1.2.2 Gauge Symmetry

We briefly mentioned that QCD is a gauge theory with gauge group $SU(3)$. This is a non-abelian group, which we will see later leads to a whole host of interesting effects. For the time being, however, we will switch to the abelian group $U(1)$, which is the group describing Quantum Electrodynamics (QED), in order to make it easier to derive some results that we can still use for the more complicated group $SU(3)$. In either case, the symmetry we require in the Lagrangian is that it is unchanged if we redefine our charges according to a group transformation, $\psi \to U\psi$. For a global redefinition where our operator $U$ does not depend on $x$, this is trivially true since the $U$s are by definition unitary. However, if we generalise to a local redefinition,

$$\begin{aligned}\mathscr{L}' &= \bar{\psi}(x)U^\dagger(x)(i\not{\partial} - m)U(x)\psi(x) \\ &= \bar{\psi}(x)(i\not{\partial} - m)\psi(x) + i\bar{\psi}(x)U^\dagger(x)\gamma^\mu \left[\partial_\mu U(x)\right]\psi(x) \neq \mathscr{L}.\end{aligned} \quad (1.9)$$

---

[1] To be more precise, the sign of $i$ chosen in the Schrödinger equation is a matter of convention and so strictly speaking it is not just the sign of the exponential that determines that this is an anti-fermion. Rather, it is because we have *both* plus and minus sign solutions and we are required to consider both of them.

The problem has appeared because of the derivative in our Lagrangian. What we would like is a different derivative term that includes the partial derivative but along with an extra part to cancel away the extra term. Equivalently, we search for a so-called *covariant derivative* that transforms as

$$D'_\mu = U(x)D_\mu U^\dagger(x). \quad (1.10)$$

Such an object can be constructed, but we have to introduce another field, $A_\mu$, that transforms non-trivially. Before we do so, however, let us take the specific example of the $U(1)$ group and explicitly evaluate the extra term. If we call the group generator $\Lambda$, then

$$U(x) = e^{iq\Lambda(x)}, \quad (1.11)$$

where $q$ is an overall scale factor, which is associated with the group charge. Then the extra term is proportional to

$$\partial_\mu U(x) = iq(\partial_\mu \Lambda(x))e^{iq\Lambda(x)}. \quad (1.12)$$

Let us define our covariant derivative in the following way:

$$D_\mu = \partial_\mu + iqA_\mu(x). \quad (1.13)$$

Then, under the gauge transformation:

$$\begin{aligned}D'_\mu \psi'(x) &= (\partial_\mu + iqA'_\mu(x))e^{iq\Lambda(x)}\psi(x) \\ &= U(x)(\partial_\mu + iq[A'_\mu(x) + (\partial_\mu\Lambda(x))])\psi(x),\end{aligned} \quad (1.14)$$

and therefore we can recover invariance so long as

$$A'_\mu(x) = A_\mu(x) - \partial_\mu\Lambda(x), \quad (1.15)$$

which we recognise precisely as a gauge transformation of the vector potential $A^\mu$. The process of constructing a covariant derivative in this manner is called the *principle of minimal coupling* and can be applied to any given gauge group. Thus, we simply replace the partial derivative in our Lagrangian with the covariant derivative to get a locally gauge invariant Lagrangian:

$$\begin{aligned}\mathscr{L}_{Dirac+Interaction} &= \bar{\psi}(x)(i\slashed{D} - m)\psi(x) \\ &= \bar{\psi}(x)(i\slashed{\partial} - m)\psi(x) + iq\bar{\psi}(x)\slashed{A}(x)\psi(x).\end{aligned} \quad (1.16)$$

From the requirement of local gauge invariance alone, we see from the second term that there is necessarily an interaction between the matter content and the gauge boson. However, our current form of the Lagrangian does not yet include a term that describes the dynamics of the gauge field itself. We know that such a term must be gauge and Lorentz invariant. Consider what would happen if we included a derivative of the field $A_\mu$ in our formalism and applied the gauge transformation;

$$\partial'_\mu A'_\nu(x) = \partial_\mu(A_\nu(x) - \partial_\nu \Lambda(x)). \quad (1.17)$$

This is clearly not gauge invariant, but the combination anti-symmetric in $\mu, \nu$ is:

$$\begin{aligned}\partial'_\mu A'_\nu(x) - \partial'_\nu A'_\mu(x) &= \partial_\mu(A_\nu(x) - \partial_\nu \Lambda(x)) - \partial_\nu(A_\mu(x) - \partial_\mu \Lambda(x)) \\ &= \partial_\mu A_\nu(x) - \partial_\nu A_\mu(x) \equiv F_{\mu\nu},\end{aligned} \quad (1.18)$$

where the last line follows from the fact that derivatives commute with each other. The object $F_{\mu\nu}$ is called the *field strength tensor*. This can also be defined in the form

$$F_{\mu\nu} = -\left(\frac{i}{q}\right)[D_\mu, D_\nu], \quad (1.19)$$

which is a form that will be useful when generalising to other gauge groups. The last requirement is to insert this into our Lagrangian in a Lorentz invariant fashion[2] and also make explicit that there are potentially many fermions that can interact with the $U(1)$ gauge bosons in this way:

$$\mathscr{L}_{QED} = \sum_{f \in flavours} \bar{\psi}_f(x)(i\slashed{D} - m)\psi_f(x) - \frac{1}{4}F_{\mu\nu}F^{\mu\nu}, \quad (1.20)$$

where we have added an arbitrary normalisation factor to our new term for convenience.

---
[2]Actually, we should really also include a term $\sim \epsilon^{\alpha\beta\mu\nu}F_{\alpha\beta}F_{\mu\nu}$ since this too is Lorentz invariant, but such a term is parity violating and this thesis will only concern itself with theories that are parity-symmetric.

### 1.2.3 QCD Lagrangian

The question of how to go from QED to QCD is equivalent to the question of what extra considerations need to be made if we work with the group $SU(3)$ rather than $U(1)$. The first difference is that $SU(3)$ has $3^2 - 1 = 8$ generators and so 8 independent rotation directions for our operator $U$. Thus, a (local) group transformation will have the form

$$U(x) = e^{ig_s \Lambda^a(x) t^a}, \tag{1.21}$$

where $t^a$ are the generators of the group. Additionally, in the fundamental representation of $SU(3)$, the Dirac spinors must be a 3-vector,

$$\psi = \begin{pmatrix} q^r \\ q^b \\ q^g \end{pmatrix}, \tag{1.22}$$

where $q$ is a specific flavour of quark (the entire theory has all of them, of course) and $r, g, b$ are colour labels, denoting the SU(3) 'colour' charge. Now we apply the principle of minimal coupling from the previous subsection to find a covariant derivative for this gauge group, being explicit with our colour space notation:

$$D_\mu = \partial_\mu \mathbb{1} + i g_s t^a A_\mu^a(x). \tag{1.23}$$

Each of the 8 gauge fields has to transform in the same way as before so as to make the covariant derivative perform its function correctly,

$$A_\mu^{a'} = A_\mu^a + \partial_\mu \Lambda^a(x), \tag{1.24}$$

and the final step is to construct the field strength tensor

$$\begin{aligned}
F_{\mu\nu} &= -\left(\frac{i}{g_s}\right)[D_\mu, D_\nu] \\
&= -\left(\frac{i}{g_s}\right)[\partial_\mu + ig_s A_\mu^a t^a, \partial_\nu + ig_s A_\nu^b t^b] \\
&= [A_\mu^a t^a, \partial_\nu] + [\partial_\mu, A_\nu^b t^b] + ig_s A_\mu^a A_\nu^b [t^a, t^b] \\
&= A_\mu^a t^a \partial_\nu - (\partial_\nu A_\mu^a) t^a + (\partial_\mu A_\nu^b) t^b - A_\nu^b t^b \partial_\mu + ig_s A_\mu^a A_\nu^b [t^a, t^b] \\
&= t^c (\partial_\mu A_\nu^c - \partial_\nu A_\mu^c) + ig_s A_\mu^a A_\nu^b (if^{abc} t^c) \\
&= t^a (\partial_\mu A_\nu^a - \partial_\nu A_\mu^a - g_s A_\mu^b A_\nu^c f^{abc}) \\
&= t^a F_{\mu\nu}^a,
\end{aligned} \quad (1.25)$$

where the $f^{abc}$ are the structure constants of $SU(3)$. The non-abelian nature has introduced an extra term alongside the derivatives in the field strength tensor and this has an interesting physical consequence: namely, that our Lagrangian will contain terms where the gauge bosons can interact with each other. This extra term has another immediate consequence: $F_{\mu\nu}$ is no longer gauge invariant. Indeed

$$\begin{aligned}
F'_{\mu\nu} &= -\left(\frac{i}{g_s}\right)[D'_\mu, D'_\nu] \\
&= -\left(\frac{i}{g_s}\right)[UD_\mu U^\dagger, UD_\nu U^\dagger] \\
&= -\left(\frac{i}{g_s}\right)U[D_\mu, D_\nu]U^\dagger \\
&= UF_{\mu\nu}U^\dagger,
\end{aligned} \quad (1.26)$$

and unlike the Abelian case, we cannot simply pass $U$ through the field strength tensor because of the non-commuting generators. We see that actually the invariant quantity is the *trace* in colour space,

$$\text{tr}(F'_{\mu\nu}) = \text{tr}(UF_{\mu\nu}U^\dagger) = \text{tr}(F_{\mu\nu}U^\dagger U) = \text{tr}(F_{\mu\nu}). \quad (1.27)$$

Remembering our need for our Lagrangian to be a Lorenz scalar as well, we are now in a position to write down the Lagrangian for QCD:

$$\mathcal{L}_{QCD} = \sum_{f \in flavours} \sum_{i,j=1}^{3} \bar{q}_f^i(x)(i\slashed{D} - m_f)_{ij} q_f^j(x) - \frac{1}{2}\text{tr}(F_{\mu\nu}F^{\mu\nu}). \quad (1.28)$$

## 1.3 From Lagrangians to Scattering Amplitudes

In the last section, we showed by considering symmetries and the equations of motion that it is possible to write down a Lagrangian. What we have not yet explained is why that was useful. Here, we provide an overview of the *path integral formalism* for QFT which will provide us with a way of computing *scattering amplitudes* in our field theory.

### 1.3.1 Path Integral Formulation

We will consider non-relativistic Quantum Mechanics as our starting point for the derivation of the Path Integral formalism. The results obtained in this regime will be immediately generalisable to a QFT. The motivating principle behind the Path Integral is to try and calculate the transition amplitude between two general quantum mechanical states. If the states are completely defined by a set of co-ordinates $\{q^i\} = q$, then the transition amplitude from a state $q_a$ at time zero to $q_b$ at time $T$ is

$$A(q_a, q_b; T) = \langle q_b; t = T | q_a; t = 0 \rangle . \tag{1.29}$$

From solving the Schrödinger equation, we know that there exists a unitary operator that takes a state at one time and evolves it to another time under the influence of a time-independent Hamiltonian operator $\hat{H}$,

$$\hat{U}(t - t_0) = e^{-i\hat{H}(t - t_0)}. \tag{1.30}$$

Our transition amplitude can then be written

$$A(q_a, q_b; T) = \langle q_b | \hat{U}(T) | q_a \rangle = \langle q_b | e^{-i\hat{H}T} | q_a \rangle . \tag{1.31}$$

Let us now break the problem up in time. We split up the time interval $T$ into $n$ equal elements $\varepsilon = T/n$ such that the operator for the full time span is made up of the continual application of an operator evolving the state in steps of $\varepsilon$, $\hat{U}(T) = \hat{U}^n(\varepsilon)$. Along with this, we will also insert $n - 1$ identity operators

$\mathbb{1} = \int dq_j \, |q_j\rangle \langle q_j|$, allowing us to write our amplitude as

$$A(q_a, q_b; T) = \int \prod_{j=1}^{n-1} (dq_j) \, \langle q_b|e^{-i\hat{H}\varepsilon}|q_{n-1}\rangle \cdots \langle q_{i+1}|e^{-i\hat{H}\varepsilon}|q_i\rangle \cdots \langle q_1|e^{-i\hat{H}\varepsilon}|q_a\rangle \, .$$
(1.32)

We now take $n$ to be large and expand one of the matrix elements of this product in $\varepsilon$,

$$\langle q_{i+1}|e^{-i\hat{H}\varepsilon}|q_i\rangle = \langle q_{i+1}|1 - i\varepsilon\hat{H} + \mathcal{O}(\varepsilon^2)|q_i\rangle \, .$$
(1.33)

Consider a Hamiltonian of the form $\hat{H} = \hat{K}(\hat{p}) + \hat{V}(\hat{q})$, that is, a Hamiltonian with a kinetic term depending only on the momenta of the system $p$ and a potential that depends only on the co-ordinates $q$. Taking just the potential term, we see the matrix element 1.33 is

$$\begin{aligned}\langle q_{i+1}|\hat{V}(\hat{q})|q_i\rangle &= V(q_i)\delta(q_i - q_{i+1}) \\ &= V\left(\frac{q_i + q_{i+1}}{2}\right) \int \frac{dp_i}{2\pi} e^{ip_i(q_{i+1}-q_i)},\end{aligned}$$
(1.34)

where the potential has been democratically evaluated at the average of the two points. For the kinetic part, we must insert a complete set of momentum eigenstates for it to act upon:

$$\begin{aligned}\langle q_{i+1}|\hat{K}(\hat{p})|q_i\rangle &= \int dp_i \, \langle q_{i+1}|\hat{K}(\hat{p})|p_i\rangle \langle p_i|q_i\rangle \\ &= \int dp_i \, K(p_i) \langle q_{i+1}|p_i\rangle \langle p_i|q_i\rangle \\ &= \int \frac{dp_i}{2\pi} K(p_i) e^{iq_{i+1}p_i} e^{-iq_i p_i} \\ &= \int \frac{dp_i}{2\pi} K(p_i) e^{ip_i(q_{i+1}-q_i)}.\end{aligned}$$
(1.35)

Therefore, our form for the Hamiltonian allows us to write

$$\langle q_{i+1}|\hat{H}(\hat{q},\hat{p})|q_i\rangle = \int \frac{dp_i}{2\pi} H\left(\frac{q_i + q_{i+1}}{2}, p_i\right) e^{ip_i(q_{i+1}-q_i)}.$$
(1.36)

Such a formula is not correct in general, however. The left-hand side depends on the non-commuting operators $\hat{q}, \hat{p}$ and the right-hand side only on the eigenvalues of these operators. If our Hamiltonian were to contain terms that depend on both of these operators in some general order then we would expect to get different physics based on how exactly those operators are ordered. We can get around this by requiring that our Hamiltonian is in *Weyl ordered* form, which is the form that by definition yields the equality 1.36. Any general Hamiltonian can be brought into this form by use of the commutation relation $[\hat{q}, \hat{p}] = i$. For example, say the Hamiltonian had a term $\hat{q}\hat{p}$, then

$$\begin{aligned}\hat{q}\hat{p} &= \frac{1}{2}(\hat{q}\hat{p} + \hat{q}\hat{p}) \\ &= \frac{1}{2}(\hat{q}\hat{p} + \hat{p}\hat{q} + i) \\ &= \hat{H}(\hat{q}\hat{p}, \hat{p}\hat{q})_{WO} + \frac{i}{2}.\end{aligned} \quad (1.37)$$

We see from this that a Weyl ordered Hamiltonian has the property that the operators $\hat{q}, \hat{p}$ appear in a symmetric fashion. We therefore have, for any Weyl ordered Hamiltonian and small $\varepsilon$,

$$\langle q_{i+1}|e^{-i\varepsilon \hat{H}(\hat{q},\hat{p})_{WO}}|q_i\rangle = \int \frac{dp_i}{2\pi} e^{-i\varepsilon H\left(\frac{q_i+q_{i+1}}{2}, p_i\right)} e^{ip_i(q_{i+1}-q_i)}. \quad (1.38)$$

Putting this in for every matrix element 1.33 in 1.32, then

$$\begin{aligned}A(q_a, q_b; T) &= \int \prod_{j=1}^{n-1}\left(\frac{dp_j dq_j}{2\pi} e^{-i\varepsilon H\left(\frac{q_j+q_{j+1}}{2}, p_i\right)} e^{ip_j(q_{j+1}-q_j)}\right) \\ &= \int \prod_{j=1}^{n-1}\left(\frac{dp_j dq_j}{2\pi}\right) \exp\left(i\varepsilon \sum_{j=1}^{n-1}\left[p_j\left(\frac{q_{j+1}-q_j}{\varepsilon}\right) - H\left(\frac{q_j+q_{j+1}}{2}, p_j\right)\right]\right).\end{aligned}$$
$$(1.39)$$

If we now formally take the limit $n \to \infty$ (and so $\varepsilon \to 0$), then $(q_{i+1} - q_i)/\varepsilon \to \dot{q}$, our discrete sum in the exponential becomes an integral with measure $dt$ and our $p_j, q_j$ integrals become *Path Integrals*:

$$A(q_a, q_b; T) = \int \mathcal{D}q(t)\mathcal{D}p(t) \exp\left(i\int_0^T dt[p\dot{q} - H(q,p)]\right). \quad (1.40)$$

The path integral can be interpreted as an integration over all paths in phase space with the boundary conditions $q(0) = q_a$ and $q(T) = q_b$ (note there is no constraint on the momenta $p$ at the endpoints). It is precisely the integration measure in 1.39 evaluated at each point in time. The integral over $p$ can be performed by considering its stationary points, which is where

$$\dot{q} - \frac{\partial H}{\partial p} = 0 \to \dot{q} = \frac{\partial H}{\partial p}. \quad (1.41)$$

We can solve this differential equation for $p$ in terms of $q, \dot{q}$ and so we are left with

$$\begin{aligned} A(q_a, q_b; T) &= \int \mathcal{D}q(t) \exp\left(i \int_0^T dt[p(q,\dot{q})\dot{q} - H(q,\dot{q})]\right) \\ &= \int \mathcal{D}q(t) \exp\left(i \int_0^T dt L(q,\dot{q})\right) \\ &= \int \mathcal{D}q(t) \exp(iS). \end{aligned} \qquad (1.42)$$

We therefore see that knowing the Lagrangian for a system in principle allows us to calculate path integrals in the theory. We can immediately generalise this result to QFT, where instead of the co-ordinates $q, \dot{q}$ we have a dependence on a field, $\phi$, and its derivative, $\partial_\mu \phi$, and our boundary conditions relate to specific configurations of the fields $\phi_a, \phi_b$:

$$A(\phi_a, \phi_b; T) = \int \mathcal{D}\phi \exp\left(i \int_o^T d^4x \mathcal{L}(\phi, \partial_\mu \phi)\right). \qquad (1.43)$$

### 1.3.2 LSZ Formula

In QFT calculations, we will work with well-defined initial ('in') and final ('out') states, which are states containing a set number of known particles. In order to construct these states, we first have to quantise our fields. The method to do this is called *canonical quantisation*; for a full description of this, the interested reader should consult the excellent discussion in [50] or [54]. The upshot is that our field $\psi$ is promoted to a quantum mechanical operator $\hat{\psi}$ that can create particles. Since we are going to work with fermions, we quantise the Dirac field of equation 1.7:

$$\hat{\psi}(x) = \int \frac{d^3\vec{p}}{(2\pi)^3 2E_{\vec{p}}} \sum_{s \in spins} \left[\hat{a}_s(\vec{p}) u^{(s)}(\vec{p}) e^{-ip \cdot x} + \hat{b}_s^\dagger(\vec{p}) v^{(s)}(\vec{p}) e^{ip \cdot x}\right], \qquad (1.44)$$

along with the corresponding equation for the conjugate field $\bar{\psi}$. We can interpret the operator $\hat{a}_s(\vec{p})$ ($\hat{b}_s^\dagger(\vec{p})$) as the operator that destroys (creates) a particle (anti-particle) with momentum $\vec{p}$ and spin $s$. We can invert this equation to find forms for these operators in terms of the fields,

$$\hat{a}_s^\dagger(\vec{p}) = \int d^3x e^{-ip\cdot x} \bar{\psi}(x)\gamma^0 u^{(s)}(\vec{p}), \tag{1.45a}$$

$$\hat{b}_s^\dagger(\vec{p}) = \int d^3x e^{-ip\cdot x} \bar{v}^{(s)}(\vec{p})\gamma^0 \psi(x). \tag{1.45b}$$

All of these results are only valid for a free theory, so we need a way to relate them to an interacting theory. If we measure our incoming state well before the interaction takes place, we can expect our field to behave as a free field. A similar statement applies to the final state long after the interaction. In other words, we can create these results for our operators as the limit of a more general, time-dependent operator

$$\hat{a}^\dagger(\vec{p}) = \lim_{t \to \pm\infty} \hat{a}^\dagger(\vec{p}, t), \tag{1.46}$$

where the operators are now *defined* through 1.45. For simplicity, we now consider a process where two fermions interact in the initial state to yield two fermions in the final state (denoted '$2 \to 2$ scattering'), but the results we obtain will be immediately generalisable to $2 \to n$ scattering. We construct our initial and final states in the following way:

$$|i\rangle = \lim_{t \to -\infty} \hat{a}^\dagger(\vec{p}_1, t)\hat{a}^\dagger(\vec{p}_2, t)|0\rangle, \tag{1.47a}$$

$$|f\rangle = \lim_{t \to +\infty} \hat{a}^\dagger(\vec{k}_1, t)\hat{a}^\dagger(\vec{k}_2, t)|0\rangle, \tag{1.47b}$$

where $|0\rangle$ is the vacuum state with no particles present. The object we are interested in computing is

$$\begin{aligned}\langle f|i\rangle &= \langle 0|\hat{a}(\vec{k}_1, \infty)\hat{a}(\vec{k}_2, \infty)\hat{a}^\dagger(\vec{p}_1, -\infty)\hat{a}^\dagger(\vec{p}_2, -\infty)|0\rangle \\ &= \langle 0|\hat{\mathcal{T}}\left(\hat{a}(\vec{k}_1, \infty)\hat{a}(\vec{k}_2, \infty)\hat{a}^\dagger(\vec{p}_1, -\infty)\hat{a}^\dagger(\vec{p}_2, -\infty)\right)|0\rangle,\end{aligned} \tag{1.48}$$

where $\hat{\mathcal{T}}$ is a time-ordering operator, placing all operators at larger times to the left of the expression. We aim to express this formula in terms of the field $\psi$. In

order to do so, consider

$$\begin{aligned}
\hat{a}^\dagger(\vec{p}, \infty) - \hat{a}^\dagger(\vec{p}, -\infty) &= \int_{-\infty}^{+\infty} dt\, \partial_0 a^\dagger(\vec{p}, t) \\
&= \int dt \int d^3x\, \partial_0 \left( e^{-ip\cdot x} \bar{\psi}(x) \gamma^0 u^{(s)}(\vec{p}) \right) \\
&= \int d^4x\, \bar{\psi}(x) \left( \overleftarrow{\partial}_0 - ip_0 \right) \gamma^0 e^{-ip\cdot x} u^{(s)}(\vec{p}) \\
&= \int d^4x\, \bar{\psi}(x) \left( \overleftarrow{\partial}_0 \gamma^0 + ip_j \gamma^j - im \right) e^{-ip\cdot x} u^{(s)}(\vec{p}) \quad (1.49) \\
&= \int d^4x\, \bar{\psi}(x) \left( \overleftarrow{\partial}_0 \gamma^0 - \overrightarrow{\partial}_j \gamma^j - im \right) e^{-ip\cdot x} u^{(s)}(\vec{p}) \\
&= \int d^4x\, \bar{\psi}(x) \left( \overleftarrow{\partial}_0 \gamma^0 + \overleftarrow{\partial}_j \gamma^j - im \right) e^{-ip\cdot x} u^{(s)}(\vec{p}) \\
&= (-i) \int d^4x\, \bar{\psi}(x) \left( i \overleftarrow{\slashed{\partial}} + m \right) e^{-ip\cdot x} u^{(s)}(\vec{p}),
\end{aligned}$$

where we used $(\slashed{p} - m)u(\vec{p}) = 0$ in the fourth line and integration by parts in the second-to-last line. Notice that the factor in the last integral is precisely the one we get from the equation of motion for $\bar{\psi}$ and so the integral would be zero in the free theory as we would expect. The upshot of this calculation is that

$$\hat{a}^\dagger(\vec{p}, -\infty) = \hat{a}^\dagger(\vec{p}, \infty) + i \int d^4x\, \bar{\psi}(x) \left( i \overleftarrow{\slashed{\partial}} + m \right) e^{-ip\cdot x} u^{(s)}(\vec{p}), \quad (1.50)$$

and, via Hermitian conjugation,

$$\hat{a}(\vec{p}, \infty) = \hat{a}(\vec{p}, -\infty) + i \int d^4x\, \bar{u}^{(s)}(\vec{p}) e^{ip\cdot x} \left( -i \overrightarrow{\slashed{\partial}} + m \right) \psi(x). \quad (1.51)$$

By substituting these results into our expression for $\langle f | i \rangle$ we get

$$\begin{aligned}
\langle f | i \rangle = \langle 0 | \hat{\mathcal{T}} \bigg[ &\left( \hat{a}^\dagger(\vec{k}_1, \infty) + i \int d^4x_1\, \bar{\psi}(x_1) \left( i \overleftarrow{\slashed{\partial}} + m \right) e^{-ik_1 \cdot x_1} u^{(r')}(\vec{k}_1) \right) \\
\times &\left( \hat{a}^\dagger(\vec{k}_2, \infty) + i \int d^4x_2\, \bar{\psi}(x_2) \left( i \overleftarrow{\slashed{\partial}} + m \right) e^{-ik_2 \cdot x_2} u^{(s')}(\vec{k}_2) \right) \\
\times &\left( \hat{a}(\vec{p}_1, -\infty) + i \int d^4y_1\, \bar{u}^{(r)}(\vec{p}_1) e^{ip_1 \cdot y_1} \left( -i \overrightarrow{\slashed{\partial}} + m \right) \psi(y_1) \right) \\
\times &\left( \hat{a}(\vec{p}_2, -\infty) + i \int d^4y_2\, \bar{u}^{(s)}(\vec{p}_2) e^{ip_2 \cdot y_2} \left( -i \overrightarrow{\slashed{\partial}} + m \right) \psi(y_2) \right) \bigg] | 0 \rangle.
\end{aligned} \quad (1.52)$$

All terms containing operators will now act on the vacuum and so drop out of the expression, since we define $\hat{a}(\vec{p}) |0\rangle = 0$. What is left is the *LSZ Formula* for

this process:

$$\langle f|i\rangle = (i)^4 \int d^4x_1 d^4x_2 d^4y_1 d^4y_2 e^{i(p_1\cdot y_1 + p_2\cdot y_2 - k_1\cdot x_1 - k_2\cdot x_2)}$$
$$\times [\bar{u}^{(r)}(p_1)(-i\overrightarrow{\partial} + m)]_\alpha [\bar{u}^{(s)}(p_2)(-i\overrightarrow{\partial} + m)]_\beta$$
$$\times \langle 0|\hat{\mathcal{T}}\left(\bar{\psi}_\alpha(y_1)\bar{\psi}_\beta(y_2)\psi_{\alpha'}(x_1)\psi_{\beta'}(x_2)\right)|0\rangle \quad (1.53)$$
$$\times [(i\overleftarrow{\partial} + m)u^{(r')}(k_1)]_{\alpha'}[(i\overleftarrow{\partial} + m)u^{(s')}(k_2)]_{\beta'},$$

where we have written out explicit spinor indices to make clear which operators act on which fields. Therefore, if we know how to calculate the time ordered product of fields then we also know how to calculate the scattering amplitude $\langle f|i\rangle$. We will now see how we can use the path integral formulation to obtain this.

### 1.3.3 Correlation Functions from Path Integrals

Consider a path integral of a general field $\phi$ of the form

$$P = \int_{\phi(-T,\vec{x})=\phi_a(\vec{x})}^{\phi(T,\vec{x})=\phi_b(\vec{x})} \mathcal{D}\phi(x)\,\phi(x_1)\phi(x_2)\exp\left(i\int_{-T}^{T} d^4x\,\mathscr{L}\right). \quad (1.54)$$

We rewrite the integral in a convenient manner:

$$\int \mathcal{D}\phi(x) = \int \mathcal{D}\phi_1(\vec{x}) \int \mathcal{D}\phi_2(\vec{x}) \int \mathcal{D}\phi(x)\delta(\phi(x_1^0,\vec{x}) - \phi_1(\vec{x}))\delta(\phi(x_2^0,\vec{x}) - \phi_2(\vec{x})). \quad (1.55)$$

This decomposition means that the main integral $\mathcal{D}\phi(x)$ is constrained at the times $x_1^0$ and $x_2^0$, but the intermediate configurations $\phi_1, \phi_2$ must be integrated over. This allows us to write our integral as

$$P = \int \mathcal{D}\phi_1(\vec{x})\phi_1(\vec{x}_1) \int \mathcal{D}\phi_2(\vec{x})\phi_2(\vec{x}_2)$$
$$\times \int \mathcal{D}\phi(x)\exp\left(i\int_{max[x_1^0,x_2^0]}^{T} d^4x\,\mathscr{L}\right)$$
$$\times \int \mathcal{D}\phi(x)\exp\left(i\int_{min[x_1^0,x_2^0]}^{max[x_1^0,x_2^0]} d^4x\,\mathscr{L}\right) \quad (1.56)$$
$$\times \int \mathcal{D}\phi(x)\exp\left(i\int_{-T}^{min[x_1^0,x_2^0]} d^4x\,\mathscr{L}\right)$$

We will take $x_2^0 > x_1^0$ for now and later on automatically generate the other possibility via use of the time ordering operator. Each of the three integrals over $\phi(x)$ is simply a transition amplitude $\langle q_b|\hat{U}(t_1 - t_0)|q_a\rangle$. Therefore

$$\begin{aligned} P = \int \mathcal{D}\phi_1(\vec{x})\phi_1(\vec{x}_1) \int \mathcal{D}\phi_2(\vec{x})\phi_2(\vec{x}_2) \\ \times \langle\phi_b(\vec{x})|e^{-i\hat{H}(T-x_2^0)}|\phi_2(\vec{x})\rangle \\ \times \langle\phi_2(\vec{x})|e^{-i\hat{H}(x_2^0-x_1^0)}|\phi_1(\vec{x})\rangle \\ \times \langle\phi_1(\vec{x})|e^{-i\hat{H}(x_1^0+T)}|\phi_a(\vec{x})\rangle \,. \end{aligned} \quad (1.57)$$

The factors $\phi_i(\vec{x}_i)$ can be interpreted as the eigenvalues of the Schrödinger operator $\hat{\phi}_S(\vec{x}_i)$ acting on the state $|\phi_i(\vec{x}_i)\rangle$. Thus

$$\begin{aligned} P = \int \mathcal{D}\phi_1(\vec{x}) \int \mathcal{D}\phi_2(\vec{x}) \\ \times \langle\phi_b(\vec{x})|e^{-i\hat{H}(T-x_2^0)}\hat{\phi}_S(\vec{x}_2)|\phi_2(\vec{x})\rangle \\ \times \langle\phi_2(\vec{x})|e^{-i\hat{H}(x_2^0-x_1^0)}\hat{\phi}_S(\vec{x}_1)|\phi_1(\vec{x})\rangle \\ \times \langle\phi_1(\vec{x})|e^{-i\hat{H}(x_1^0+T)}|\phi_a(\vec{x})\rangle \,. \end{aligned} \quad (1.58)$$

We can now use the completeness relation $\int \mathcal{D}\phi_i(\vec{x}) |\phi_i(\vec{x})\rangle \langle\phi_i(\vec{x})| = \mathbb{1}$ on the $\phi_1, \phi_2$ states and are left with

$$P = \langle\phi_b(\vec{x})|e^{-i\hat{H}(T-x_2^0)}\hat{\phi}_S(\vec{x}_2)e^{-i\hat{H}(x_2^0-x_1^0)}\hat{\phi}_S(\vec{x}_1)e^{-i\hat{H}(x_1^0+T)}|\phi_a(\vec{x})\rangle \,. \quad (1.59)$$

At this point, we recognise that we can relate the Heisenberg operator to the Schrödinger operators via $\hat{\phi}_H(x) = e^{i\hat{H}t}\hat{\phi}_S(\vec{x})e^{-i\hat{H}t}$ and therefore

$$P = \langle\phi_b(\vec{x})|e^{-i\hat{H}T}\hat{\mathcal{T}}\left(\hat{\phi}_H(x_2)\hat{\phi}_H(x_1)\right)e^{-i\hat{H}T}|\phi_a(\vec{x})\rangle \,, \quad (1.60)$$

where the time ordering operator $\hat{\mathcal{T}}$ has been inserted so that we are also including the possibility that $x_1^0 > x_2^0$. The last step is to take the limit where $T$ becomes large. Doing this naively, however, will leave us with an ill-defined limit, so we take the limit $T \to \infty(1 - i\delta)$ with $\delta$ small and positive. Then

$$e^{-i\hat{H}T}|\phi_a\rangle = \sum_n e^{-i\hat{H}T}|n\rangle\langle n|\phi_a\rangle \to \sum_n e^{-iE_n(\infty(1-i\delta))}|n\rangle\langle n|\phi_a\rangle = |0\rangle\langle 0|\phi_a\rangle \,, \quad (1.61)$$

where in the last step we have assumed that the vacuum state has zero energy and thus is the only one that survives out of the sum. Therefore

$$\lim_{T \to \infty(1-i\delta)} P = \langle 0|\hat{\mathcal{T}} \left( \hat{\phi}_H(x_2)\hat{\phi}_H(x_1) \right) |0\rangle \times \langle 0|\phi_a\rangle \langle \phi_b|0\rangle. \qquad (1.62)$$

Comparing this to equation 1.53, we see that we have reproduced the part depending on the time ordered products of fields up to the factors of the overlap between the states $\phi_a, \phi_b$ and the vacuum. We can simply divide these out by evaluating the path integral without the factors of the field. We therefore have the result

$$\langle 0|\hat{\mathcal{T}} \left( \hat{\phi}_H(x_2)\hat{\phi}_H(x_1) \right) |0\rangle = \lim_{T \to \infty(1-i\delta)} \frac{\int \mathcal{D}\phi(x)\, \phi(x_1)\phi(x_2) \exp\left(i \int_{-T}^{T} d^4x\, \mathcal{L}\right)}{\int \mathcal{D}\phi(x) \exp\left(i \int_{-T}^{T} d^4x\, \mathcal{L}\right)}. \qquad (1.63)$$

### 1.3.4 Calculating a Scattering Amplitude

We are almost at the point where we can calculate a scattering amplitude. We have the LSZ formula as a 'master equation' and we have shown how the time-ordered product of fields can be calculated by path integrals. What remains, then, is a method for calculating such integrals. To do this, we define a *generating functional*

$$\mathcal{Z}_0 \equiv \int \mathcal{D}\phi\, e^{i \int d^4x (\mathcal{L}_0 + J\phi)}, \qquad (1.64)$$

where $\mathcal{L}_0$ is a free Lagrangian, the Lagrangian for a field with no interactions (i.e. the field satisfies the free equation of motion at all times). $J$ is a source term that will turn out to be very useful for our calculations. To demonstrate the procedure, we will take a simple explicit example of the Lagrangian for a real scalar field,

$$\mathcal{L}_{0,RS} = \frac{1}{2}\partial^\mu \phi \partial_\mu \phi - \frac{1}{2}m^2\phi^2. \qquad (1.65)$$

We saw in the previous section that we had to analytically continue the time to get a well-defined limit. This can also be achieved by making the substitution $m^2 \to m^2 - i\delta$. Performing a Fourier transform on the source and field, we can

write our free action as

$$\begin{aligned} S_{0,RS}(J) &\equiv \int d^4x (\mathscr{L}_{0,RS} + J\phi) \\ &= \frac{1}{2} \int \frac{d^4k}{(2\pi)^4} \left( \tilde{\phi}(k)(k^2 - m^2 + i\delta)\tilde{\phi}(-k) + \tilde{J}(k)\tilde{\phi}(-k) + \tilde{J}(-k)\tilde{\phi}(k) \right). \end{aligned}$$
(1.66)

We shift our variable to $\tilde{\eta}(k) = \tilde{\phi}(k) + \frac{1}{k^2-m^2+i\delta}\tilde{J}(k)$ in order to complete the square and therefore arrive at

$$S_{0,RS} = \frac{1}{2} \int \frac{d^4k}{(2\pi)^4} \left( \tilde{\eta}(k)(k^2 - m^2 + i\delta)\tilde{\eta}(-k) - \frac{\tilde{J}(k)\tilde{J}(-k)}{k^2 - m^2 + i\delta} \right). \quad (1.67)$$

The only $\eta$ dependence is in the first quadratic term. We recognise this as a Gaussian functional integral over $\eta$ and so will correspond to the overall normalisation of $\mathcal{Z}_0$, which can be adjusted by properly defining the measure of the functional integral such that $\mathcal{Z}_0(J = 0) = 1$. In fact, from the previous section we know that the relevant quantity is the ratio of correlation functions and so any overall factors like this normalisation would cancel out in any case. For simplicity and clarity, we choose to do the adjustment and so

$$\mathcal{Z}_0 = \exp\left( -\frac{i}{2} \int \frac{d^4k}{(2\pi)^4} \frac{\tilde{J}(k)\tilde{J}(-k)}{k^2 - m^2 + i\delta} \right). \quad (1.68)$$

In configuration space, we can write this result as

$$\mathcal{Z}_0 = \exp\left( -\frac{1}{2} \int d^4x \int d^4y \, J(x) D_F(x-y) J(y) \right), \quad (1.69)$$

with

$$D_F(x-y) = \int \frac{d^4k}{(2\pi)^4} e^{-ik(x-y)} \frac{i}{k^2 - m^2 + i\delta}. \quad (1.70)$$

This object is called the *Feynman propagator* and will appear often in calculations. The last ingredient is to devise a technique for calculating a general correlation function from the generating functional. This is achieved with *functional derivatives*, defined such that

$$\frac{\delta}{\delta J(x)} J(y) = \delta^4(x-y), \quad \frac{\delta}{\delta J(x)} \int d^4y \, J(y)\phi(y) = \phi(x). \quad (1.71)$$

We can then compute a two-point function in the free theory,

$$
\begin{aligned}
\int \mathcal{D}\phi\, \phi(x_1)\phi(x_2) e^{i\int d^4x \mathscr{L}_0} &= \frac{1}{i}\frac{\delta}{\delta J(x_1)}\frac{1}{i}\frac{\delta}{\delta J(x_2)} \int \mathcal{D}\phi\, e^{i\int d^4x(\mathscr{L}_0+J\phi)}\bigg|_{J=0} \\
&= \frac{1}{i}\frac{\delta}{\delta J(x_1)}\frac{1}{i}\frac{\delta}{\delta J(x_2)} \mathcal{Z}_0(J)\bigg|_{J=0}.
\end{aligned}
\quad (1.72)
$$

Such a procedure can be extended to an arbitrary number of fields in our correlation function. After taking the derivatives, this expression yields $D_F(x_1 - x_2)$ and so we see why the nomenclature 'propagator' is used: the two-point correlation function in the free theory is interpreted as the propagation of a particle between the two points.

## 1.4 Feynman Rules (in QCD)

Though there are more aspects to consider in order to be completely accurate, we will instead leave the general QFT discussion here and move onto the main method for how particle physics calculations are performed: the evaluation of correlation functions in an interaction Lagrangian. For the reader interested in the fuller story, we defer to [50] or [54].

We can always split a general Lagrangian $\mathscr{L}$ into a 'free' part $\mathscr{L}_0$, which we have seen can be solved exactly, and an interacting part $\mathscr{L}_I$. This means our generating functional will always have the form

$$
\begin{aligned}
\mathcal{Z}(J) &= \int \mathcal{D}\phi\, e^{i\int d^4x(\mathscr{L}+J\phi)} \\
&= \int \mathcal{D}\phi\, e^{i\int d^4x(\mathscr{L}_0+\mathscr{L}_I+J\phi)}.
\end{aligned}
\quad (1.73)
$$

As an example, we go back to our simple Lagrangian with a real scalar field (equation 1.67) and introduce an interaction Lagrangian of the form $\mathscr{L}_I = g\phi^3$. Since any factors of the field can be generated by taking functional derivatives, instead of adding this interaction to the free Lagrangian we can equivalently define

$$
\mathcal{Z}_1 \propto \exp\left(ig \int d^4y \left(\frac{1}{i}\frac{\delta}{\delta J(y)}\right)^3\right) \int \mathcal{D}\phi\, e^{i\int d^4x(\mathscr{L}_0+J\phi)}.
\quad (1.74)
$$

This equation cannot be solved exactly, but if $g$ is small enough we can expand the exponential and order-by-order perform the functional derivatives, stopping at some point. In this way, we are considering the interaction term as a perturbation to the free Lagrangian and hence the method is called *perturbation theory*. It should be clear that as the calculation proceeds to higher and higher orders in perturbation theory, the number of terms to deal with gets uncontrollably large. Thankfully, for many processes it is generally true that the first few orders tend to be enough to give us a good description and so higher order terms are not 'needed'. However, there are some instances where perturbation theory breaks down, which we explore in detail in Chapter 2, and that provides the main motivation for the work in this thesis.

Overlooking that for now, we would appear to have a formula which we can generally apply. This is the case, but having to keep track of all the terms that come from our functional derivatives and distinguishing which of these terms are the same as others is a very time-consuming and error-prone task. Luckily, *Feynman rules* and *Feynman graphs* provide us with an excellent, intuitive tool to deal with perturbation theory.

The Feynman rules are a set of instructions for drawing a Feynman diagram, which itself represents a mathematical equation that can be written down and evaluated yielding part or possibly all of the transition amplitude $\langle f|i\rangle$ at a given order in perturbation theory. The Feynman rules themselves will depend on the Lagrangian of any given theory we are interested in. To demonstrate the procedure for generating Feynman rules and diagrams, let us consider a simple Lagrangian of the form

$$\mathscr{L}_{example} = \frac{1}{2}\partial_\mu\phi\partial^\mu\phi - \frac{1}{2}m^2\phi^2 + \frac{g}{3!}\phi^3. \tag{1.75}$$

The process then proceeds schematically as follows:

1. Identify the entire field content of the Lagrangian. In this case, we have only one field $\phi$. For each field, identify the propagator terms, which are terms that contain exactly two instances of the field. For our Lagrangian, these are the first two terms. If such a collection of terms exists, we can represent the propagation of a field from one point to another diagrammatically by some line. With our example, we will simply represent the propagation of $\phi$ from one point to another by a solid black line.

2. All other terms represent interactions, with the number of fields in each term corresponding to the number of fields present in the interaction. Here, we have only one interaction term with three instances of the field $\phi$, which means our only allowed interaction is between three particles of the field $\phi$.

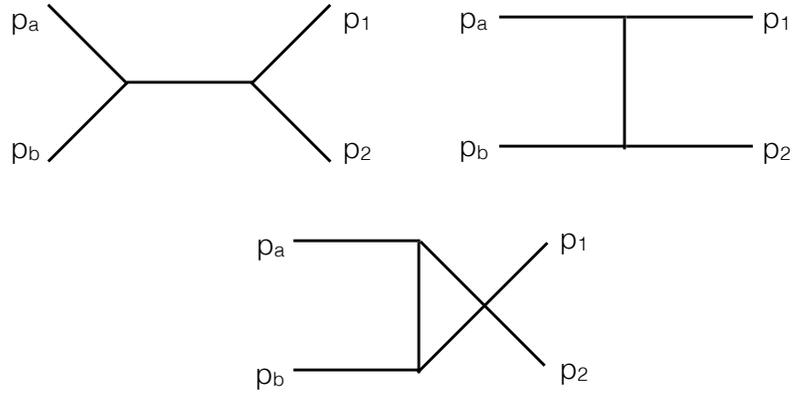

**Figure 1.2** *Example Feynman diagrams. The top-left diagram is called the s-channel, the top-right the t-channel and the bottom the u-channel. All momenta flow from left to right.*

3. Consider the type of process to be described. For instance, here we could be interested in calculating the transition amplitude for two particles of the field $\phi$ to scatter off each other into again two particles of the field $\phi$ (more compactly, a $2 \to 2$ process).

4. Assuming for the time being we are interested in the lowest order of perturbation theory, draw all possible diagrams with the lowest number of interaction points for the process (also sometimes known as *tree level*), subject to the propagation and interaction rules discussed. We will use the convention where time is flowing from left to right through the diagram. It is almost always more useful to work in momentum space, so also label the particles with some momentum values.

For our example theory, the diagrams at leading order, as specified by following these rules, are shown in figure 1.2.

The topology of these diagrams is commonly seen in many physical theories as well; they have the special names *s*-channel, *t*-channel and *u*-channel. These names come from the *Mandelstam variables* which refer to the invariant mass present in the intermediate propagator in each diagram, fixed by momentum conservation at each vertex; $s = (p_a + p_b)^2$, $t = (p_a - p_1)^2$ and $u = (p_a - p_2)^2$.

Now that we have seen the general procedure for drawing diagrams, we need mathematical rules to convert these diagrams into calculable expressions. These are the famous *Feynman rules* and while there is a general way of deriving them for any given theory we will jump straight to the rules we need for the theory of QCD that the rest of the thesis depends upon:

1. *External Fermion Lines.* For an external quark, associate a factor $u(p)$ for incoming and $\bar{u}(p)$ for outgoing. For external anti-quarks, associate a factor $\bar{v}(p)$ for incoming and $v(p)$ for outgoing. To ensure consistent multiplication in the spinor indices, follow a quark line backwards through the diagram.

2. *External Gluon Lines.* Associate a factor $\varepsilon_\mu(p)$ for incoming and $\varepsilon_\mu^*(p)$ for outgoing. These objects are the so-called *polarisation vectors* of the gluons and arise from solving their equations of motion.

3. *Internal propagators.* For each internal propagator, associate the relevant factor as detailed in table 1.1.

4. *Vertices.* For each vertex, associate the relevant factor as detailed in table 1.1.

5. *Unconstrained momentum.* Beyond first order in perturbation theory, Feynman graphs will contain momenta that are not fixed by momentum conservation, resulting in what are called *loops*. For each such momenta $k$, include an integration $\int \frac{d^4k}{(2\pi)^4}$.

6. *Extra signs.* Each fermion or anti-fermion loop comes with a factor of (-1) and each anti-fermion line that flows from the initial state to the final state also comes with a factor of (-1). These rules result from the fact the fermion operators anti-commute; an explicit demonstration of precisely why we need these extra considerations will not be shown.

Following these rules and adding together all diagrams results in the quantity $iM$, where $M$ is the *matrix element* or *amplitude* $\langle f|i\rangle$ at the calculated order in the coupling expansion. Later on, we will see exactly how this relates to a physical quantity.

### 1.4.1 $qQ \to qQ$ at Leading Order

As an explicit exercise in performing QCD calculations with Feynman rules, let us take the example of the elastic scattering of two quarks of different flavours. This process is particularly simple because there is only one contributing Feynman diagram, shown in figure 1.3.

| Diagrammatic Element | Description | Feynman Rule |
|---|---|---|
| 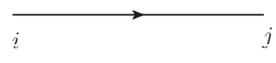 | Quark Propagator with fundamental colour index $i \to j$, momentum $p$ and mass $m_f$ | $\dfrac{i\delta_{ij}}{\not{p}-m_f} = \dfrac{i\delta_{ij}(\not{p}+m_f)}{p^2-m_f^2}$ |
| 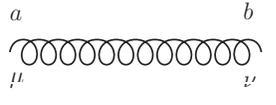 | Gluon Propagator (taken in Feynman gauge) with adjoint colour index $a \to b$ and momentum $p$ | $\dfrac{-i\delta^{ab}\eta^{\mu\nu}}{p^2}$ |
| 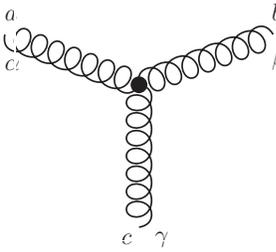 | Three-Gluon Vertex. All momenta taken as incoming, with $k_1$ associated with the gluon with adjoint index $a$, $k_2$ with $b$ and $k_3$ with $c$ | $-g_s f^{abc}(\eta^{\alpha\beta}(k_1-k_2)^\gamma + \eta^{\beta\gamma}(k_2-k_3)^\alpha + \eta^{\gamma\alpha}(k_3-k_1)^\beta)$ |
| 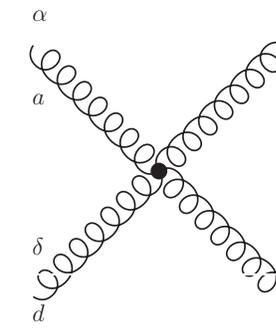 | Four-Gluon Vertex with adjoint indices $a,b,c,d$ | $-ig_s^2(f^{ade}f^{cde}(\eta^{\alpha\gamma}\eta^{\beta\delta}-\eta^{\alpha\delta}\eta^{\beta\gamma}) + f^{ace}f^{bde}(\eta^{\alpha\beta}\eta^{\gamma\delta}-\eta^{\alpha\delta}\eta^{\gamma\beta}) + f^{ade}f^{bce}(\eta^{\alpha\beta}\eta^{\delta\gamma}-\eta^{\alpha\gamma}\eta^{\delta\beta}))$ |
| 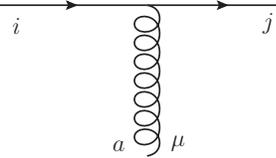 | Quark-Gluon Vertex with fundamental indices $i,j$ and adjoint index $a$ | $-ig_s \gamma^\mu t^a_{ij}$ |

**Table 1.1**  *QCD Feynman Rules for propagators and vertices.*

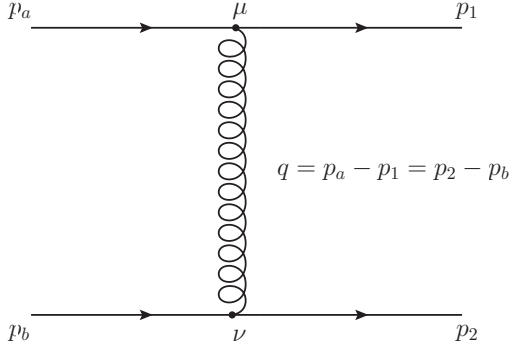

**Figure 1.3** *Two quarks of different flavour scattering via the exchange of a gluon in the t-channel.*

Applying the Feynman rules, we arrive at the expression for the amplitude:

$$
\begin{aligned}
iM &= \bar{u}(p_2)(-ig_s t^b \gamma^\nu) u(p_b) \left( \frac{-i\eta_{\mu\nu}\delta^{ab}}{(p_a - p_1)^2} \right) \bar{u}(p_1)(-ig_s t^a \gamma^\mu) u(p_a) \\
&= \left( \frac{t^b \delta^{ab} t^a i g_s^2}{(p_a - p_1)^2} \right) \bar{u}(p_2)\gamma_\mu u(p_b) \bar{u}(p_1)\gamma^\mu u(p_a) \\
&= \left( \frac{t^a \delta^{ab} t^b i g_s^2}{\hat{t}} \right) \bar{u}(p_2)\gamma_\mu u(p_b) \bar{u}(p_1)\gamma^\mu u(p_a),
\end{aligned}
\qquad (1.76)
$$

where we have for now suppressed the colour and spin dependence of the quark spinors. In general, there is no control over the colours and spins of the particles in our experiment and so we should take this into account, by averaging over all initial state spins/colours and summing over all final state spins/colours. From the general laws of Quantum Mechanics [30], we know that any physical quantity must be proportional to $|M|^2$. Since each helicity and colour configuration are physically distinguishable (that is, they do not interfere with each other quantum mechanically, we are just ignorant of what the states are), this sum/average must be done at the $|M|^2$ level:

$$
|M|^2 = \frac{1}{9} \times \frac{1}{4} \times \frac{g_s^4}{\hat{t}^2} \times \sum_{colours} \sum_{spins} |\tilde{M}|^2, \qquad (1.77)
$$

where we have extracted some factors that don't depend on spin or colour out of $\tilde{M}$, which represents the pre-averaged matrix element. We will begin by performing the spin sum. It will be useful to keep track of spinor indices in this calculation, which are implicitly summed over, and to introduce the notation

$u(p) \equiv u_p$ to get

$$\sum_{spins} |\tilde{M}|^2 = \sum_{spins} [\bar{u}_2^{i_1} \gamma_\mu^{i_1 j_1} u_b^{j_1}][\bar{u}_1^{i_2} \gamma^{\mu,i_2 j_2} u_a^{j_2}][\bar{u}_2^{i_3} \gamma_\nu^{i_3 j_3} u_b^{j_3}]^\dagger [\bar{u}_1^{i_4} \gamma^{\nu,i_4 j_4} u_a^{j_4}]^\dagger$$
$$= \sum_{spins} [\bar{u}_2^{i_1} \gamma_\mu^{i_1 j_1} u_b^{j_1}][\bar{u}_1^{i_2} \gamma^{\mu,i_2 j_2} u_a^{j_2}][\bar{u}_b^{i_3} \gamma_\nu^{i_3 j_3} u_2^{j_3}][\bar{u}_a^{i_4} \gamma^{\nu,i_4 j_4} u_1^{j_4}], \quad (1.78)$$

where we have used $(\gamma^0)^\dagger = \gamma^0$ and $(\gamma^\mu)^\dagger = \gamma^0 \gamma^\mu \gamma^0$. At this point we introduce the important result

$$\sum_{spins} u_p \bar{u}_p = \slashed{p} + m, \quad (1.79)$$

which amounts to the statement that the set of $u$ functions is complete and, for simplicity, we will take the mass of the quarks in our calculation to be zero. This is a good approximation if the momenta of the particles involved in the scattering are much larger than the value of the mass. We will always assume this is the case unless otherwise stated in the entirety of this thesis. Using this result to rewrite our amplitude and being careful with spinor indices, we see

$$\sum_{spins} |\tilde{M}|^2 = [\slashed{p}_b^{j_1 i_3} \gamma_\nu^{i_3 j_3} \slashed{p}_2^{j_3 i_1} \gamma_\mu^{i_1 j_1}][\slashed{p}_a^{j_2 i_4} \gamma^{\nu,i_4 j_4} \slashed{p}_1^{j_4 i_2} \gamma^{\mu,i_2 j_2}]$$
$$= \text{tr}[\gamma_\alpha \gamma_\nu \gamma_\beta \gamma_\mu] p_b^\alpha p_2^\beta \times \text{tr}[\gamma^\rho \gamma^\nu \gamma^\sigma \gamma^\mu] p_{a,\rho} p_{1,\sigma}. \quad (1.80)$$

What remains is to evaluate the traces of the products of gamma matrices. There are a whole host of so-called *trace theorems* [50], of which we quote one result:[3]

$$\text{tr}[\gamma^\alpha \gamma^\nu \gamma^\beta \gamma^\mu] = 4(\eta^{\alpha\nu} \eta^{\beta\mu} - \eta^{\alpha\beta} \eta^{\nu\mu} + \eta^{\alpha\mu} \eta^{\nu\beta}). \quad (1.81)$$

Applying this result to our calculation, we arrive at

$$\sum_{spins} |\tilde{M}|^2 = 16(p_{b,\nu} p_{2,\mu} - p_b \cdot p_2 \eta_{\nu\mu} + p_{b,\mu} p_{2,\nu}) \times (p_a^\nu p_1^\mu - p_a \cdot p_1 \eta^{\nu\mu} + p_a^\mu p_1^\nu)$$
$$= 32 \left( (p_b \cdot p_a)(p_1 \cdot p_2) + (p_b \cdot p_1)(p_2 \cdot p_a) \right), \quad (1.82)$$

which is a simple result. The final ingredient of the amplitude is the colour sum. Since this is the first time we are performing such a calculation, it is instructive

---
[3]This result only applies in a 4-dimensional spacetime, which we have in this problem, but can be generalised to $D$ dimensions if required.

to do this in detail. We write out the colour indices in full [53]:

$$
\begin{aligned}
\sum_{colours} |\tilde{M}|^2 &\sim \sum_{colours} [t^a_{q_1,q_a} \delta^{ab} t^b_{q_2,q_b}][t^c_{q_1,q_a} \delta^{cd} t^d_{q_2,q_b}]^\dagger \\
&= \sum_{a,b,c,d,q_1,q_a,q_2,q_b} [t^a_{q_1,q_a} \delta^{ab} t^b_{q_2,q_b}][t^c_{q_a,q_1} \delta^{cd} t^d_{q_b,q_2}] \\
&= \sum_{a,c,q_1,q_a,q_2,q_b} [t^a_{q_1,q_a} t^a_{q_2,q_b}][t^c_{q_a,q_1} t^c_{q_b,q_2}] \\
&= \sum_{a,c} \mathrm{tr}(t^a t^c) \mathrm{tr}(t^a t^c) \\
&= \sum_{a,c} \frac{1}{2}\delta^{ac} \frac{1}{2}\delta^{ac} \\
&= \sum_a \frac{1}{4}\delta^{aa} \\
&= 2.
\end{aligned}
\tag{1.83}
$$

Combining the results, we find an expression for the full amplitude:

$$
\begin{aligned}
|M|^2 &= \frac{16 g_s^4}{9\hat{t}^2} \left( (p_b \cdot p_a)(p_1 \cdot p_2) + (p_b \cdot p_1)(p_2 \cdot p_a) \right) \\
&= g_s^4 \times \frac{4}{9} \left( \frac{\hat{s}^2 + \hat{u}^2}{\hat{t}^2} \right).
\end{aligned}
\tag{1.84}
$$

As a matter of interest, this result is the same as the one we would get from electron-muon scattering, except with a different coupling strength and no overall colour factor. We should expect this because our process did not involve any parts where the gluon interacts with itself and so behaves similarly to a photon in this interaction. If we were to do a next-to-leading order calculation, however, this would no longer be true and QCD loop calculations are, in general, difficult to compute.

## 1.5 Cross-Sections in Proton-Proton Collisions

In a collider experiment, the physical quantity is the *cross-section*, not the squared matrix element itself. The two are intimately related, however, by *Fermi's Golden Rule*, which in this case states

$$
d\hat{\sigma} = S \times \frac{|M|^2}{F} \times (2\pi)^4 \delta^{(4)}(p_a + p_b - \sum_{f=1}^n p_f) \times \prod_{i=1}^n \frac{d^3 \vec{p}_i}{2E_i (2\pi)^3}.
\tag{1.85}
$$

$|M|^2$ is the matrix element squared for the process of interest and contains all the dynamical information. $F = 4\sqrt{(p_a \cdot p_b)^2 - m_a m_b}$ is a factor that accounts for the flux of incoming particles. The delta function ensures momentum conservation in the process. The integral measures are for the *phase space* integrals over the final particles and, finally, $S$ is a statistical factor that serves to avoid over-counting in processes with indistinguishable outgoing particles – for each group of $s$ identical particles in the final state, $S$ gains a factor $1/s!$. In theory, then, we can insert any matrix element calculated at a certain order in our perturbation theory and perform the phase space integrals to yield the cross-section for any process we desire. In practice, this is not as simple as it sounds. For instance, as the number of final state particles increases, an analytic integration over their momenta becomes more and more difficult (the matrix elements as well become more difficult to compute – a point we will revisit later).

Furthermore, in QCD we cannot collide two quarks together with a well-defined energy, as our matrix element calculation might suggest, because of a property called *confinement* – because of how the strong force interacts, it is impossible to observe a free quark or gluon[4]. Instead, we have to collide hadrons together (for the LHC, specifically protons) which are a dynamic soup of quarks, anti-quarks and gluons (from here on, *partons*). However, if the energy of the collider is large enough (which it certainly is in the LHC), then we can model a proton-proton collision as a collision of two partons within each of the protons, each of which carrying some fraction of the total proton momentum. This treatment is essentially a probabilistic one and has led to the development of *Parton Distribution Functions* (PDFs) [51], which in itself is an area of intense research. For our purposes though, we need not discuss in detail the field of PDF research and instead just use the property that we can write our total proton-proton cross-section as some convolution

$$d\sigma_{pp} = \sum_{f_a, f_b} \int_0^1 dx_a \int_0^1 dx_b f_a(x_a) f_b(x_b) \times d\hat{\sigma}_{partonic}, \qquad (1.86)$$

where we interpret $x_a, x_b$ as the fraction of the total momentum carried by the parton from each of the protons and $f_a(x_a), f_b(x_b)$ as the value of the PDF for a parton of flavour $a, b$ carrying a momentum fraction $x_a, x_b$.

---

[4]Though confinement has been phenomenologically established, it is not understood from a purely theoretical viewpoint. This is essentially because it would involve calculating without the use of perturbation theory, which is a very tough task. It is therefore an outstanding problem set by the Clay Mathematics Institute to prove confinement.

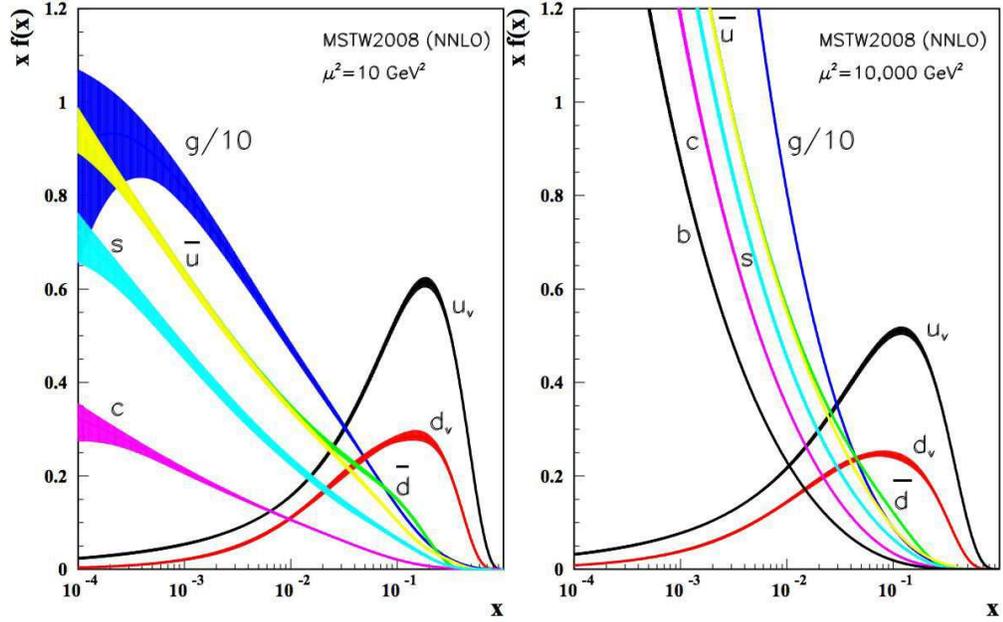

**Figure 1.4** *An example of a PDF set and how the value of the PDF changes depending on the scale probed. Figure from a review by the PDG [48].*

This current formula does not explain all the necessary considerations, however. For example, we should expect that the form of the distribution functions has some dependence on the energy scale of the scattering, which we will call $Q^2$. The reason for this is that, as $Q^2$ is increased, one is able to probe the interior of the proton more and more precisely and thus the likelihood of picking out a certain parton of a specific momentum will depend on this scale. An example is shown in figure 1.4 where the behaviour of the PDF with respect to $x$ is clearly different depending on the scale. A better equation would then be

$$d\sigma_{pp} = \sum_{f_a, f_b} \int_0^1 dx_a \int_0^1 dx_b f_a(x_a, Q^2) f_b(x_b, Q^2) \times d\hat{\sigma}_{partonic}, \tag{1.87}$$

for some value of $Q^2$. What that value should be, however, is somewhat mysterious – given a hard scattering, what is the relevant scale at which the PDF is probed? This ambiguity leads to the calculation of *scale variations*, which we will return to in a later chapter. Before that, we must also consider how confinement affects the final state of our scattering.

The partonic cross-section would lead us to believe that there are free partons after the scattering that we could detect, which goes against the principle of confinement. Indeed, we observe that the final state of partonic scatterings consists of objects called *jets*, which are 'cones' of hadrons and other particles caused by the *hadronisation* of the partons produced in a scattering. A further discussion of the difference between partons and jets will also be discussed towards the end of chapter 2, but let us for now consider some 'n-jet function' that 'passes' an event (i.e. has value equal to 1) if there are $n$ jets present in the final state and 'fails' (has value 0) otherwise. Such a function is called an *exclusive* function, because only events with exactly $n$ jets will contribute, but we could also imagine an *inclusive* function that will 'pass' events with at least $n$ jets. We could therefore write

$$d\sigma_{pp \to n-jet}^{inc/exc} = \sum_{f_a, f_b} \int_0^1 dx_a \int_0^1 dx_b f_a(x_a, Q^2) f_b(x_b, Q^2) \times d\hat{\sigma}_{partonic} \times \mathcal{J}(\text{n-jet}^{inc/exc}). \tag{1.88}$$

Generally speaking, it is not known how many jets $n$ will be produced by a scattering leaving $m$ partons in the final state and this (along with the more fundamental question of how to even properly *define* a jet) is again an area of significant research. The up-shot is that, for many reasons, the calculation of QCD cross-sections is difficult and not without ambiguity. We will later see how, despite this, we can still get physically relevant results from the theory.

## 1.6 Spinor Helicity Formalism

We conclude the chapter by discussing the *spinor helicity formalism* for the calculation of amplitudes. The formalism makes the expression of amplitudes involving massless particles[5] less cumbersome and we will make repeated use of it throughout the remainder of this thesis. *Helicity* is the projection of a particle's spin along its momentum vector and can have two values: 'negative' when the spin is anti-aligned and 'positive' when it is aligned. We would clearly like to describe this projection using some operator. This can be achieved by introducing a new gamma matrix $\gamma^5 = i\gamma^0 \gamma^1 \gamma^2 \gamma^3$, which can be used to project out helicity states

---

[5]The formalism can also be extended to include massive particles, but this is not its primary use and we will always be working with massless entities (either effectively or exactly) in this work.

as follows:

$$u_{\pm}(p_i) = (1 \pm \gamma^5)u(p_i) \equiv |i^{\pm}\rangle \tag{1.89a}$$

$$v_{\mp}(p_i) = (1 \pm \gamma^5)v(p_i) \equiv |i^{\pm}\rangle \tag{1.89b}$$

$$\bar{u}_{\pm}(p_i) = \bar{u}(p_i)(1 \mp \gamma^5) \equiv \langle i^{\pm}| \tag{1.89c}$$

$$\bar{v}_{\mp}(p_i) = \bar{v}(p_i)(1 \mp \gamma^5) \equiv \langle i^{\pm}|. \tag{1.89d}$$

We also define the basic spinor products as

$$\bar{u}_-(p_i)u_+(p_j) = \bar{v}_+(p_i)v_-(p_j) = \langle i^-|j^+\rangle \equiv \langle ij\rangle \tag{1.90a}$$

$$\bar{u}_+(p_i)u_-(p_j) = \bar{v}_-(p_i)v_+(p_j) = \langle i^+|j^-\rangle \equiv [ij]. \tag{1.90b}$$

The final object we will need to deal with is a *current* with the form $\bar{u}\gamma^{\mu}u$. Because $\gamma^5$ anti-commutes with the other gamma matrices, only currents where the two spinors have the same helicity are non-zero:

$$\bar{u}_{\pm}(p_i)\gamma^{\mu}u_{\pm}(p_j) = \bar{v}_{\mp}(p_i)\gamma^{\mu}v_{\mp}(p_j) = \langle i^{\pm}|\mu|j^{\pm}\rangle \equiv J_{ij}^{\pm,\mu}. \tag{1.91}$$

There are many identities that these objects satisfy [31] and we list here a select few that we will use repeatedly in our calculations:

$$\langle ij\rangle = -\langle ji\rangle \tag{1.92a}$$

$$[ij] = -[ji] \tag{1.92b}$$

$$\langle ij\rangle^* = [ji] \tag{1.92c}$$

$$\langle i^+|\mu|j^+\rangle^{\dagger} = \langle j^+|\mu|i^+\rangle \tag{1.92d}$$

$$\langle i^{\pm}|\mu|i^{\pm}\rangle = 2p_i^{\mu} \tag{1.92e}$$

$$\langle i^+|\mu|j^+\rangle = \langle j^-|\mu|i^-\rangle \tag{1.92f}$$

$$\langle i^+|\mu|j^+\rangle \langle k^+|\mu|l^+\rangle = 2\langle jl\rangle [ki] \tag{1.92g}$$

$$\langle ij\rangle [ji] = 2p_i \cdot p_j = s_{ij} \tag{1.92h}$$

$$\slashed{p}_i = |i^+\rangle\langle i^+| + |i^-\rangle\langle i^-| \tag{1.92i}$$

$$\langle ij\rangle \langle kl\rangle = \langle ik\rangle \langle jl\rangle + \langle il\rangle \langle kj\rangle. \tag{1.92j}$$

Given their importance to this thesis, we devote some space here to the derivation of some of these results. The easiest way to do so is to pick an explicit representation for the spinors and gamma matrices and work in individual components. It is instructive to work in *light cone coordinates* where we make the substitution $p^{\pm} = E \pm p_z$ and parametrise momenta transverse to the beam axis as $p_\perp = p_x + ip_y$. For outgoing particles with four-momentum $p$, we use

$$u^+(p) = \begin{pmatrix} \sqrt{p^+} \\ \sqrt{p^-} \frac{p_\perp}{|p_\perp|} \\ 0 \\ 0 \end{pmatrix} \tag{1.93a}$$

$$u^-(p) = \begin{pmatrix} 0 \\ 0 \\ \sqrt{p^-} \frac{p_\perp^*}{|p_\perp|} \\ -\sqrt{p^+} \end{pmatrix}. \tag{1.93b}$$

For incoming particles with 4-momentum $p$ moving along the positive light cone direction, we use:

$$u^+(p) = \begin{pmatrix} \sqrt{p^+} \\ 0 \\ 0 \\ 0 \end{pmatrix} \tag{1.94a}$$

$$u^-(p) = \begin{pmatrix} 0 \\ 0 \\ 0 \\ -\sqrt{p^+} \end{pmatrix}. \tag{1.94b}$$

For incoming particles with 4-momentum $p$ moving in the negative light cone direction, we use:

$$u^+(p) = \begin{pmatrix} 0 \\ -\sqrt{p^-} \\ 0 \\ 0 \end{pmatrix} \tag{1.95a}$$

$$u^-(p) = \begin{pmatrix} 0 \\ 0 \\ -\sqrt{p^-} \\ 0 \end{pmatrix}. \tag{1.95b}$$

We also use the following representation of the gamma matrices:

$$\gamma^0 = \begin{pmatrix} 0 & 0 & 1 & 0 \\ 0 & 0 & 0 & 1 \\ 1 & 0 & 0 & 0 \\ 0 & 1 & 0 & 0 \end{pmatrix} \tag{1.96a}$$

$$\gamma^1 = \begin{pmatrix} 0 & 0 & 0 & -1 \\ 0 & 0 & -1 & 0 \\ 0 & 1 & 0 & 0 \\ 1 & 0 & 0 & 0 \end{pmatrix} \tag{1.96b}$$

$$\gamma^2 = \begin{pmatrix} 0 & 0 & 0 & i \\ 0 & 0 & -i & 0 \\ 0 & -i & 0 & 0 \\ i & 0 & 0 & 0 \end{pmatrix} \tag{1.96c}$$

$$\gamma^3 = \begin{pmatrix} 0 & 0 & -1 & 0 \\ 0 & 0 & 0 & 1 \\ 1 & 0 & 0 & 0 \\ 0 & -1 & 0 & 0 \end{pmatrix}. \tag{1.96d}$$

With our conventions defined, we can move on to some derivations. An identity that is used often is $\langle i^\pm | \mu | i^\pm \rangle = 2p_i^\mu$ so this would be a sensible one to prove. If we have an incoming particle with momentum $p$, then the first element of the product is (we take positive helicity particles)

$$\langle i^+ | 0 | i^+ \rangle = (u_i^+)^\dagger \gamma^0 \gamma^0 u_i^+ = (u_i^+)^\dagger u_i^+ = p_i^+. \tag{1.97}$$

By inspection, we see that such a product will only be non-zero if the multiplication $\gamma^0 \gamma^\mu$ applied to $u_i^+$ is such that there is a non-zero component in the first component of the resulting spinor. In other words, the product $\gamma^0 \gamma^\mu$ must have a non-zero entry in the top-left. Explicit calculation of $\gamma^0 \gamma^1$ and $\gamma^0 \gamma^2$ shows this not to be the case, and so

$$\langle i^+ | 1 | i^+ \rangle = \langle i^+ | 2 | i^+ \rangle = 0. \tag{1.98}$$

Finally, the product $\gamma^0 \gamma^3$ does have a non-zero entry in the top-left which is equal to 1 and thus

$$\langle i^+ | 3 | i^+ \rangle = \langle i^+ | 0 | i^+ \rangle = p_i^+. \tag{1.99}$$

Converting $p_i^+$ to $E + p_z$ and remembering that the particle is massless and has no transverse component if it is incoming (i.e. $E = |p_z|$), then indeed we see that $\langle i^\pm|\mu|i^\pm\rangle = 2p_i^\mu$. The calculation for outgoing particles is longer because of the presence of the transverse term, but again this relationship is seen to hold. The other identity we will make particularly regular use of is $\langle ij\rangle[ji] = s_{ij}$. To prove this, let us take $p_i$ to be incoming and along the positive direction and $p_j$ to be incoming and along the negative direction. Direct calculation then yields

$$\langle ij\rangle = \bar{u}_i^- u_j^+ = \sqrt{p_i^+ p_j^-}$$
$$[ji] = \bar{u}_j^+ u_i^- = \sqrt{p_i^+ p_j^-}, \qquad (1.100)$$

and thus $\langle ij\rangle[ji] = p_i^+ p_j^- = (E_i + E_i)(E_j + E_j) = 4E_i E_j = s_{ij}$. Again, the calculation involving outgoing particles is longer but the result still holds.

As an actual practical demonstration of the technique, let us repeat the calculation in the previous section in this new language. We will once again sum over all spins (equivalent to summing over all helicities), but for clarity we work first with the case that all particles have positive helicity. Then

$$\begin{aligned} iM_{++++} &= \langle 2^+|\,(-ig_s t^b \gamma^\nu)\,|b^+\rangle \left(\frac{-i\eta_{\mu\nu}\delta^{ab}}{(p_a - p_1)^2}\right) \langle 1^+|\,(-ig_s t^a \gamma^\mu)\,|a^+\rangle \\ &= \left(\frac{t^b \delta^{ab} t^a i g_s^2}{\hat{t}}\right) \langle 2^+|\mu|b^+\rangle \langle 1^+|\mu|a^+\rangle \\ &= \left(\frac{t^b \delta^{ab} t^a i g_s^2}{\hat{t}}\right) 2\,\langle ba\rangle\,[12]. \end{aligned} \qquad (1.101)$$

Since there are two helicity states for each particle, we would expect there to be $2^4 = 16$ different helicity configurations we would have to calculate. However, since the currents $J_{2b}$ and $J_{1a}$ disappear if the quark helicity is not conserved, we only have 4 non-zero configurations. Furthermore, our relation $\langle i^+|\mu|j^+\rangle = \langle j^-|\mu|i^-\rangle$ reduces this to only two independent configurations. For example, the configuration where all particles have negative helicity is the same as the one with all positive helicity, except that we need to take the Hermitian conjugate of the currents. Once we take the absolute value squared of this quantity, this conjugation is irrelevant and so the amplitude contains no new information. Therefore, the only other matrix element we need to calculate is the one where the two incoming quarks have opposite helicities:

$$\begin{aligned} iM_{+-+-} &= \left(\frac{t^b \delta^{ab} t^a i g_s^2}{\hat{t}}\right) \langle 2^-|\mu|b^-\rangle \langle 1^+|\mu|a^+\rangle \\ &= \left(\frac{t^b \delta^{ab} t^a i g_s^2}{\hat{t}}\right) \langle b^+|\mu|2^+\rangle \langle 1^+|\mu|a^+\rangle \\ &= \left(\frac{t^b \delta^{ab} t^a i g_s^2}{\hat{t}}\right) 2 \langle 2a\rangle [1b]. \end{aligned} \qquad (1.102)$$

The colour sum and average must still be performed, of course, but other than that we can simply take the modulus squared, sum and average over the helicities:

$$\begin{aligned} |M|^2 &= \frac{1}{4} \times \frac{2}{9} \times \left(|M_{++++}|^2 + |M_{+-+-}|^2 + |M_{-+-+}|^2 + |M_{----}|^2\right) \\ &= \frac{1}{18} \times \left(2|M_{++++}|^2 + 2|M_{+-+-}|^2\right) \\ &= g_s^4 \times \frac{4}{9} \left(\frac{\langle ba\rangle [12][ab]\langle 21\rangle + \langle 2a\rangle [1b][a2]\langle b1\rangle}{\hat{t}^2}\right) \\ &= g_s^4 \times \frac{4}{9} \left(\frac{s_{ab}s_{12} + s_{a2}s_{1b}}{\hat{t}^2}\right) \\ &= g_s^4 \times \frac{4}{9} \left(\frac{\hat{s}^2 + \hat{u}^2}{\hat{t}^2}\right), \end{aligned} \qquad (1.103)$$

where in the last line we used momentum conservation to equate $s_{ab}$ with $s_{12}$ and $s_{a2}$ with $s_{1b}$ and then substituted in the relevant Mandelstam variable. This is, of course, the same result as before except we did not have to go through the trouble of evaluating the traces of gamma matrices, which is a procedure that can be quite error-prone. Furthermore, we have gained some physical insight into why the terms $\hat{s}$ and $\hat{u}$ appear in the numerator; the former comes from the helicity configuration where the quarks have the same helicity and the latter from the configuration where they are opposite. This simple example showed how easy the formalism is to work with and in the next chapter we will see how it is also powerful, in the sense that we will be able to use it to express matrix elements to all orders.

# Chapter 2

# High Energy Jets

We have shown how we can calculate physical quantities in a QFT by treating the interaction terms in a Lagrangian as a perturbation to the free Lagrangian. The expansion parameter is the strength of the coupling, usually denoted by $g_s$ in QCD, and if this parameter is small enough[1], we can expect that the first few orders of perturbation theory should give us an accurate result. This implicitly assumes, however, that each term in the perturbative series is itself quite small. If these terms were too large, then we would run the risk of the perturbative series being non-convergent and so never be able to have a good description of the process. The natural question to ask is whether it is generally true that the terms in the perturbative series are small enough. We will see in this chapter that this is not always the case and how a technique known as *resummation* can take this into account.

## 2.1 The Problem with Perturbation Theory in the High Energy Limit

In order to proceed, we must define what is meant by the High Energy limit. For a $2 \to n$ process, we define it as the limit where the invariant mass between any two particles is large and where all transverse momenta in the problem are fixed

---

[1]At the level of the physical cross-section, because of the combination of phase space and Fermi's Golden Rule factors, it is often said that the relevant parameter is actually $\alpha_s = g_s^2/4\pi$, which is even smaller.

and much smaller. Formally;

$$\forall_{i,j}: \ s_{ij} \to \infty, \ |p_{i\perp}| \sim |p_{j\perp}|, \tag{2.1}$$

where we have defined $p_\perp = p_x + ip_y$ and $i,j$ run over all final state partons. This limit is also referred to as the *Multi-Regge Kinematic* (MRK) limit. We will investigate what effect this limit has on our calculations and why we should be worried about our perturbative expansion. In order to do so, let us first make some statements about the Mandelstam variables in a $2 \to 2$ process. We introduce the *rapidity* of a particle

$$y = \frac{1}{2} \ln \left( \frac{E + p_z}{E - p_z} \right), \tag{2.2}$$

which is a convenient parameter to work with in this limit. Given the rapidity of a particle and its angle with respect to the beam line (defined as the $z$-axis), we can parametrise an outgoing momentum in the following fashion:

$$p_i = |p_{i\perp}|(\cosh(y_i), \cos(\phi_i), \sin(\phi_i), \sinh(y_i)). \tag{2.3}$$

This allows for the following evaluation of the Mandelstam variable $\hat{s}$:

$$\hat{s} = (p_1 + p_2)^2 = 2|p_{1\perp}||p_{2\perp}|\left(\cosh(\Delta y) - \cos(\Delta \phi)\right), \tag{2.4}$$

with $\Delta y = y_1 - y_2$ and $\Delta \phi = \phi_1 - \phi_2$. Since the MRK limit takes $\hat{s}$ to infinity while keeping the transverse momentum fixed, we must have that $\cosh(\Delta y)$ becomes large and so $|\Delta y| \to \infty$. In other words, the MRK limit is the limit where all pairs of outgoing particles are separated by a large rapidity gap. Since we require that the transverse components of the momenta are much smaller in comparison to $\hat{s}$, then we must have that $p_{a,z} \approx p_{1,z}$, $p_{b,z} \approx p_{2,z}$, $E_a \approx E_1$ and $E_b \approx E_2$. We therefore have that, in the MRK limit,

$$\hat{s} \approx |p_{1\perp}||p_{2,\perp}|\exp(\Delta y) \tag{2.5a}$$
$$\hat{t} = (p_a - p_1)^2 \approx -|p_{1\perp}|^2 \tag{2.5b}$$
$$\hat{u} = (p_a - p_2)^2 \approx (p_1 - p_2)^2 = -\hat{s}. \tag{2.5c}$$

We conclude from this that (since all transverse momenta are assumed to have approximately equal magnitude) it is the quantity $\frac{\hat{s}}{-\hat{t}} \approx \exp(\Delta y)$ that is the relevant variable for high energy scattering. Alternatively, we take $\Delta y$ to be the relevant variable and relate this to $\log(\frac{\hat{s}}{-\hat{t}})$ – this is the basis for what is known as the *Leading Logarithmic* approach to scattering amplitudes.

### 2.1.1 $qQ \to qQ$ at LO and NLO in the High Energy Limit

Since we already have the full leading order result as given in equation 1.84, we can simply apply the MRK limit to that. This gives

$$|M_{qQ \to qQ}^{MRK}|^2 = g_s^4 \times \frac{8}{9} \times \frac{\hat{s}^2}{|p_{1\perp}|^2 |p_{2\perp}|^2}. \tag{2.6}$$

We see that we lose some information about the amplitude (we 'lost' $\hat{u}$, which we saw came from the scattering of opposite helicity quarks, and approximated the full $\hat{t}$) but it is the correct expression in the relevant limit, as shown in figure 2.1, where we plot the full LO result and MRK limit of the $ud \to ud$ amplitude. The momenta are parametrised such that $p_1 = (40\cosh(\Delta), 40, 0, 40\sinh(\Delta))$ and $p_2 = (40\cosh(-\Delta), -40, 0, 40\sinh(-\Delta))$. We also deduce that $M \sim \hat{s}$, a scaling relationship that we will return to later.

For the next-to-leading order calculation, there are a number of diagrams that need to be taken into account, a selection of which are shown in figure 2.2. If we were interested in the full NLO expression, we would have to work out the contribution from each diagram and sum them up along with the LO calculation. However, we are now interested in expressions that are relevant in the MRK regime only, so we should identify which diagrams are leading in that limit[2]; namely, diagrams that have a dependence on the rapidity gap between the two extremal partons. Therefore, we can expect that only loops that have 'knowledge' of this can give rise to an expression leading in $\log(\frac{\hat{s}}{-\hat{t}})$ – hence the terminology 'Leading Logarithmic' (LL). Self-energy diagrams of the quarks, such as the one on the top right of figure 2.2, clearly cannot be LL since the momentum of the other quark line does not enter into it and cannot influence the loop integral. A similar argument holds for the vertex correction diagrams, such as the one in the top left. For the self-energy diagram of the gluon (middle-right of the diagram) the propagators entering the loop are of the order of the transverse scale, which we keep fixed in comparison to growing $\hat{s}$ and so again cannot be leading. This leaves only the bottom two diagrams to be calculated, of which we need only do one because we can relate one to the other (up to a colour factor we can read off) via crossing symmetry. We choose to calculate the bottom-left diagram.

---

[2] This seems like a gauge dependent statement and indeed it is, but the argument that follows is valid for all covariant gauges, which are the only ones we will use.

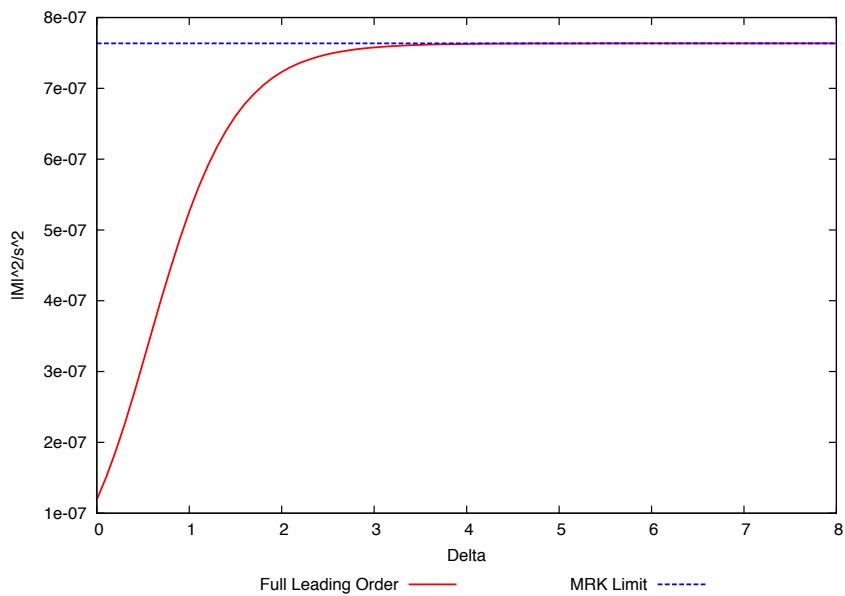

**Figure 2.1** *Comparison of the full Leading Order calculation and the MRK limit of the process $ud \to ud$. A high value of $\Delta$ means that the final state particles are well-separated in rapidity.*

Feynman diagrams that involve loops are, in general, difficult to calculate. The loop momentum we need to integrate over appears in a number of fermionic and gluonic propagators, the former leading to some cumbersome spinor algebra. Fortunately, there is a way to simplify the calculation by use of the *Cutkosky rules* [24]. We can define the matrix $S$ which encodes all possible ways a state $|i\rangle$ can evolve to a state $|f\rangle$, with elements

$$S_{fi} = \delta_{fi} + i(2\pi)^4 \delta^4 \left( \sum_{i \in initial} p_i - \sum_{j \in final} p_j \right) M_{fi}, \qquad (2.7)$$

where $\delta_{fi}$ represents a process where no interaction occurs and $M_{fi}$ is the scattering matrix element we have been working with before. In matrix notation, this can be written $S = \mathbb{1} + iT$ for the appropriate definition of $T$. Clearly, since the probability that an in state ends up in a particular out state, summed over all possible out states, must be unity, the matrix $S$ must be unitary, $S^\dagger S = \mathbb{1}$. This immediately leads to the non-trivial requirement

$$2\,\mathrm{Im}(T) = T^\dagger T. \tag{2.8}$$

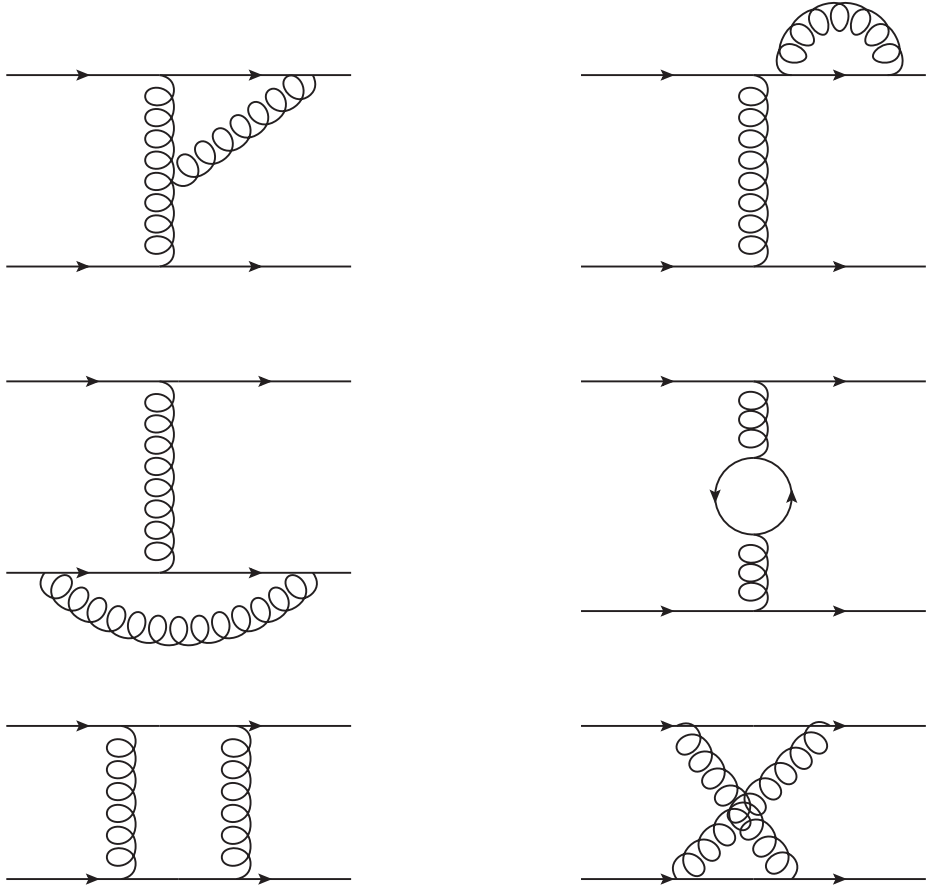

**Figure 2.2** *A selection of NLO diagrams for $qQ \to qQ$. We need only be interested in the bottom two.*

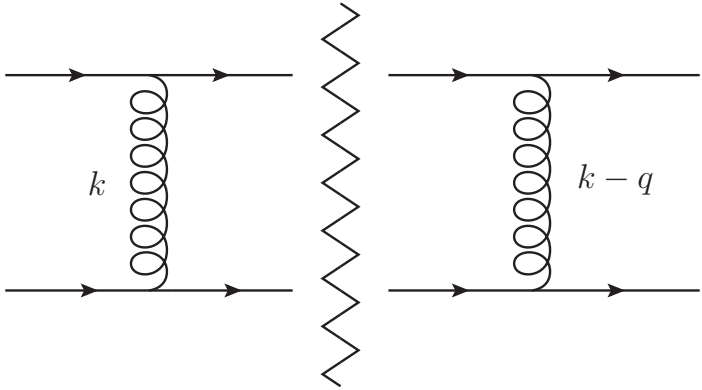

**Figure 2.3** *By 'cutting' our NLO diagram like this, we can find the imaginary part of the amplitude by considering a product of LO diagrams.*

We can project out a certain initial and final state

$$\begin{aligned}
2\text{Im}(T_{fi}) &= \langle f|T^\dagger T|i\rangle \\
&= \sum_k \langle f|T^\dagger|k\rangle \langle k|T|i\rangle \\
&= \sum_k T^*_{kf} T_{ki} \\
&= \sum_k (2\pi)^4 \delta^4\left(\sum_i p_i - \sum_k p_k\right) M^*_{kf} M_{ki}.
\end{aligned} \qquad (2.9)$$

where in the second line we have inserted a complete set of intermediate states $|k\rangle$.

This is a general statement about the full scattering amplitude. However, if we break this down into an order-by-order expression in the coupling, the statement allows one to relate diagrams at next-to-leading order (the left-hand side) to the product of other diagrams of leading order (the right-hand side). Using the Cutkosky rules, then, we can diagrammatically represent the imaginary part of the amplitude we are trying to work out as something like figure 2.3: we 'cut' the diagram vertically, setting the quark propagators crossing the cut on-shell and treating the process as a product of two leading order processes. Algebraically, this yields

$$\text{Im}(M_{NLO,1}) = \frac{1}{2}\int d(\text{phase space})M_{LO}(k)M_{LO}^\dagger(k-q)$$
$$= \frac{1}{2}\int \frac{d^4l}{(2\pi)^3}\frac{d^4l'}{(2\pi)^3}\delta(l^2)\delta(l'^2)(2\pi)^4\delta^4(p_a-p_b-l-l')M_{LO}(k)M_{LO}^\dagger(k-q)$$
$$= \frac{1}{8\pi^2}\int d^4k\,\delta((p_a-k)^2)\delta((p_b+k)^2)M_{LO}(k)M_{LO}^\dagger(k-q). \tag{2.10}$$

In the MRK limit, the leading order matrix element depended on $\hat{s}$ and $k_\perp$ as

$$M_{MRK} \sim g_s^2 \frac{2\hat{s}}{|k_\perp|^2}, \tag{2.11}$$

so it would be useful to write our integration over $k^\mu$ in terms of an explicit integration over $k_\perp$. We can do this by using a Sudakov parametrisation

$$k^\mu = \rho p_a^\mu + \lambda p_b^\mu + k_\perp^\mu, \tag{2.12}$$

which allows us to switch our integration variables to $\rho$, $\lambda$ and $k_\perp$, with the relation $d^4k = \frac{1}{2}\hat{s}\,d\rho\,d\lambda\,d^2k_\perp$. The High Energy Limit tells us that in this parametrisation, both $\rho$ and $\lambda$ are much smaller than 1. Thus

$$\text{Im}(M_{NLO,1}) = \frac{g_s^4\hat{s}}{4\pi^2}\int d\rho\,d\lambda\,d^2k_\perp\,\delta(-\hat{s}(1-\rho)\lambda + |k_\perp|^2)$$
$$\times \delta(\hat{s}(1+\lambda)\rho + |k_\perp|^2)\frac{\hat{s}}{|k_\perp|^2}\frac{\hat{s}}{|k_\perp - q_\perp|^2}$$
$$\approx \frac{g_s^4\hat{s}^3}{4\pi^2}\int d\rho\,d\lambda\,d^2k_\perp\,\delta(-\hat{s}\lambda + |k_\perp|^2)\delta(\hat{s}\rho + |k_\perp|^2)\frac{1}{|k_\perp|^2}\frac{1}{|k_\perp - q_\perp|^2}$$
$$= \frac{g_s^4\hat{s}}{4\pi^2}\int d\rho\,d\lambda\,d^2k_\perp\,\delta(-\lambda + |k_\perp|^2/\hat{s})\delta(\rho + |k_\perp|^2/\hat{s})\frac{1}{|k_\perp|^2}\frac{1}{|k_\perp - q_\perp|^2}$$
$$= \frac{g_s^4\hat{s}}{4\pi^2}\int d^2k_\perp\,\frac{1}{|k_\perp|^2}\frac{1}{|k_\perp - q_\perp|^2}. \tag{2.13}$$

We restate our postulate now that this amplitude is logarithmically enhanced in $\hat{s}/\hat{t}$, such that:

$$\ln\left(\frac{\hat{s}}{\hat{t}}\right) = \ln\left|\frac{\hat{s}}{\hat{t}}\right| - i\pi, \tag{2.14}$$

where we have used the fact that $\hat{t}$ is negative. Given this, we can immediately construct our real part from the imaginary part:

$$\text{Re}(M_{NLO,1}) = \ln\left|\frac{\hat{s}}{\hat{t}}\right| \frac{-g_s^4 \hat{s}}{4\pi^3} \int d^2k_\perp \frac{1}{|k_\perp|^2} \frac{1}{|k_\perp - q_\perp|^2}. \tag{2.15}$$

Kinematically, the other diagram we need to calculate is equivalent to this diagram under the exchange $\hat{s} \to \hat{u}$. In the MRK limit, $\hat{u} \approx -\hat{s}$ and so we can actually combine these two graphs into one by treating the colour factors properly:

$$\begin{aligned}
C_{NLO,1} - C_{NLO,2} &= t^b_{q_2 q_\alpha} t^a_{q_\alpha q_b} t^b_{q_1 q_\beta} t^a_{q_\beta q_a} - t^a_{q_2 q_\alpha} t^b_{q_\alpha q_b} t^b_{q_1 q_\beta} t^a_{q_\beta q_a} \\
&= [t^b, t^a]_{q_2 q_b} t^b_{q_1 q_\beta} t^a_{q_\beta q_a} \\
&= \frac{i f^{bac}}{2} t^c_{q_2 q_b} \left([t^b, t^a]_{q_1 q_a} + \{t^b, t^a\}_{q_1 q_a}\right) \\
&= \frac{i f^{bac}}{2} t^c_{q_2 q_b} [t^b, t^a]_{q_1 q_a} \\
&= \frac{-f^{bac} f^{bad}}{2} t^c_{q_2 q_b} t^d_{q_1 q_a} \\
&= \frac{-C_A \delta^{cd}}{2} t^c_{q_2 q_b} t^d_{q_1 q_a} \\
&= \frac{-C_A}{2} C_{Tree},
\end{aligned} \tag{2.16}$$

where $C_A = N = 3$ is a constant associated to the $SU(3)$ group. We see that the colour factor here is simply a constant multiplied by the tree level colour factor. With a small amount of manipulation and using $\hat{t} \approx -q_\perp^2$, we can then write the leading part of the NLO amplitude as proportional to the tree-level amplitude (using $|\hat{s}/\hat{t}| = -\hat{s}/\hat{t}$),

$$M_{NLO} = M_{LO} \ln\left(\frac{\hat{s}}{-\hat{t}}\right) \hat{\alpha}(q_\perp^2), \tag{2.17}$$

with

$$\hat{\alpha}(q_\perp^2) = \frac{C_A \alpha_s}{4\pi^2} \int d^2k_\perp \frac{-q_\perp^2}{k_\perp^2 (k_\perp - q_\perp)^2}. \tag{2.18}$$

This form of the amplitude clearly shows the dependence on the large logarithm. Even though $\hat{\alpha}$ contains a factor of $\alpha_s$, which should be small in order for our perturbation theory to make sense, the logarithm is large enough to overcome this suppression such that the product $\alpha_s \ln\left(\frac{\hat{s}}{-\hat{t}}\right)$ is of order 1. Therefore, this correction is as important as the leading order contribution and should not be ignored. In fact, terms proportional to $\alpha_s^n \ln^{n-1}\left(\frac{\hat{s}}{-\hat{t}}\right)$ continue to appear at *every order* [38] and so, by the same argument, we should not ignore these contributions either. We therefore find ourselves in need of an all-order treatment of scattering amplitudes in the High Energy Limit. We will now see how the High Energy Jets (HEJ) framework achieves this.

## 2.2 HEJ Amplitudes

We have seen how taking the full MRK limit on the $qQ \to qQ$ amplitude allowed us to derive a result for the NLO virtual corrections to the process. The simplicity of that procedure should give us hope to find a way of generalising the result to obtain the virtual corrections at all orders. If we want to describe all LHC processes, however, we have a few more things to consider beforehand. Firstly, we need to account for different parton types in the incoming state. It is not clear *a priori* whether $qg \to qg$ and $gg \to gg$ should behave as nicely in this limit, since the LO contribution contains more diagrams than the $t$-channel one alone. Secondly, we should account for extra final states; that is, generalise to a $2 \to n$ process. This will also be important when we want to calculate real corrections to our amplitudes, which is a subject we have not yet touched upon. We will tackle each of these concerns individually.

### 2.2.1 $qg \to qg$ in the High Energy Limit

We can demonstrate how the High Energy Limit relates quarks to gluons by taking the limit on the LO $qg \to qg$ amplitude and comparing it to the $qQ \to qQ$ amplitude. There are a total of 3 diagrams we must consider, which are shown in figure 2.4. In analogy with how we calculated the $qQ \to qQ$ amplitude, we will consider the helicity configuration $q^+g^+ \to q^+g^+$. Since we have an external gluon now, we also have gluon polarisation vectors in our calculation. Such vectors have a gauge redundancy and so we must choose a certain gauge to work in. If we take the incoming gluon to have momentum $p_b$ and the outgoing to have momentum $p_2$, it turns out to be useful to use the following parametrisation for the vectors

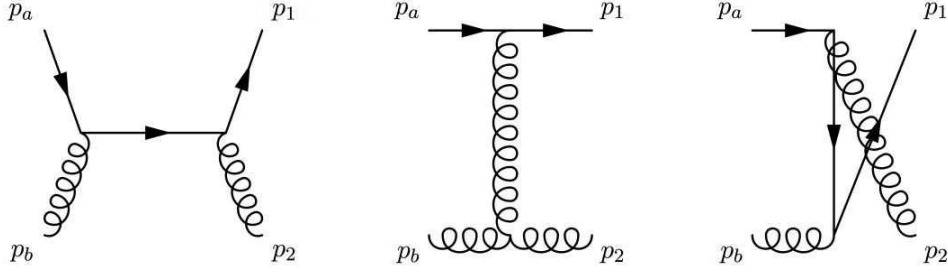

**Figure 2.4**  The three diagrams (s, t, and u channels) for $qg \to qg$ scattering at LO.

[7, 31]:

$$\varepsilon_{2\rho}^{+*} = \frac{\langle 2^+|\rho|b^+\rangle}{\sqrt{2}\,\langle b2\rangle}, \tag{2.19a}$$

$$\varepsilon_{2\rho}^{-*} = -\frac{\langle b^+|\rho|2^+\rangle}{\sqrt{2}\,[b2]}, \tag{2.19b}$$

$$\varepsilon_{b\rho}^{+} = -\frac{\langle 2^+|\rho|b^+\rangle}{\sqrt{2}\,[2b]}, \tag{2.19c}$$

$$\varepsilon_{b\rho}^{-} = \frac{\langle b^+|\rho|2^+\rangle}{\sqrt{2}\,\langle 2b\rangle}. \tag{2.19d}$$

We conduct the calculation by moving from left to right in figure 2.4 and thus begin with the $s$-channel diagram. The Feynman rules give us

$$\begin{aligned}
M_s &= \bar{u}_1^+(-ig_s\gamma^\nu T_{1q}^2)\left(\frac{i(\slashed{p}_a + \slashed{p}_b)}{\hat{s}}\right)(-ig_s\gamma^\mu T_{qa}^b)u_a^+\varepsilon_{2\nu}^{+*}\varepsilon_{b\mu}^{+}\\
&= \frac{ig_s^2 T_{1q}^2 T_{qa}^b}{2\hat{s}\,\langle b2\rangle[2b]}\langle 2^+|\nu|b^+\rangle\langle 2^+|\mu|b^+\rangle\langle 1^+|\gamma^\nu(\slashed{p}_a+\slashed{p}_b)\gamma^\mu|a^+\rangle \\
&= \frac{-ig_s^2 T_{1q}^2 T_{qa}^b}{\hat{s}\hat{t}}\langle 2^+|\nu|b^+\rangle\langle 2^+|\mu|b^+\rangle p_a^\mu\langle 1|\nu|a\rangle,
\end{aligned} \tag{2.20}$$

where in the last line we have used the completeness relation to expand the $\slashed{p}$ terms along with some of our spinor helicity identities. The $t$-channel diagram gives

$$\begin{aligned}
M_t &= \bar{u}_1^+(-ig_s\gamma^\mu T_{1a}^g)u_a\left(\frac{-i\eta_{\mu\nu}}{\hat{t}}\right)\\
&\quad f^{g2b}(-g_s)(\eta^{\nu\rho}(2p_2-p_b)^\sigma - \eta^{\rho\sigma}(p_2+p_b)^\nu + \eta^{\sigma\nu}(2p_b-p_2)^\rho)\varepsilon_{2\rho}^{+*}\varepsilon_{b\sigma}^{+},
\end{aligned} \tag{2.21}$$

but, since $p_{2/b}$ dotted with any of the polarisation vectors is zero and the two polarisation vectors themselves dotted together (with the chosen helicities) is also zero, we see that

$$M_t = 0. \tag{2.22}$$

It is important to realise that this result only holds because of the gauge we chose; it is the final result, which is the sum of all diagrams, that is the gauge invariant quantity. Thus, though it may seem surprising at first that the $t$-channel diagram is zero given that we are trying to show that the High Energy Limit is dominated by $t$-channel poles, it is simply a consequence of our gauge choice. The final diagram is the u-channel

$$\begin{aligned} M_u &= \bar{u}_1^+(-ig_s\gamma^\nu T_{1q}^b)\left(\frac{i(\not{p}_a - \not{p}_2)}{\hat{u}}\right)(-ig_s\gamma^\mu T_{qa}^2)u_a^+ \varepsilon_{b\nu}^{+*}\varepsilon_{2\mu}^+ \\ &= \frac{ig_s^2 T_{1q}^b T_{qa}^2}{2\hat{u}\langle b2\rangle [2b]} \langle 2^+|\nu|b^+\rangle \langle 2^+|\mu|b^+\rangle \langle 1^+|\gamma^\nu(\not{p}_a - \not{p}_2)\gamma^\mu|a^+\rangle \\ &= \frac{-ig_s^2 T_{1q}^b T_{qa}^2}{\hat{u}\hat{t}} \langle 2^+|\nu|b^+\rangle \langle 2^+|\mu|b^+\rangle\, p_a^\mu\, \langle 1|\nu|a\rangle\,. \end{aligned} \tag{2.23}$$

We see therefore that

$$M^{LO}_{q^+g^+ \to q^+g^+} = \frac{-ig_s^2}{\hat{t}} \langle 2^+|\nu|b^+\rangle \langle 2^+|\mu|b^+\rangle\, p_a^\mu \langle 1^+|\nu|a^+\rangle \left(\frac{T_{1q}^2 T_{qa}^b}{\hat{s}} + \frac{T_{1q}^b T_{qa}^2}{\hat{u}}\right). \tag{2.24}$$

We recall now that in the High Energy Limit, $\hat{u} \approx -\hat{s}$. We can also approximate that

$$\langle 2^+|\mu|b^+\rangle\, p_a^\mu \approx 2p_b \cdot p_a = \hat{s}. \tag{2.25}$$

Applying the limit to our result we see that

$$M^{LO,HE}_{q^+g^+ \to q^+g^+} = \frac{g_s^2 f^{2bc} T_{1a}^c}{\hat{t}} \langle 2^+|\nu|b^+\rangle \langle 1^+|\nu|a^+\rangle\,, \tag{2.26}$$

and therefore the amplitude is expressible as proportional to a $t$-channel pole. Furthermore, by taking the absolute square of this along with the colour sum, we find that

$$|M^{LO,HE}_{q^+g^+ \to q^+g^+}|^2 = \frac{C_A}{C_F}|M^{LO,HE}_{q^+Q^+ \to q^+Q^+}|^2, \tag{2.27}$$

where, for the $SU(3)$ group, we have $C_A = 3$ and $C_F = \frac{4}{3}$. We therefore have that the High Energy Limit relates quarks to gluons in a very simple way: namely, via the multiplication of a colour factor. We find that this result holds for the other 'helicity conserving' amplitudes (i.e., the amplitudes where both the quark and gluon do not flip helicities in the scattering) and that the 'helicity non-conserving' amplitudes are identically zero, so we can in fact generalise this to

$$|M_{qg \to qg}^{LO,HE}|^2 = \frac{C_A}{C_F}|M_{qQ \to qQ}^{LO,HE}|^2. \tag{2.28}$$

Of course, this equality only holds in the full High Energy Limit. In [7], it was seen that instead of using this colour factor, it is possible to derive a momentum-dependent colour factor that improves the description away from the strict High Energy Limit. In essence, the $qg \to qg$ calculation was performed with High Energy considerations whilst taking care not to immediately approximate the gluon momentum as $p_2^- \sim p_b^-$. Keeping these factors separate like this and one can extract the so-called *Colour Acceleration Multiplier* at the end of the calculation:

$$\tilde{C}_A = \frac{1}{2}\left(C_A - \frac{1}{C_A}\right)\left(\frac{p_b^-}{p_2^-} + \frac{p_2^-}{p_b^-}\right) + \frac{1}{C_A}. \tag{2.29}$$

It is clear that this tends to $C_A$ in the limit where $p_2^- \sim p_b^-$. We will employ this factor rather than the strict $C_A$ in the rest of this thesis (and, indeed, in the HEJ formalism itself).

### 2.2.2 Regge Theory and the Connection to High Energy $2 \to n$ Amplitudes

It is instructive here to take a short detour into the realm of Regge theory, which relates to properties of general scattering processes. In particular, before the advent of a full QCD theory, the relativistic form of Regge theory allowed for some predictions to be made about strongly interacting particles. More specifically, it predicts that in the High Energy Limit the amplitude of a given scattering process should behave as the Mandelstam variable $s$ raised to the power of the spin of the particle exchanged in the $t$-channel (denoted by $\alpha(t)$),

$$M_{Regge} \sim s^{\alpha(t)}. \tag{2.30}$$

A more complete description of the history of the use of Regge theory applied to strong dynamics can be found in [38]. For our purposes, we need only remark that the expression generalises if there is instead a 'chain' of exchanged particles in the $t$-channel such as that shown in figure 2.5, and we describe such a configuration

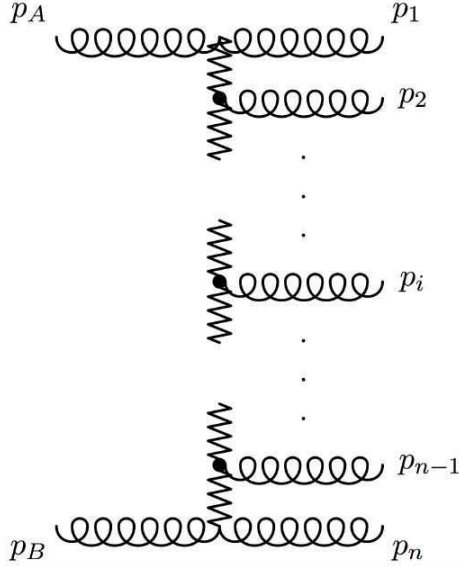

**Figure 2.5** *A schematic representation of an example MRK amplitude. Taken from [6].*

as 'Multi-Regge Kinematical':

$$M_{Multi-Regge} \sim s_{12}^{\alpha(t_1)} s_{23}^{\alpha(t_2)} \cdots s_{n-1,n}^{\alpha(t_{n-1})}. \tag{2.31}$$

We have already seen in $qQ \to qQ$ scattering that the amplitude did indeed behave as $s^1$ in the MRK limit. Since our only two choices of particles involved in $t$-channel exchange are the spin-one gluon and the spin-half quark, clearly all leading amplitudes in this limit must only have gluons exchanged. The only thing that can be emitted from a t-channel gluon is another gluon (any quark must be accompanied by an anti-quark, with an intermediate t-channel quark propagator) and so we need only consider extra gluon emissions from our base $2 \to 2$ processes to capture all the leading contributions. Such configurations are also referred to as 'Fadin-Kuraev-Lipatov' or simply FKL configurations [45], named after the three scientists who developed the formalism. Use of this fact and the study of amplitudes within the strict MRK limit [6, 26] show that the analytical form of the amplitude is

$$|M^{MRK}_{f_1 f_2 \to f_1 g \ldots g f_2}|^2 = \left(\frac{4\hat{s}^2}{N_c^2 - 1}\right) \frac{g^2 C_{f_1}}{|p_{1\perp}|^2} \left(\prod_{i=2}^{n-1} \frac{4g^2 C_A}{|p_{i\perp}|^2}\right) \frac{g^2 C_{f_2}}{|p_{n\perp}|^2}, \tag{2.32}$$

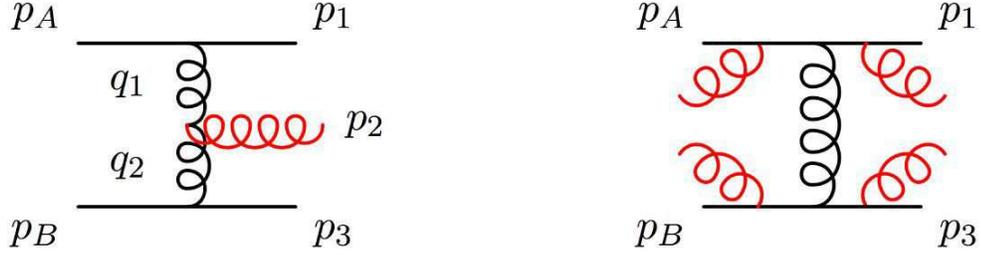

**Figure 2.6** *The five possibilities for an extra gluon emission. The diagram on the right-hand side should be seen as four separate diagrams. Taken from [6].*

where $C_{f_1}, C_{f_2}$ are $C_F, C_A$ depending on whether the incoming particles are quarks or gluons respectively. However, as we saw when we took the explicit limit of the $qQ \to qQ$ amplitude, this limit is only physically useful for a small amount of the available phase space that the LHC explores. On the other hand, the analytic expression is remarkably simple and at high gluon multiplicities it is much more practical to calculate than a full LO expression. Ideally, we would like to have something that bridges the gap between these two points; an expression that follows more closely the full LO matrix element in the LHC phase space whilst at the same time being simple enough such that we can evaluate it for processes with large numbers of final state gluons. The formulation of such amplitudes is one of the cornerstones of HEJ.

### 2.2.3   $qQ \to qgQ$ in the High Energy Limit

A good basis for finding a simple formulation for gluon emission away from the full MRK limit is to consider an extra emission in our $qQ$ scattering process, since the only possible choice there is $qQ \to qgQ$. There are a total of five diagrams for this, which are shown in 2.6. Before we begin the calculation, let us first take a detour to discuss the *Eikonal Approximation*.

The Eikonal Approximation is a way of simplifying the structure of vertex functions in the limit where one of the momenta entering the vertex becomes very small. We will show that this approximation is also valid for situations where a momentum is not small in an absolute sense but small compared to other momenta in the problem. This is best seen by considering an example. Take a quark-quark-gluon vertex such as the one shown in figure 2.7, where the quark line connects to a larger diagram on the left-hand side and the gluon is a final state particle. The Feynman rules for such a part of the diagram give an expression that includes the term $\bar{u}(p)\gamma^\mu(\slashed{p}+\slashed{k})/(p+k)^2$. If we imagine that our quark is not deflected much by this emission, then we can take the limit $k \ll p$. This gives

$$\begin{aligned}
\frac{\bar{u}(p)\gamma^\mu(\slashed{p}+\slashed{k})}{(p+k)^2} &\approx \frac{\bar{u}(p)\gamma^\mu \slashed{p}}{2p\cdot k} \\
&= \frac{\bar{u}(p)\gamma^\mu \gamma^\alpha p_\alpha}{2p\cdot k} \\
&= \frac{\bar{u}(p)(2\eta^{\mu\alpha} - \gamma^\alpha \gamma^\mu)p_\alpha}{2p\cdot k} \\
&= \bar{u}(p)\frac{p^\mu}{p\cdot k},
\end{aligned} \qquad (2.33)$$

where in the last line we have used $\bar{u}(p)\slashed{p} = 0$ by virtue of the Dirac equation. A similar result holds if we replace our quarks with gluons as well, so the approximation is 'blind' to the spin of the particle emitting the gluon, an effect that allows us to very easily relate quarks and gluons in the High Energy Limit even as the final state multiplicity increases. We then see that the leading terms in the High Energy limit are equivalent to the leading terms in the Eikonal Limit.

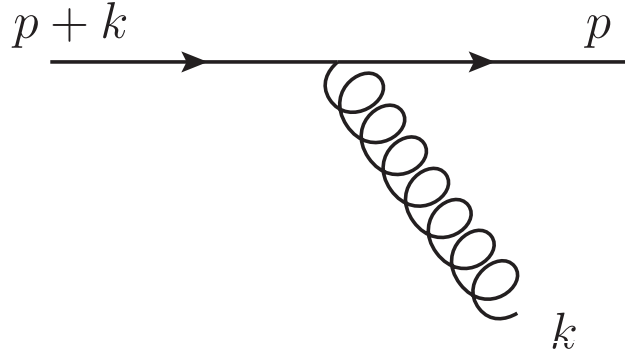

**Figure 2.7**  *The emission of a gluon from a quark line. By taking the limit $k \ll p$, we obtain the Eikonal rule.*

Using this approximation, we can calculate the four diagrams with a gluon emitted from the extremal legs straightforwardly. The result of adding all these contributions together gives

$$M_{eik} = M_{qQ\to qQ}(ig_s)\varepsilon^*_\rho \left( C_1 \frac{p_1^\rho}{p_1\cdot p_2} - C_2 \frac{p_a^\rho}{p_a\cdot p_2} + C_3 \frac{p_3^\rho}{p_3\cdot p_2} - C_4 \frac{p_b^\rho}{p_b\cdot p_2} \right). \qquad (2.34)$$

The form of the approximation makes it clear that the (kinematic part of the) amplitude must be proportional to the $qQ \to qQ$ amplitude. The remaining diagram, however, involves emission from the $t$-channel exchanged gluon and so it is not clear if this will also be simply proportional to the $qQ \to qQ$ result. The relevant part of this diagram is

$$M_{3g} \sim \langle 1|\mu|a\rangle\,(-g_s)(\eta^{\mu\rho}(q_1+p_2)^\nu + \eta^{\rho\nu}(-p_2+q_2)^\mu + \eta^{\nu\mu}(-q_2-q_1)^\rho)\,\langle 3|\nu|b\rangle\,\varepsilon^*_\rho, \tag{2.35}$$

where $q_1 = p_a - p_1$ and $q_2 = p_a - p_1 - p_2 = p_3 - p_b$. In the High Energy Limit, we have already discussed how the extremal partons in the amplitude have momenta similar to the incoming particles; in other words, $p_1 \sim p_a$ and $p_3 \sim p_b$. In turn, this means that $\langle 1|\mu|a\rangle \sim 2p_a^\mu$ and $\langle 3|\nu|b\rangle \sim 2p_b^\nu$. Applying these approximations to our amplitude, we find

$$\begin{aligned}M_{3g} &\sim (-g_s)\varepsilon^*_\rho(2p_b^\rho(s_{1a}+2s_{a2}) - 2p_a^\rho(s_{3b}+2s_{2b}) + 2\hat{s}(-q_2-q_1)^\rho)\\ &\approx (-2\hat{s}g_s)\varepsilon^*_\rho\left(2p_b^\rho\frac{s_{a2}}{\hat{s}} - 2p_a^\rho\frac{s_{2b}}{\hat{s}} + (q_1+q_2)^\rho\right),\end{aligned} \tag{2.36}$$

since $s_{1a} \ll s_{2a}$ and $s_{3b} \ll s_{b2}$. Reinstating the factors from the rest of the diagram and rewriting the $2\hat{s}$ we factorised outside as $\langle 1|\mu|a\rangle\,\langle 3|\mu|b\rangle$, then

$$M_{3g} = M_{qQ \to qQ} C_t (-g_s)\varepsilon^*_\rho\left(2p_b^\rho\frac{s_{a2}}{\hat{s}} - 2p_a^\rho\frac{s_{2b}}{\hat{s}} + (q_1+q_2)^\rho\right), \tag{2.37}$$

and therefore the $qQ \to qQ$ amplitude does indeed completely factor out in this limit. We now turn our attention to the colour factors

$$\begin{aligned}C_1 - C_2 &= t^g_{b2}t^e_{1q}t^g_{qa} - t^g_{b2}t^g_{1q}t^e_{qa}\\ &= if^{gec}t^c_{a1}t^g_{b2}\\ &= -iC_t,\end{aligned} \tag{2.38}$$

and similarly $C_3 - C_4 = iC_t$. The sum of all the diagrams is then proportional to the same colour factor and to the $qQ \to qQ$ amplitude. Putting everything together, we can then write

$$M_{qQ \to qgQ} = g_s^3 C_t \varepsilon^*_\rho \frac{\langle 1|\mu|a\rangle\,\langle 3|\mu|b\rangle}{q_1^2 q_2^2} V^\rho(q_1, q_2), \tag{2.39}$$

where

$$V^\rho(q_1, q_2) = -(q_1+q_2)^\rho + p_a^\rho\left(\frac{q_1^2}{p_2 \cdot p_a} + 2\frac{p_2 \cdot p_b}{p_a \cdot p_b}\right) - p_b^\rho\left(\frac{q_2^2}{p_2 \cdot p_b} + 2\frac{p_2 \cdot p_a}{p_a \cdot p_b}\right) \tag{2.40}$$

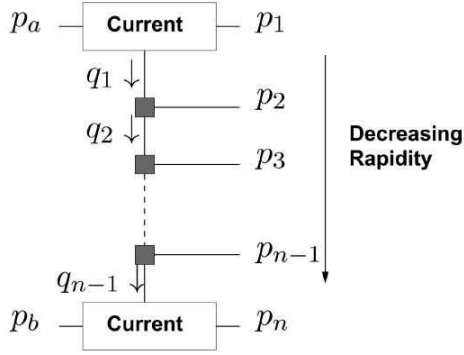

**Figure 2.8** *A schematic view of a HEJ amplitude. Taken from [3].*

is an *effective* or *Lipatov* vertex describing the emission of a gluon. Since this vertex was derived by approximating $p_a$ with $p_1$ and $p_b$ with $p_3$, we can symmetrise this vertex in these momenta to 'undo' this approximation somewhat[3]. The important point is that we can now construct an amplitude with any number of extra gluons by simply inserting the relevant number of Lipatov vertices. As previously discussed, the difference between amplitudes with initial state quarks and amplitudes with initial state gluons is an overall colour factor, so a general $2 \to n$ amplitude takes the form

$$|\bar{M}^t_{f_1 f_2 \to f_1 g \ldots g f_2}|^2 = \frac{1}{4(N_C^2 - 1)} ||S_{f_1 f_2 \to f_1 f_2}||^2 \left(g_s^2 C_{f_1} \frac{1}{\hat{t}_1}\right) \left(g_s^2 C_{f_2} \frac{1}{\hat{t}_{n-1}}\right) \prod_{i=1}^{n-2} \left(\frac{-g_s^2 C_A}{\hat{t}_i \hat{t}_{i+1}} V^\mu(q_i, q_{i+1}) V_\mu(q_i, q_{i+1})\right),$$

(2.41)

where we have introduced the notation $||S_{f_1 f_2 \to f_1 f_2}||^2$ to represent the sum over helicities of the modulus squared of the contraction of currents that appears at tree-level. We have also used the result $\sum_{pol} \varepsilon_\mu \varepsilon_\nu^* \to -\eta_{\mu\nu}$ to contract the Lipatov vertex. In the full MRK limit, this is proportional to $\hat{s}^2$ as we saw earlier, but we are able to keep more information about the process by keeping the full dependence like this (for example, we will reproduce *exactly* the full LO $qQ \to qQ$ calculation in this manner, not just the MRK limit thereof). We can represent a HEJ amplitude pictorially as shown in figure 2.8. We see clearly how this expression has the simplicity of the full MRK expression whilst at the same time being more flexible, as it is able to describe better the behaviour of the matrix element away from this limit.

---

[3]This symmetrised version is used in practice but adds nothing to the discussion here.

## 2.3 Resummation Technique

So far, we have derived an approximation to the LO matrix element for a certain subset of $2 \to n$ processes. We will now show that the form of this approximation allows for the inclusion of the High Energy leading logarithmic terms of the amplitude *to all orders in perturbation theory.*

### 2.3.1 Lipatov Ansatz

Earlier, we saw explicitly that the Leading Logarithmic contribution to the NLO correction for $qQ \to qQ$ scattering was given by multiplying the LO result by a logarithm and a factor as given by equation 2.18. One of the postulates of the *Lipatov Ansatz* is that this result exponentiates, such that the virtual corrections to this process to all orders are given by

$$M^{LO+virt}_{qQ \to qQ} = M^{LO}_{qQ \to qQ} \times \exp\left[\hat{\alpha}(q_\perp)\Delta y\right], \qquad (2.42)$$

where we have related $\Delta y$ to the logarithm as explained at the beginning of section 2.1. The ansatz then goes further to say that this exponentiation holds in the MRK limit for any number of $t$-channel propagators present at the leading order, such that the virtual corrections are obtained by making the substitution

$$\frac{1}{\hat{t}_i} \to \frac{1}{\hat{t}_i} \exp\left[\hat{\alpha}(q_{i,\perp})\Delta y_{i,i+1}\right]. \qquad (2.43)$$

In the appropriate limit, this ansatz has been proved to the sub-leading level [17, 35–37]. However, by inspection of the $\hat{\alpha}$ function given in equation 2.18, we see that it is clearly divergent; for small values of $k_\perp$ in the integration region, the integrand tends to infinity. Such a divergence is called an *infrared divergence* and is a common problem in particle physics amplitudes involving massless particles. The problem arises because virtual corrections are not the only corrections to an amplitude we need to consider. We can also have the case where a real emission has taken place but at an energy scale small enough that only a detector with infinite resolution could have detected it. Since such a detector cannot exist, there is no way of differentiating between a 'pure' $qQ \to qQ$ scattering and another one accompanied by the emission of a low energy (or 'soft') gluon. Such contributions are also divergent, but in such a way that when combined with the divergence in the virtual corrections, the divergences cancel. We will show precisely how this happens in the next part. For now, we choose to use the technique of *dimensional regularisation* to rewrite our $\hat{\alpha}$ function in a more convenient way that makes the divergence clear. The formalism works by extending the dimensionality of the two-dimensional perpendicular component integral to $2 + 2\varepsilon$ dimensions. This will allow us to write the result as an expansion in $\varepsilon$, which will manifest our divergence as a term proportional to $\varepsilon^{-1}$. We must eventually take the limit $\varepsilon \to 0$, but we will see how this prescription allows us to remove the divergences from our theory in a neat and systematic way before we do so. A further note is that this dimensional shift also slightly alters the overall energy dimension of the integral and so we must introduce a scale $\mu$ to absorb this effect. The upshot is that we perform the integral in this convention and end up with

$$\hat{\alpha}(q_{i\perp}, \varepsilon) = -g^2 C_A \frac{\Gamma(1-\varepsilon)}{(4\pi)^{2+\varepsilon}} \frac{2}{\varepsilon} \left(\frac{|q_{i\perp}|^2}{\mu^2}\right)^\varepsilon. \tag{2.44}$$

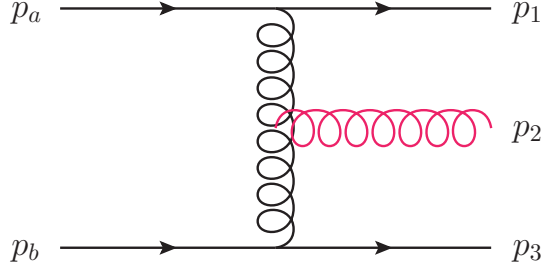

**Figure 2.9** *If the momentum $p_2$ is small enough, we cannot detect the gluon emission and so treat it as a real correction to the $2 \to 2$ process.*

### 2.3.2 Combining Real and Virtual Corrections at All Orders

As already discussed, the real correction to the amplitude occurs when the momentum of an outgoing gluon becomes very small. Since the emissions are controlled entirely by the effective vertex, we then need to find the limit of the function as the momenta becomes small. It is a simple exercise to show that, if we denote the emitted momentum by $p_2$ as in figure 2.9, the soft limit is

$$\lim_{p_2 \to 0} \frac{-V(q_1, q_2) \cdot V(q_1, q_2)}{\hat{t}_1 \hat{t}_2} = \frac{4}{|p_{2\perp}|^2}. \tag{2.45}$$

Therefore, the first real correction to $qQ \to qQ$ is

$$|M^{HE,1RC}_{qQ \to qQ}|^2 = \frac{||S_{qQ \to qQ}||^2}{4(N_c^2 - 1)} \frac{g^2 C_F}{\hat{t}_1} \frac{g^2 C_F}{\hat{t}_2} \left(\frac{4g^2 C_A}{|p_{2\perp}|^2}\right). \quad (2.46)$$

We need to regularise the soft divergence, so let us integrate this over the soft part of phase space (again, using dimensional regularisation) by introducing some soft transverse scale $\lambda$:

$$\begin{aligned}
&\mu^{-2\varepsilon} \int_{soft} \frac{d^{3+2\varepsilon} p_2}{(2\pi)^{3+2\varepsilon} 2 E_2} \left(\frac{4g^2 C_A}{|p_{2\perp}|^2}\right) \\
&= \mu^{-2\varepsilon} \int_0^\lambda \frac{d^{2+2\varepsilon} p_{2\perp}}{(2\pi)^{2+2\varepsilon}} \int_{y_1}^{y_3} \frac{dy_2}{4\pi} \left(\frac{4g^2 C_A}{|p_{2\perp}|^2}\right) \\
&= \mu^{-2\varepsilon} \frac{4g^2 C_A}{(2\pi)^{2+2\varepsilon}(4\pi)} (y_3 - y_1) \int_0^\lambda \frac{d^{2+2\varepsilon} p_{2\perp}}{|p_{2\perp}|^2} \\
&= \frac{g^2 C_A}{\pi (2\pi)^{2+2\varepsilon}} (y_3 - y_1) \frac{1}{\varepsilon} \frac{\pi^{1+\varepsilon}}{\Gamma(1+\varepsilon)} \left(\frac{\lambda^2}{\mu^2}\right)^\varepsilon.
\end{aligned} \quad (2.47)$$

By virtue of the Lipatov ansatz, the virtual corrections to the process are given by

$$|M^{HE,VC}_{qQ \to qQ}|^2 = \frac{||S_{qQ \to qQ}||^2}{4(N_c^2 - 1)} \frac{g^2 C_F}{\hat{t}_1} \frac{g^2 C_F}{\hat{t}_2} \exp\left[2\hat{\alpha}(q_{1\perp}, \varepsilon)(y_1 - y_3)\right], \quad (2.48)$$

and so the first virtual correction is simply the exponential expanded to first order:

$$\begin{aligned}
|M^{HE,1VC}_{qQ \to qQ}|^2 &= \frac{||S_{qQ \to qQ}||^2}{4(N_c^2 - 1)} \frac{g^2 C_F}{\hat{t}_1} \frac{g^2 C_F}{\hat{t}_2} [2\hat{\alpha}(q_{1\perp}, \varepsilon)(y_1 - y_3)] \\
&= \frac{||S_{qQ \to qQ}||^2}{4(N_c^2 - 1)} \frac{g^2 C_F}{\hat{t}_1} \frac{g^2 C_F}{\hat{t}_2} \left[-4(y_3 - y_1) g^2 C_A \frac{\Gamma(1-\varepsilon)}{(4\pi)^{2+\varepsilon}} \frac{1}{\varepsilon} \left(\frac{|q_{1\perp}|^2}{\mu^2}\right)^\varepsilon\right].
\end{aligned} \quad (2.49)$$

We must now expand both of these results in $\varepsilon$, add them together and then take the limit $\varepsilon \to 0$. Doing so yields

$$\begin{aligned}
\lim_{\varepsilon \to 0} |M^{HE,1VC+1RC}_{qQ \to qQ}|^2 &= \lim_{\varepsilon \to 0} \frac{||S_{qQ \to qQ}||^2}{4(N_c^2 - 1)} \frac{g^2 C_F}{\hat{t}_1} \frac{g^2 C_F}{\hat{t}_2} [\omega_0 + \mathcal{O}(\varepsilon)] \\
&= \frac{||S_{qQ \to qQ}||^2}{4(N_c^2 - 1)} \frac{g^2 C_F}{\hat{t}_1} \frac{g^2 C_F}{\hat{t}_2} \omega_0,
\end{aligned} \quad (2.50)$$

with

$$\omega_0 = \frac{g^2 C_A}{4\pi^2} \ln\left(\frac{\lambda^2}{|q_{1\perp}|^2}\right). \quad (2.51)$$

The argument clearly continues to higher orders and so we can immediately generalise the result to yield a full HEJ amplitude:

$$|M^{HEJ}_{f_1f_2 \to f_1...f_2}|^2 = \frac{||S_{f_1f_2 \to f_1f_2}||^2}{4(N_c^2 - 1)} \frac{g^2 C_{f_1}}{\hat{t}_1} \frac{g^2 C_{f_2}}{\hat{t}_{n-1}}$$
$$\prod_{i=1}^{n-2} \frac{-g^2 C_A V(q_i, q_{i+1}) \cdot V(q_i, q_{i+1})}{\hat{t}_i \hat{t}_{i+1}} \quad (2.52)$$
$$\prod_{j=1}^{n-1} \exp\left[\omega_0(q_{j\perp})(y_{j+1} - y_j)\right].$$

The final comment to make about the HEJ amplitude is that we can extend it to also include the emission of a Higgs, a $W^\pm$ [3] or a $Z$ boson [5] along with the jets. The only difference is the form of the current contraction $S_{f_1f_2 \to f_1f_2}$. For the pure jets case, this is simply $j_\mu j^\mu$, the contraction of two pure quark currents. We can also define a current for the emission of, for example, a $W$ boson as $j_W^\mu$, which will depend on the momenta of the decay products of the $W$ along with the momentum of the quark line. For a $W$ process, then, we would have an object $S_{f_1f_2 \to f_1f_2e\nu} = j_W^\mu j_\mu$. The rest of the derivation then proceeds as before.

## 2.4 Monte Carlo Implementation

We now have an expression for an all-order matrix element that is free of divergences in four spacetime dimensions. For physical relevance, we must now integrate the matrix element over the entire detector phase space of the LHC. The general expression for the integrand has already been shown in section 1.5. The way we do this complicated integral is via the technique of *Monte Carlo Integration* and we dedicate this section to explaining the process within the context of our implementation.

### 2.4.1 The Motivation Behind Monte Carlo

Monte Carlo integration is a numerical integration technique that provides an estimate for a (usually very complicated or even analytically undoable) integral via use of random numbers. An insightful example of how it works is to consider how it can be used to estimate the value of $\pi$. Figure 2.10 shows a simple setup for how to achieve this. The area of the blue square is $(2R)^2$ and the area of the red circle is $\pi R^2$. The ratio of the area of the circle to the area of the square is then $\pi/4$. If we were to pick a random point in the square (since we have only two dimensions here, this is the same as randomly sampling a pair of $x$ and $y$ coordinates) then there is a probability of $\pi/4$ that it also falls within the circle. Therefore, given $N$ total choices of points within the square (or 'trials') of which $M$ also fall inside the circle,

$$\pi \approx \frac{4M}{N}. \tag{2.53}$$

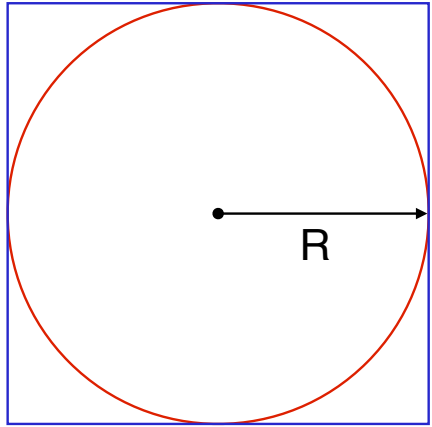

**Figure 2.10** *A square with sides of length $2R$ enclosing a circle of radius $R$. We can estimate the value of $\pi$ by picking random points in the square and seeing whether or not they also lie in the circle.*

Of course, if we do not conduct many trials $N$, our approximation will not be a good one because we have simply not sampled enough of the available space. As we increase our number of trials, however, the approximation gets closer to the real value of $\pi$, as shown[4] in table 2.1. This is a consequence of the *law of large numbers*, on which Monte Carlo techniques depend.

To make clear the relation to integral problems, we will discuss this problem again from a different point of view. We can think of the square which contains the circle as a two-dimensional 'volume' (more usually called area, of course, but we will introduce the more general term now) which bounds our system with value $V_2 = 4R^2$. Within this volume, we are trying to calculate another volume; namely, the area of the circle. Taking the origin of our co-ordinate system to be

---

[4]In fact, we tend to 'bounce' around the true value of $\pi$ because the nature of our approximation means we constantly switch between overestimating and underestimating the true value.

| Number of trials | Estimate of $\pi$ |
|---|---|
| $10^0$ | 4 |
| $10^1$ | 2 |
| $10^2$ | 3.6 |
| $10^3$ | 3.192 |
| $10^4$ | 3.1644 |
| $10^5$ | 3.1404 |
| $10^6$ | 3.141828 |
| $10^7$ | 3.14139320 |
| $10^8$ | 3.14145576 |
| $10^9$ | 3.14161324 |

**Table 2.1** *Estimates of $\pi$ via a simple Python Monte Carlo program for different numbers of trials.*

the centre of the circle, then the area of the circle is given by

$$V_C = \int_0^R dr\, r \int_0^{2\pi} d\theta \equiv \int_{V_C} d\Omega. \tag{2.54}$$

This is a particularly simple integral with value $V_C = \pi R^2$. In the initial formulation of the problem, we took this to be known but were ignorant on what the value of $\pi$ numerically is. Instead, let us imagine that we did not know how to perform the integral at all. We always know, however, that an integral (which we generalise to one depending on many variables) can be related to the average value of its integrand:

$$\langle f(\vec{x}) \rangle_{x \in V} = \frac{\int_V d\Omega f(\vec{x})}{V}. \tag{2.55}$$

The average can be estimated by simply sampling the integrand at $N$ random points and then dividing by $N$. So long as the distribution of random points in the volume is flat, the *Central Limit Theorem* guarantees that

$$\int_V f(\vec{x}) d\Omega \approx V \langle f \rangle \pm V \sqrt{\frac{\langle f^2 \rangle - \langle f \rangle^2}{N}}, \tag{2.56}$$

with

$$\langle f \rangle = \frac{1}{N} \sum_{i=1}^N f(\vec{x_i}), \tag{2.57a}$$

$$\langle f^2 \rangle = \frac{1}{N} \sum_{i=1}^N f^2(\vec{x_i}). \tag{2.57b}$$

The error given by the Central Limit Theorem is an estimate of the error only – there is no guarantee that the error is distributed as Gaussian. We clearly see that the error scales as $N^{-\frac{1}{2}}$, which is rather slowly convergent, but the important point is that it is *completely independent of the dimensionality* of the problem.

The last step is to increase the domain of integration to the larger area $V_2$ that encapsulates entirely $V_C$:

$$\int_{V_C} d\Omega\, f = \int_{V_2} d\Omega\, f|_{V_C} \approx V_2 \langle f|_{V_C}\rangle \pm V_2 \sqrt{\frac{\langle f|_{V_C}^2\rangle - \langle f|_{V_C}\rangle^2}{N}}. \tag{2.58}$$

Then, since in our original problem the integral we are interested in is simply the area integral, $f = 1$ if any chosen random point is within $V_C$ and 0 otherwise. Thus, given $N$ trials of which $M$ lie inside the circle (we drop the error estimate now for brevity),

$$\int_{V_2} d\Omega\, 1|_{V_C} = V_C \approx V_2 \frac{M}{N}$$
$$\therefore \frac{V_C}{V_2} \approx \frac{M}{N}, \tag{2.59}$$

which is the statement we started with.

We could imagine extending this experiment in two ways. Firstly, we could be interested in the behaviour of a more complicated function within a volume (i.e., a more complicated $f$). Secondly, we could also think about taking the problem into three dimensions by considering a sphere enclosed within a cube. Indeed, mathematically speaking, we could extend this into as many dimensions as we desire, to the point where we no longer have such an obvious geometric interpretation of the problem. It is precisely the arena of high-dimensionality integration problems with complicated integrands where Monte Carlo techniques come into their own.

### 2.4.2 Monte Carlo in High Energy Jets

We recall from section 1.5 that the integral we are trying to perform looks like

$$\sigma_{pp\to n-jet}^{inc/exc} = \sum_{f_a, f_b} \int_0^1 dx_a \int_0^1 dx_b f_a(x_a, Q^2) f_b(x_b, Q^2) \times \hat{\sigma}_{partonic} \times \mathcal{J}(\text{n-jet}^{inc/exc}), \tag{2.60}$$

with

$$\hat{\sigma}_{partonic} = S \times \frac{|M|^2}{F} \times (2\pi)^4 \delta^{(4)}(p_a + p_b - \sum_{f=1}^{n} p_f) \times \prod_{i=1}^{n} \int \frac{d^3\vec{p}_i}{2E_i(2\pi)^3}, \quad (2.61)$$

and so the problem is clearly suited to evaluation by Monte Carlo methods. The general process is as follows:

1. Generate a number of partons for the final state.

2. Pick the flavours for the incoming state.

3. Generate the momenta for all partons.

4. Perform a basic cluster into jets to check whether this event will give something consistent with cuts (a set of kinematical constraints imposed on the analysis). If not, throw away the point here and start again.

5. Calculate the matrix element and multiply by other factors in the integrand.

6. Cluster into jets and check to see if they pass all imposed cuts on the process. If it does, add the calculated point to the estimation of the average.

7. Repeat.

Such a technique is perfectly acceptable and results can be achieved this way. However, if we are simply generating random numbers flatly in these steps then we are being needlessly inefficient. For example, we see from graphs of Parton Distribution Functions that there are clear areas where the value of it is much larger than at other points. If we simply generate flatly then we equally sample these two regions which will clearly hurt our convergence. Indeed, there is a technique to combat this called *importance sampling*.

Importance sampling is aimed at reducing as much as possible the value of $\sqrt{\langle f \rangle^2 - \langle f^2 \rangle} \equiv \sigma_{MC}$, also known as the *variance*, which we saw was directly related to the estimate of the error. To see how this achieved, let us consider a one-dimensional integral and utilise our freedom to multiply and divide by another (well-behaved) distribution $q(x)$:

$$\int_a^b dx f(x) = \int_a^b dx\, q(x) \frac{f(x)}{q(x)} = \int_a^b dx\, q(x) h(x). \quad (2.62)$$

We can then perform a change of variables

$$\int_a^b dx\, q(x)h(x) = \int_{Q(a)}^{Q(b)} dy\, h\left(Q^{(-1)}(y)\right), \qquad (2.63)$$

with $dQ(x)/dx = q(x)$. If we normalise $q(x)$ such that it integrates to unity over the integration domain, then the fundamental theorem for Monte Carlo integration shows

$$\int_a^b dx f(x) = \int_a^b dx\, q(x) \frac{f(x)}{q(x)} \approx \left\langle \frac{f(x)}{q(x)} \right\rangle \pm \sqrt{\frac{\langle f^2(x)/q^2(x)\rangle - \langle f(x)/q(x)\rangle^2}{N}}. \qquad (2.64)$$

Thus if we pick $q(x) = 1/(b-a)$, we arrive back at equation 2.56 (a one-dimensional volume is simply the length of the line), but we are not bound to make this choice – we should instead pick a $q(x)$ that minimises as much as possible the second term. There are a few subtleties and difficulties in doing this, but the upshot is that this is best achieved by sampling from a distribution $q(x)$ that is close to $f(x)$ in shape. This is done at many points in the HEJ program, including:

- Picking the incoming partons in such a way to more often sample those with a higher PDF value. The LHAPDF package [18] which contains a whole range of different PDF sets to choose from provides the value of $f/x$, so what we actually optimise for is the quantity $x \times f/x$.

- Generating transverse momenta skewed towards the lower end of the spectrum since the cross-section falls off rapidly with $p_\perp$.

- Generating the rapidity of particles to more often create 'valid' configurations, where we keep the FKL rapidity ordering.

With such considerations, the variance of our estimate is greatly reduced and stable results can be obtained fairly quickly, depending on the precise nature of the analysis.

### 2.4.3 Concerning Partons and Jets

In the previous subsection it was remarked that during our Monte Carlo program we check to see how the partons we generate correspond to observed jets. How this is done is worthy of a longer discussion. The process by which hard partons produced by the scattering culminate in jets of hadrons is an inherently non-perturbative process; there is no such effect that arises naturally from the rules of our perturbation theory. The implication is clear: perturbation theory must break down at some point. The cause of this is that the value of the expansion parameter $\alpha_s$ is not a constant but 'runs' with the energy scale as shown in figure 2.11. At small energy scales, the parameter becomes large enough that a perturbative theory no longer makes sense. Thankfully, we can separate (or 'factorise') these low energy processes away from our high energy perturbative process so long as we have a defined way of estimating how many final state jets appear given a certain set of hard external partons. This has led to the development of *jet algorithms*.

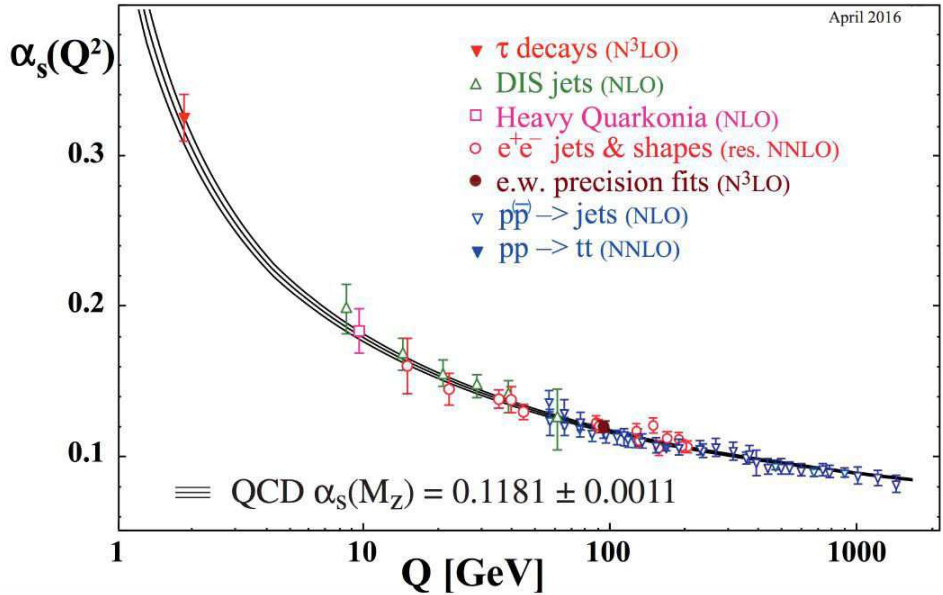

**Figure 2.11** *The value of $\alpha_s$ as evaluated at different energy scales $Q^2$. At low energies, the value becomes large enough that a perturbative treatment of QCD is no longer valid. Taken from [49].*

There are two broad categories of algorithms: 'Cone-Type' and 'Sequential Clustering' [9]. The first type are heavily disfavoured by the theory community because of their tendency to be infrared unsafe (meaning they depend on the low energy physics of the underlying theory) and so we will focus on the second type. The general process for this type of algorithm is as follows:

1. Define some distance measure $d_{ij}$ that can be calculated between each pair

of hard partons $i, j$ and a distance $d_{iB}$ between all particles and the beam line.

2. Compute all distances $d_{ij}$ and $d_{iB}$ and find the smallest.

3. If the smallest distance is a $d_{ij}$, then combine the momenta of the partons $i, j$ and recalculate all distances. If the smallest is a $d_{iB}$, remove parton $i$ from the process and call it a 'jet'.

4. Repeat until all partons are clustered into jets.

The natural question, of course, is what we should choose for the distances $d_{ij}$ and $d_{iB}$. The general form is

$$\begin{aligned} d_{ij} &= \min\left(k_{i\perp}^{2p}, k_{j\perp}^{2p}\right) \frac{\Delta_{ij}}{R} \quad \text{with} \quad \Delta_{ij} = \sqrt{(y_i - y_j)^2 - (\phi_i - \phi_j)^2} \\ d_{iB} &= k_{i\perp}^{2p}. \end{aligned} \quad (2.65)$$

The parameter $R$ scales the $d_{ij}$ with respect to $d_{iB}$ such that any pair of final jets $a, b$ are at least separated by $\Delta_{ab} = R$. For this reason, we often refer to $R$ as a 'jet radius'. The value of $p$ can be chosen to govern the relative power of energy and geometrical scales in the distance parameter. There are three main algorithms which have three different choices for $p$: the $k_T$ algorithm with $p = 1$, the Cambridge/Aachen with $p = 0$ and the anti-$k_T$ with $p = -1$ [9]. There are reasons why one might want to use one over the other and so one would wish for a simple way of changing between conventions in any computer program one wants to run. Fortunately, the *FastJet* library [19] does precisely this. For the user, the only requirement is to specify which algorithm is to be used at the start of the calculation. Any further calculations involving the jet objects are easily handled within the defined classes in the package.

### 2.4.4 Scale Variations

It has already been discussed that the value of a parton distribution function is dependent on the energy scale at which the proton is probed. We have now seen that the value of the strong coupling constant $\alpha_s$ is also dependent on the energy scale. In principle, these two scales are different, and called the *factorisation scale* $\mu_f$ and *renormalisation scale* $\mu_r$ respectively. Clearly, we need to make a choice for these scales before we can do any calculations. However, it is not clear at all what we should choose. We can imagine at least that the scales do not differ much and so we could simply consider $\mu_f = \mu_r$. Still, we find there is no one good answer to this question and we are left with the task of calculating *scale variations* in order to provide an estimate of uncertainty to our calculation arising from this. With HEJ, we by default allow for the input of four different scale choices:

1. Fixed scale;

2. Maximum jet $p_\perp$;

3. $H_T/2$, which is the scalar sum of all parton $p_\perp$ divided by 2;

4. Invariant mass of the jets.

When calculating scale variations, we take the scale as defined and then vary independently $\mu_f$ and $\mu_r$ around it by factors of $2, \sqrt{2}, 1, \frac{1}{\sqrt{2}}, \frac{1}{2}$. Since there are five choices for each scale, we overall have 25 different results. However, since we expect $\mu_f$ and $\mu_r$ to be close to each other, we remove the points where $\mu_f/\mu_r > \sqrt{2}$ or $\mu_r/\mu_f > \sqrt{2}$, which leaves us with 19 options. These will form bands around a central prediction, which we then use as our estimation for the effect of scale variation for our results.

## 2.5 Experimental Analyses

We have spent the best part of this chapter arguing from a theoretical standpoint that the systematic treatment of High Energy logarithms is required for the accurate description of LHC data. We would like to end this chapter by providing plots from actual LHC analyses to show that the experimental data actually agrees with that statement. HEJ has been used in a wide range of experimental analyses [10, 12–14, 21, 22, 25] and we present a few select plots for discussion here. To understand them all fully, we first need to explain a couple of other features of HEJ we have not yet presented.

Firstly, HEJ includes a multiplicative matching to full LO calculations for matrix elements up to and including four final state jets. After a process has been clustered into final state jets, the matrix element is recalculated with the jet momenta[5] along with the full LO matrix element (currently provided by *MadGraph* [2]) and the process is multiplied by the ratio of the full LO result to our matrix element. We also include full LO matrix elements for non-FKL processes, which we saw are sub-leading in the high energy limit but important in other regions of phase space. Including matching in this way allows HEJ to be as competitive as other approaches in areas far away from the high energy limit it is designed for.

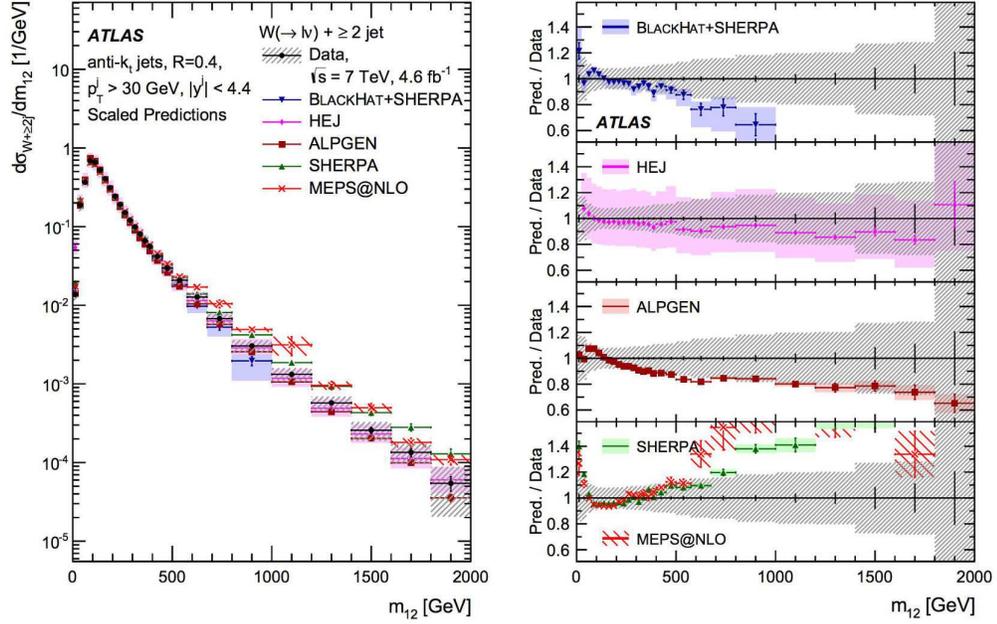

**Figure 2.12** *Differential cross-section in $m_{12}$ bins in the ATLAS study [14]*

Secondly, it is possible to interface HEJ with the *parton shower* ARIADNE [4]. A parton shower is designed to simulate the hadronisation of the event and as such is a completely distinct problem from that of calculating the hard scattering matrix element that HEJ provides. The specifics of the technique are interesting [42] but beyond the scope of this thesis. The important point is that a combined HEJ+ARIADNE program can describe the high energy behaviour of the hard scattering element along with the soft behaviour of the parton shower.

Figure 2.12 shows a range of predictions for a $W$ plus at least two jet event, binned in the invariant mass between the most forward and backward jet. The tail of this distribution is precisely where we expect the high energy logarithms to become important. The failure of the fixed order approaches to describe this region well clearly shows that this is indeed the case. The error band on the HEJ prediction is large and comes from the scale variations; the scale variation bands are not included on the other generators in the plot and would be just as large. The exception is the BlackHat line: their scale variation is shown and it is smaller than HEJ's because it is an NLO calculation, whereas HEJ only matches to LO. It is clear from the flatness of HEJ's line in the ratio plot that it is the only prediction tracking the data. Indeed, further investigation in [5] showed that the scale variation has the effect of an overall normalisation and has no bearing on the shape.

---

[5]Since some partons may not make it into final state jets, momentum conservation may not necessarily hold here. This is solved by distributing the momenta of partons that are not included into the jets in such a way as to keep it as close as possible to the original jet momenta.

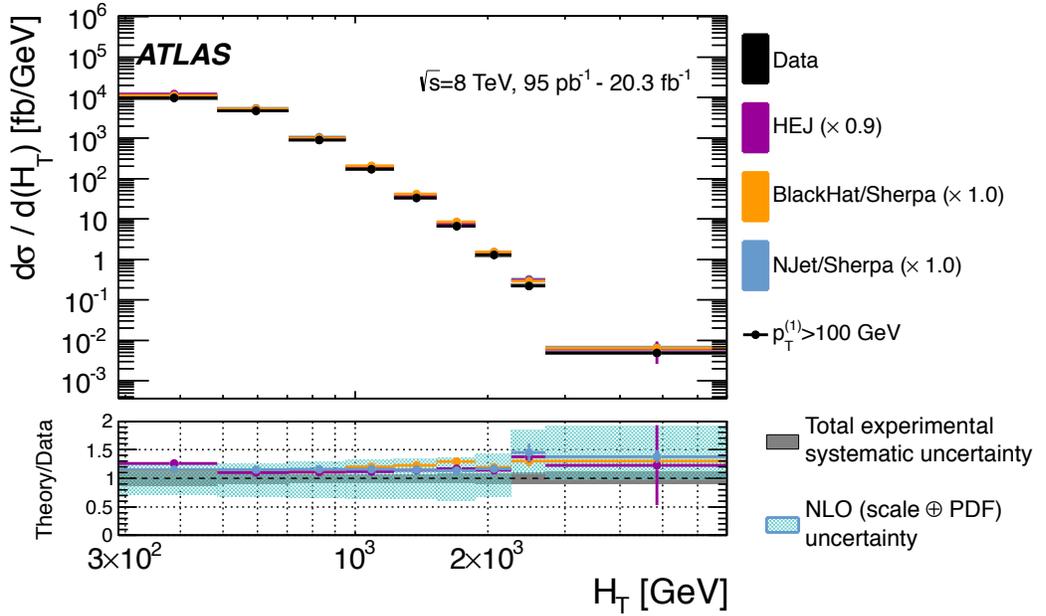

**Figure 2.13**  *Differential cross-section in HT (the scalar sum of all jet momenta) bins in a four jet ATLAS study [13].*

Figure 2.13 shows the cross-section for a process involving at least four final state jets binned in $H_T$, which is the scalar sum of all transverse momenta. The interesting feature of this plot is that HEJ continues to match the data out to high values of $H_T$, which correspondingly means high values of $p_\perp$ for (at least one of) the jets. The High Energy Limit that inspired HEJ holds only when the $p_\perp$ of the final state particles are all much smaller than the centre of mass energy and therefore should not be expected to provide a good description in the regime probed here. We see that HEJ keeps enough of the full process and this, along with the matching procedure, allows for it to be a fully completive description in a large part of the LHC phase space.

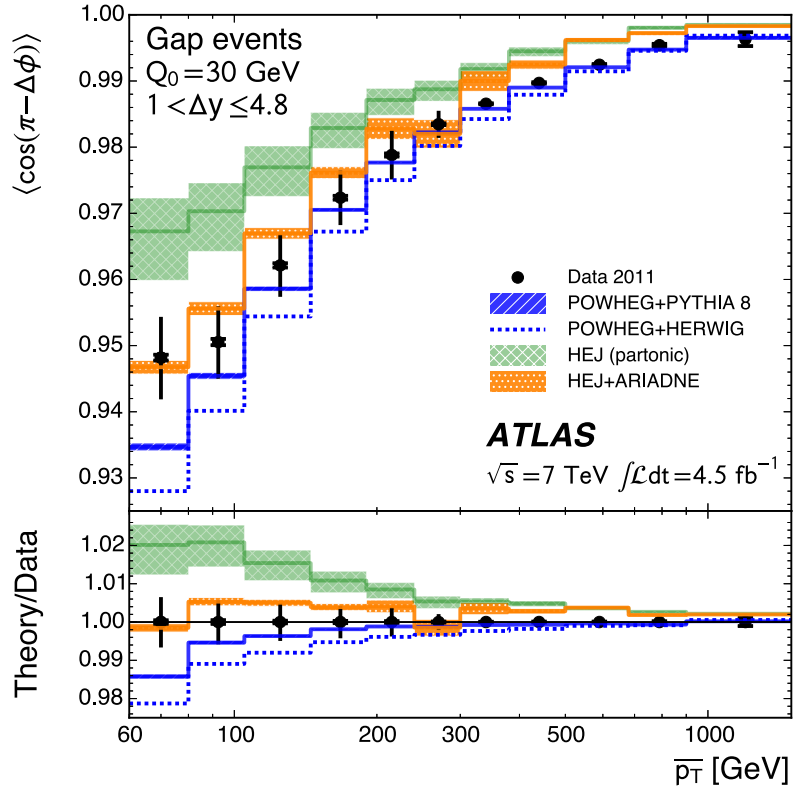

**Figure 2.14**  *Plot of an azimuthal decorrelation observable in average $p_\perp$ in the ATLAS analysis [12].*

Figure 2.14 is a plot from an analysis involving a jet veto. A scale (in this case, 30 GeV) is chosen whereby any extra jet emission in a rapidity gap between a defined two-jet system (here, the most forward and backward jets) is vetoed. Such a measurement is useful for testing perturbative predictions on the absence of activity in the gap; for example, we see from our HEJ amplitude that a large $\Delta y$ corresponds to a larger value of the resummation exponential. The predictions for both HEJ and HEJ+ARIADNE are shown for an observable related to the angular decorrelation of the dijet system. It is clear in this case that the best description of data comes from having both the high energy and the parton shower effects included; in the first few bins the partonic HEJ, which only includes the former, overshoots, and the POWHEG line, which only includes the latter, undershoots. The orange line (HEJ + ARIADNE), containing both, tracks the data excellently.

All of these analyses were performed during the first run of the LHC with a centre-of-mass energy of either 7 or 8 TeV. Run II will provide us with data for events at a centre of mass energy of 13 or 14 TeV and so we expect that the effect of these high energy logarithms will become even more prevalent in future analyses.

# Chapter 3

# Beyond Leading Log with High Energy Jets

Now that the driving principles behind HEJ have been explained, we move on to discuss recent improvements to the formalism. We begin by discussing how the form of our amplitudes allows us to also capture terms that are Next-to-Leading Logarithmic (NLL) in the perturbative series. We will go into the motivations for why we should do this before presenting a full description of how some of these terms are derived and incorporated. We then finalise the chapter by presenting new results clearly showing the beneficial effect of the addition of these contributions, both in terms of the HEJ program and in comparison to real LHC data.

## 3.1 Motivations for NLL

There are a few reasons one might consider trying to go beyond LL with the formalism. Firstly, it has been proved that the Lipatov ansatz is valid at the NLL level [36] and so any sub-leading amplitude that still factors out into a product of $t$-channel poles can be resummed in the same way as before. The full form of the trajectory $\hat\alpha$, which we derived at the leading order in section 2.1.1, of course becomes more complicated [34], but the important point is that the form of the Lipatov ansatz (equation 2.43) still holds. Secondly, by having access to these sub-leading terms we also expect to reduce the scale variation bands on our calculation, since these variations are directly related to how much control we have over higher order terms. In order to claim full NLL accuracy, we must have:

1. Calculations of FKL amplitudes but with the rapidity ordering of one emitted gluon disturbed (one gluon is allowed to be emitted with a rapidity that is outside of an extremal parton). We call such contributions 'unordered' processes.

2. Corrections to the Lipatov vertex for the emission of gluons.

3. Calculations of inherently non-FKL amplitudes (those which cannot be drawn with only $t$-channel gluon exchanges) that contribute at the NLL level in the jet cross-section.

The mathematics of the first point had already been completed and the author's contribution to this project was to incorporate these routines in the HEJ program's pure jet production section. The second point has not yet been attempted but remains a long-term goal of the collaboration. This thesis will consider the first and final points, providing a brief overview of the former and a complete description and derivation of the author's individual work for the latter. Including such considerations will mark an important step towards full NLL HEJ and, in the case of the addition of the non-FKL subprocesses, it extends the applicability of HEJ since we will now have a LL description of these events; in conclusion, we will be able to resum more of the contribution to the cross-section.

## 3.2 Unordered Emissions in Pure Jets

The derivation of the amplitude for an unordered (or simply 'uno') contribution revolves around extending our current formalism to include the effect of a gluon being emitted outside the FKL strongly rapidity-ordered chain. We can express this by introducing a modified current object that depends on the rapidity of this unordered parton and the parton next to it in rapidity that does satisfy the ordering. For example, the amplitude for $qQ \to gqQ$ would behave as

$$M^{uno}_{qQ \to gqQ} \sim \frac{j^{\mu}_{uno}(p_a, p_1, p_{uno}) j_{\mu}(p_b, p_2)}{\hat{t}_2}, \qquad (3.1)$$

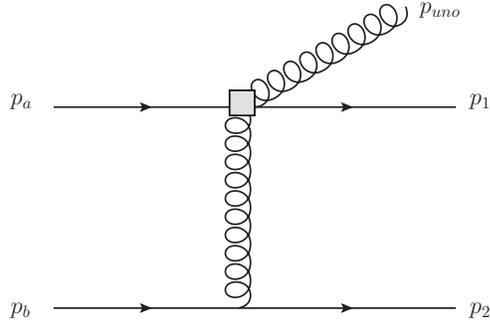

**Figure 3.1** *A schematic view of an unordered emission amplitude.*

where $y_{uno} \sim y_1$ and $y_1 \gg y_2$. Throughout the following sections, we will make clear where $t$-channel poles appear by employing the notation $\hat{t}_i = p_a - \sum_{j=1}^{j=i} p_j$ and such invariants will *always* be factored out of the final expression. For other instances where the use of a $t$-channel propagator still makes sense, we use the notation $t_{ij}$, which is to be interpreted as $(p_i - p_j)^2 = -s_{ij}$. Returning to the unordered amplitude, we effectively collapse the gluon emission to a point along the usual current such that there is only one suitable $t$-channel pole to be resummed, as opposed to the two we would get if the gluon were emitted in the FKL ordering – it is thus clear to see why this is a sub-leading contribution. Diagrammatically, we can represent this as shown in figure 3.1. We should also keep in mind that discussing the colour properties of this amplitude will be more complicated than the FKL case and the result will have to be treated more carefully in this regard. To derive the form of the uno current, we will recalculate the $qQ \to gqQ$ amplitude but with the consideration that the rapidity of the gluon is no longer far away from the rapidity of the forward quark current. What this will essentially mean is that the kinematic arguments for dropping some terms, as was done in section 2.2.3, will no longer be valid. We will therefore start by writing the full LO result for this amplitude (where we will write $p_{uno} = p_g$ for

brevity):

$$\begin{aligned}
M^{LO}_{qQ\to gqQ} =& (ig_s)^3 T^c_{1i} T^d_{ia} T^d_{2b} \varepsilon_\nu(p_g) \frac{\langle 1|\nu|g\rangle \langle g|\mu|a\rangle + 2p_1^\nu \langle 1|\mu|a\rangle}{s_{1g}\hat{t}_2} \langle 2|\mu|b\rangle \\
& - (ig_s)^3 T^d_{1i} T^c_{ia} T^d_{2b} \varepsilon_\nu(p_g) \frac{2p_a^\nu \langle 1|\mu|a\rangle - \langle 1|\mu|g\rangle \langle g|\nu|a\rangle}{s_{ag}\hat{t}_2} \langle 2|\mu|b\rangle \\
& + (ig_s)^3 T^c_{2i} T^d_{ib} T^d_{1a} \varepsilon_\nu(p_g) \frac{\langle 2|\nu|g\rangle \langle g|\mu|b\rangle + 2p_2^\nu \langle 2|\mu|b\rangle}{s_{2g} t_{a1}} \langle 1|\mu|a\rangle \\
& - (ig_s)^3 T^d_{2i} T^c_{ib} T^d_{1a} \varepsilon_\nu(p_g) \frac{2p_b^\nu \langle 2|\mu|b\rangle - \langle 2|\mu|g\rangle \langle g|\nu|b\rangle}{s_{bg} t_{a1}} \langle 1|\mu|a\rangle \\
& - g_s^3 f^{dec} T^d_{1a} T^e_{2b} \varepsilon_\nu(p_g) \frac{\langle 1|\rho|a\rangle \langle 2|\mu|b\rangle}{t_{a1}\hat{t}_2} (2p_g^\mu \eta^{\nu\rho} - 2p_g^\rho \eta^{\mu\nu} - (q_1+q_2)^\nu \eta^{\mu\rho}),
\end{aligned} \quad (3.2)$$

where $q_1 = p_a - p_1 = p_2 - p_b + p_g$ and $q_2 = p_2 - p_b = p_a - p_1 - p_g$. With the full expression available, we can investigate which terms we can still drop in this new limit $y_g \sim y_1 \gg y_2$. We see that the first term in the third line and the second term in the fourth are the only ones we can drop because (depending on helicities) the $\mu$ contraction will give something that scales as $\sqrt{s_{ag}}$ or $\sqrt{s_{g1}}$, which are now small invariants in comparison to all other scales. By dropping these terms, all remaining terms are proportional to $\langle 2|\mu|b\rangle$ and so by comparison to equation 3.1, the sum of the terms multiplying this current will give us our unordered current. However, to be truly consistent with the factorised picture, we must factorise out the colour factor $T^d_{2b}$ from this amplitude as well. This is already the case for the first, second and last lines and now that we have dropped terms for the other two lines we can use the (still valid) limit $p_2 \sim p_b$ to yield

$$\begin{aligned}
& -ig_s^3 \langle 1|\mu|a\rangle \langle 2|\mu|b\rangle \varepsilon_\nu(p_g) \left(\frac{2p_2^\nu}{t_{a1}s_{2g}} - \frac{2p_b^\nu}{t_{a1}s_{bg}}\right) \\
& \approx -ig_s^3 \langle 1|\mu|a\rangle \langle 2|\mu|b\rangle \varepsilon_\nu(p_g) \frac{1}{t_{a1}}\frac{2p_b^\nu}{s_{bg}} T^d_{1a}(T^c_{2i}T^d_{ib} - T^d_{2i}T^c_{ib}) \\
& = g_s^3 \langle 1|\mu|a\rangle \langle 2|\mu|b\rangle \varepsilon_\nu(p_g) \frac{1}{t_{a1}}\frac{2p_b^\nu}{s_{bg}} f^{cde} T^d_{1a} T^e_{2b} \\
& = g_s^3 \langle 1|\mu|a\rangle \langle 2|\mu|b\rangle \varepsilon_\nu(p_g) f^{cde} T^d_{1a} T^e_{2b} \frac{1}{t_{a1}} \left(\frac{p_b^\nu}{s_{bg}} + \frac{p_2^\nu}{s_{2g}}\right),
\end{aligned} \quad (3.3)$$

where in the last line we have restored the symmetry between $p_2$ and $p_b$. Our factorised amplitude is then

$$M^{uno,fact}_{qQ\to gqQ} = -g_s^3 \frac{\langle 2|\mu|b\rangle}{\hat{t}_2} T^d_{2b} \left(iT^c_{1i}T^d_{ia}U_1^{\mu\nu} + iT^d_{1i}T^c_{ia}U_2^{\mu\nu} + f^{ecd}T^e_{1a}L^{\mu\nu}\right), \quad (3.4)$$

where

$$\begin{aligned}
U_1^{\mu\nu} &= \frac{1}{s_{1g}}(j_{1g}^\nu j_{ga}^\mu + 2p_1^\nu j_{1a}^\mu) \\
U_2^{\mu\nu} &= -\frac{1}{s_{ag}}(2j_{1a}^\mu p_a^\nu - j_{1g}^\mu j_{ga}^\nu) \\
L^{\mu\nu} &= \frac{1}{t_{a1}}\left(-2p_g^\mu j_{1a}^\nu + 2p_g \cdot j_{1a}\eta^{\mu\nu} + (q_1+q_2)^\nu j_{1a}^\mu + q_2^2 j_{1a}^\mu\left(\frac{p_2^\nu}{s_{2g}} + \frac{p_b^\nu}{s_{gb}}\right)\right).
\end{aligned} \quad (3.5)$$

Gauge invariance of this expression has been checked by replacing the polarisation vector with the gluon momentum and seeing that the expression gives zero, in accordance with the Ward Identity. This current can then be used as a basis for all unordered amplitudes, with further emissions included via the Lipatov vertex and other incoming states accounted for by multiplications of $C_F$ and $\tilde{C}_A$ as appropriate. One thing to note, however, is that the leg that the unordered current is dependent on cannot be a gluon, since a trivial rewriting of momenta in that case will lead back to the FKL ordering.

## 3.3 Calculations of NLL Partonic Subprocesses

It was discussed in chapter 2 that the dominant amplitudes in the High Energy Limit are given by those that involve the maximal number of gluon exchanges in the $t$-channel by analysis of Regge Theory. To access the sub-leading partonic configurations, we simply replace one gluon propagator by one quark propagator in these FKL amplitudes (whilst keeping the strict rapidity ordering). There are two distinct possibilities we can imagine: we can either replace the first or last propagator in the chain, or one in the middle along the chain. We can assign the nomenclature 'extremal' and 'central' to the two cases respectively. The simplest case of an extremal process is $qg \to qQ\bar{Q}$ and for the central case it is $qq' \to qQ\bar{Q}q'$. From analysis of these amplitudes, we can derive the 'building blocks' that will allow us to build up other, related amplitudes by multiplication of Lipatov vertices and colour factors.

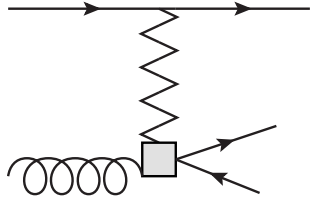

**Figure 3.2** *Schematic diagram involving $g \to q\bar{q}$ impact factor.*

### 3.3.1 Calculation of $qg \to qQ\bar{Q}$ in the High Energy Limit

The ultimate aim of this calculation will be to factorise the $qg \to qQ\bar{Q}$ amplitude into an expression of the form

$$M_{qg \to qQ\bar{Q}} \sim \frac{\langle 1|\mu|a\rangle \, Q^{\mu\nu}(p_2, p_3, p_b)\varepsilon_\nu(p_b)}{\hat{t}_1}, \qquad (3.6)$$

where $Q^{\mu\nu}$ is an effective vertex that encapsulates the effect of the emission of a quark/anti-quark pair at the end of the rapidity chain. We can express this equation in the form a diagram; see figure 3.2. Given the momentum dependence of the vertex and how it looks schematically, we can also interpret $Q^{\mu\nu}$ as a $g \to Q\bar{Q}$ impact factor.

The technique for this is as follows: we will first study the complete amplitude for $qg \to qQ\bar{Q}$, for which there are five contributing diagrams as shown in figure 3.3. After we have the full LO expression, we will make some approximations based on the High Energy behaviour of the process to bring the entire amplitude into the desired form. During this approximation stage, we must remember to take care to maintain gauge invariance since our final expression will be applied in all of phase space. We remind the reader at this point that, in the massless quark limit, $u^\pm(p) = v^\mp(p)$ and so the notation $\langle p|\mu|k\rangle$ can refer to $\bar{u}_p\gamma^\mu v_k$ or $\bar{u}_p\gamma^\mu u_k$ interchangeably with no practical need to distinguish.

We will begin with the diagram shown in the top left of figure 3.3 and then proceed from left to right and top to bottom. Using the Feynman rules, we see

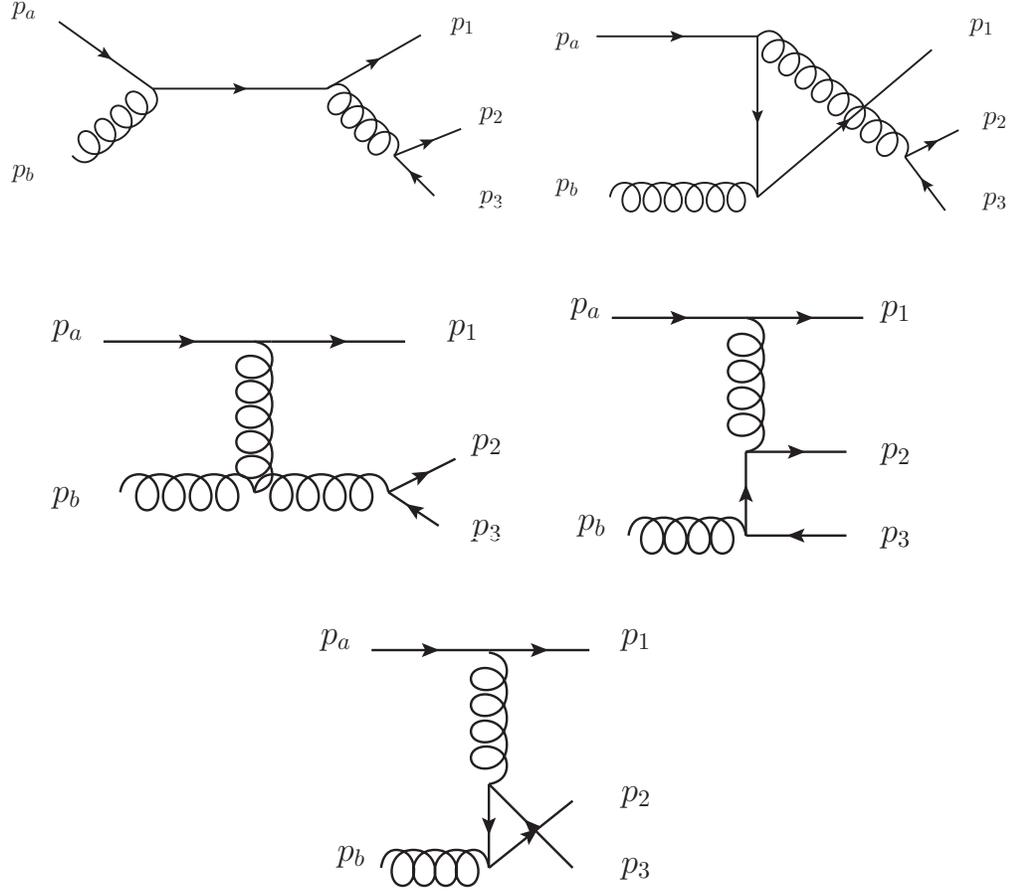

**Figure 3.3** *All LO graphs for $qg \to qQ\bar{Q}$.*

this contribution is

$$iM_1 = \bar{u}_1(-ig_s\gamma^\nu T^g_{1q})\frac{i(\slashed{p}_a + \slashed{p}_b)}{s_{ab}}(-ig_s\gamma^\mu T^b_{qa})u_a\varepsilon_\mu(p_b)\frac{-i\eta^{\nu\sigma}}{s_{23}}\langle 2|\sigma|3\rangle(-ig_s T^g_{23})$$
$$= \frac{ig_s^3 T^g_{1q}T^b_{qa}T^g_{23}}{s_{ab}s_{23}}\left[\bar{u}_1\gamma^\nu(\slashed{p}_a + \slashed{p}_b)\gamma^\mu u_a\right]\langle 2|\nu|3\rangle\,\varepsilon_\mu(p_b).$$

(3.7)

For the diagram involving a $u$-channel quark propagator, the expression is very similar:

$$iM_2 = \frac{-ig_s^3 T^b_{1q}T^g_{qa}T^g_{23}}{s_{1b}s_{23}}\left[\bar{u}_1\gamma^\mu(\slashed{p}_1 - \slashed{p}_b)\gamma^\nu u_a\right]\langle 2|\nu|3\rangle\,\varepsilon_\mu(p_b). \qquad (3.8)$$

The next diagram involves a three-gluon vertex and by invoking the Feynman

rules we get

$$iM_3 = [\bar{u}_1(-ig_s\gamma^\mu T^g_{1a})u_a]\frac{-i\eta_{\mu\nu}}{\hat{t}_1}\left[\bar{u}_2(-ig_s\gamma^\rho T^{g'}_{23})v_3\right]\frac{-i\eta_{\beta\rho}}{s_{23}}(-g_s f^{gg'b})V_{3g}^{\nu\beta\alpha}\varepsilon_\alpha(p_b), \tag{3.9}$$

where

$$\begin{aligned}V_{3g}^{\nu\beta\alpha} &= (q_1+q_2)^\alpha\eta^{\nu\beta} - (q_2+p_b)^\nu\eta^{\beta\alpha} + (p_b-q_1)^\beta\eta^{\alpha\nu}\\ &= (2p_2+2p_3)^\alpha\eta^{\nu\beta} - (2p_b)^\nu\eta^{\beta\alpha} + (2p_b)^\beta\eta^{\alpha\nu},\end{aligned} \tag{3.10}$$

once terms that are zero when contracted with terms outside of the three-gluon vertex are removed. Algebraic manipulation of this expression leads to

$$iM_3 = \frac{-g_s^3 T^g_{1a}T^{g'}_{23}f^{gg'b}}{\hat{t}_1 s_{23}}\langle 1|\nu|a\rangle\langle 2|\beta|3\rangle V_{3g}^{\nu\beta\alpha}\epsilon_\alpha(p_b). \tag{3.11}$$

Finally, we have the last two diagrams that involve both a $t$-channel gluon and quark propagator. For the first, we have

$$iM_4 = [\bar{u}_1(-ig_s\gamma^\mu T^g_{1a})u_a]\frac{-i\eta_{\mu\nu}}{\hat{t}_1}\left[\bar{u}_2(-ig_s\gamma^\nu T^g_{2q})\frac{-i(\slashed{p}_3-\slashed{p}_b)}{t_{3b}}(-ig_s\gamma^\rho T^b_{q3})v_3\right]\varepsilon_\rho(p_b). \tag{3.12}$$

Note the minus sign in the propagator; this is because the Feynman rule for the quark propagator requires that the momentum flows in the same direction as the charge. This can be written as

$$iM_4 = \frac{-ig_s^3 T^g_{1a}T^b_{q3}T^g_{2q}}{\hat{t}_1 t_{3b}}\langle 1|\nu|a\rangle\left[\bar{u}_2\gamma^\nu(\slashed{p}_3-\slashed{p}_b)\gamma^\rho v_3\right]\varepsilon_\rho(p_b). \tag{3.13}$$

A similar analysis for the last diagram (no minus sign from the fermion propagator this time) gives

$$iM_5 = \frac{ig_s^3 T^g_{1a}T^b_{2q}T^g_{q3}}{\hat{t}_1 t_{b2}}\langle 1|\nu|a\rangle\left[\bar{u}_2\gamma^\rho(\slashed{p}_2-\slashed{p}_b)\gamma^\nu v_3\right]\varepsilon_\rho(p_b). \tag{3.14}$$

We now have expressions that, when summed, will give the exact, LO result for the process $qg \to qQ\bar{Q}$. Now we must approximate in order to factor out the $t$-channel pole. No approximation is required for $M_3$, $M_4$ or $M_5$ since the $t$-channel pole is immediately explicit. $M_1$ and $M_2$, however, need special attention. The problematic part is the square brackets of, for example, $M_1$, which we can rewrite using the completeness relation:

$$\bar{u}_1\gamma^\nu(\slashed{p}_a+\slashed{p}_b)\gamma^\mu u_a = \langle 1|\nu|a\rangle 2p_a^\mu + \langle 1|\nu|b\rangle\langle b|\mu|a\rangle, \tag{3.15}$$

where we continue our notation of not assigning a helicity index to the spinor brackets to indicate that the expansion is valid for both negative and positive helicities. The $\nu$ index is contracted with the quark current $\langle 2|\nu|3\rangle$. Depending on the helicity choices, the second term after contraction varies either as $\sqrt{s_{b3}s_{12}}$ or $\sqrt{s_{13}s_{b2}}$. Similarly, the first term varies either as $\sqrt{s_{a3}s_{12}}$ or $\sqrt{s_{13}s_{a2}}$. The relative size of these terms is then $\sqrt{s_{b3}/s_{a3}}$ or $\sqrt{s_{b2}/s_{a2}}$. In the High Energy limit, it is clear that both $s_{a3}$ and $s_{a2}$ are large. Also, since we are dealing with the case where the $Q\bar{Q}$ pair is emitted close in rapidity to one end of the chain, we can reasonably assume that $s_{b3}$ and $s_{b2}$ do not have to be large. We can therefore drop the second term with respect to the first and so

$$iM_1 \approx \frac{ig_s^3 T_{1q}^g T_{qa}^b T_{23}^g}{s_{ab}s_{23}} \left[2p_a^\mu \langle 1|\nu|a]\right] \langle 2|\nu|3\rangle \, \varepsilon_\mu(p_b). \tag{3.16}$$

A similar argument holds for $M_2$ and so

$$iM_2 \approx \frac{-ig_s^3 T_{1q}^b T_{qa}^g T_{23}^g}{s_{1b}s_{23}} \left[2p_1^\mu \langle 1|\nu|a]\right] \langle 2|\nu|3\rangle \, \varepsilon_\mu(p_b). \tag{3.17}$$

We can now take the limit $p_a \sim p_1$, which allows us to combine these two diagrams by using the colour commutator result

$$(T_{1q}^g T_{qa}^b - T_{1q}^b T_{qa}^g)T_{23}^g = if^{gbc}T_{1a}^c T_{23}^g. \tag{3.18}$$

This is the same colour factor as that of the diagram involving a $t$-channel gluon exchange under the relabelling $g \to c$ and $g' \to g$. Because of this, we will from now on call the result of this $iC_t$. Thus

$$i(M_1 + M_2) \to \frac{-g_s^3 C_t}{s_{ab}s_{23}} 2p_a^\mu \langle 1|\nu|a\rangle \langle 2|\nu|3\rangle \, \varepsilon_\mu(p_b), \tag{3.19}$$

and so we have now factored out the quark current in all the amplitudes. If we now combine all the amplitudes together, then we obtain

$$\begin{aligned}Q^{\mu\nu} &= -\frac{C_1}{t_{b3}}\left(\bar{u}_2\gamma^\mu(\slashed{p}_3 - \slashed{p}_b)\gamma^\nu v_3\right) + \frac{C_2}{t_{b2}}\left(\bar{u}_2\gamma^\nu(\slashed{p}_2 - \slashed{p}_b)\gamma^\mu v_3\right) \\ &+ i\frac{C_t}{s_{23}}\left(\frac{2\,p_a^\nu\, q_1^2}{s_{ab}}\langle 2|\mu|3\rangle + V_{3g}^{\mu\rho\nu}\langle 2|\rho|3\rangle\right),\end{aligned} \tag{3.20}$$

where we have relabelled some Lorentz indices to conform with equation 3.6 and

$$C_1 = T^g_{1a} T^b_{q3} T^g_{2q}, \tag{3.21a}$$

$$C_2 = T^g_{1a} T^b_{2q} T^g_{q3}, \tag{3.21b}$$

$$C_t = f^{gbc} T^c_{1a} T^g_{23}. \tag{3.21c}$$

We should check at this point that our expression is indeed still gauge invariant after having made these approximations. The simplest way is to make use of the Ward Identity, which implies $Q^{\mu\nu} p_{b,\nu} = 0$ when gauge invariance is satisfied. Explicitly

$$\begin{aligned}
Q^{\mu\nu} p_{b,\nu} &= -\frac{C_1}{t_{b3}} \left( \bar{u}_2 \gamma^\mu \slashed{p}_3 \slashed{p}_b v_3 \right) + \frac{C_2}{t_{b2}} \left( \bar{u}_2 \slashed{p}_b \slashed{p}_2 \gamma^\mu v_3 \right) + i \frac{C_t}{s_{23}} \left( q_1^2 \langle 2|\mu|3] + V^{\mu\rho\nu}_{3g} p_{b\nu} \langle 2|\rho|3] \right) \\
&= -\frac{C_1}{t_{b3}} s_{3b} \langle 2|\mu|3] + \frac{C_2}{t_{b2}} s_{2b} \langle 2|\mu|3] + i \frac{C_t}{s_{23}} (q_1^2 \langle 2|\mu|3] + (s_{2b} + s_{3b}) \langle 2|\mu|3] \\
&\quad - 2 p^\mu_b p^\rho_b \langle 2|\rho|3] + 2 p^\mu_b p^\rho_b \langle 2|\rho|3]) \\
&= \frac{\langle 2|\mu|3]}{s_{23}} (C_1 s_{23} - C_2 s_{23} + i C_t (q_1^2 + (s_{2b} + s_{3b}))) \\
&\quad + i \langle 2|\rho|3] \frac{C_t}{s_{23}} (-2 p^\mu_b p^\rho_b + 2 p^\mu_b p^\rho_b) \\
&= i C_t \frac{\langle 2|\mu|3]}{s_{23}} (-s_{23} + (-s_{2b} - s_{3b} + s_{23} + s_{2b} + s_{3b})) \\
&= 0,
\end{aligned} \tag{3.22}$$

where we have used the result $C_1 - C_2 = -i C_t$. We therefore conclude that the effective vertex is gauge invariant in all of phase space. It may seem, however, that the effective vertex is not truly factorised since there is a clear instance of $p_a$ in the vertex. Because the complete term goes as $p^\mu_a / s_{ab}$, it is actually independent of $p_a$[1]. We still have freedom to make a gauge choice for our calculations, however, and a good choice will be the gauge where the gluon polarisation vector is orthogonal to $p_a$, so that

$$\begin{aligned}
Q^{\mu\nu}_{gauge} &= -\frac{C_1}{t_{b3}} \left( \bar{u}_2 \gamma^\mu (\slashed{p}_3 - \slashed{p}_b) \gamma^\nu v_3 \right) + \frac{C_2}{t_{b2}} \left( \bar{u}_2 \gamma^\nu (\slashed{p}_2 - \slashed{p}_b) \gamma^\mu v_3 \right) \\
&\quad + i \frac{C_t}{s_{23}} \left( (2 p_2 + 2 p_3)^\nu \eta^{\mu\rho} - 2 p^\mu_b \eta^{\nu\rho} + 2 p^\rho_b \eta^{\nu\mu} \right) \langle 2|\rho|3],
\end{aligned} \tag{3.23}$$

completing our calculation for the basis HEJ amplitude in the case of an extremal

---
[1] An easy way to convince oneself of this is to work in light-cone co-ordinates and see that the $p^+_a$ scale factors out on the top and bottom of the expression.

$Q\bar{Q}$ process. Since there are only two independent colour factors in this expression we could rewrite the result in terms of $C_1$ and $C_2$ only, but the author is of the opinion that keeping the terms along with their 'natural' colour factors like this is more computationally beneficial and easier to understand. One other interesting point to note is that this effective vertex can be shown to be related to the unordered vertex via crossing symmetry. The full calculation of this is shown in Appendix A.

### 3.3.2 Verifications of the Extremal $Q\bar{Q}$ Vertex

In order to check the derivation of this vertex, we will explicitly investigate how the amplitudes that contain it behave in the MRK limit. We discussed before that we expect them to be suppressed by the invariant $s_{Q\bar{Q}}$ with respect to the leading FKL configurations at the $|M|^2$ level. Since the FKL amplitudes behaved as $\hat{s}^2$ at the $|M|^2$ level, we should see a systematic suppression of these new amplitudes if we plot $|M|^2/\hat{s}^2$ and furthermore multiplication of $s_{Q\bar{Q}}$ should combat this. In these amplitudes, we have a more complicated colour structure than we did before, since the effective vertex depends on three colour factors (although only two are actually independent, of course). At the $|M|^2$ level, we must deal with this correctly when performing the colour sum. This is done by splitting up the vertex into sub-vertices, each one of which is associated with one colour factor. We choose to represent this as

$$Q^{\mu\nu} = C_1 Q_1^{\mu\nu} + C_2 Q_2^{\mu\nu} + C_t Q_t^{\mu\nu}, \qquad (3.24)$$

which allows us to calculate the squared amplitude in the following way:

$$|M_{qg \to qQ\bar{Q}}|^2 \sim \frac{1}{24}\bigg(|C_1|^2 Q_1 \cdot Q_1^\dagger + |C_2|^2 Q_2 \cdot Q_2^\dagger + |C_t|^2 Q_t \cdot Q_t^\dagger$$
$$+ 2Re(C_1 C_2^\dagger Q_1 \cdot Q_2^\dagger) + 2Re(C_1 C_t^\dagger Q_2 \cdot Q_t^\dagger) + 2Re(C_2 C_t^\dagger Q_2 \cdot Q_t^\dagger)\bigg), \qquad (3.25)$$

where the pre-factor comes from the colour averaging and the colour sum can be explicitly performed. The helicity sum is also performed explicitly and the average brings about a further factor of 1/4. In order to check how our matrix element performs against the full leading order result (taken from MadGraph), we plot the value of $|M|^2/\hat{s}^2$ through a slice of phase space. The momenta are

chosen such that

$$\begin{aligned}
p_1 &= (40\cosh(\Delta), 0, 40, 40\sinh(\Delta)), \\
p_2 &= (40\sqrt{2}, 40, -40, 0), \\
p_3 &= (40\cosh(-\Delta), -40, 0, 40\sinh(-\Delta)),
\end{aligned} \quad (3.26)$$

where $\Delta$ is the rapidity of the extremal jets. Therefore, increasing $\Delta$ corresponds to approaching the high energy limit. The results are plotted in figure 3.4. We see that the two calculations follow each other very closely and we also see the suppression at large $\Delta$ as we expect. In figure 3.5, we multiply this result by $s_{Q\bar{Q}}$ and see that the results tend to a finite constant. Hence, the amplitude is behaving how Regge theory predicts.

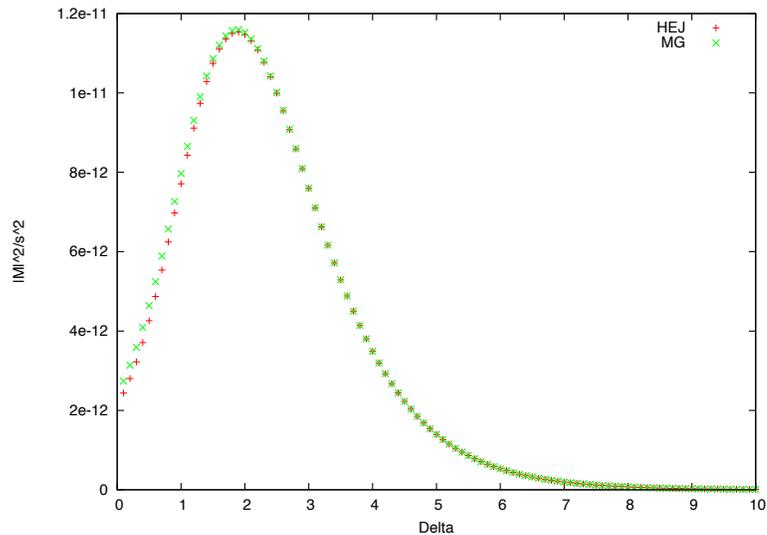

**Figure 3.4**  *Effective vertex approach to the $qg \to qQ\bar{Q}$ amplitude (red) compared to the full LO (green).*

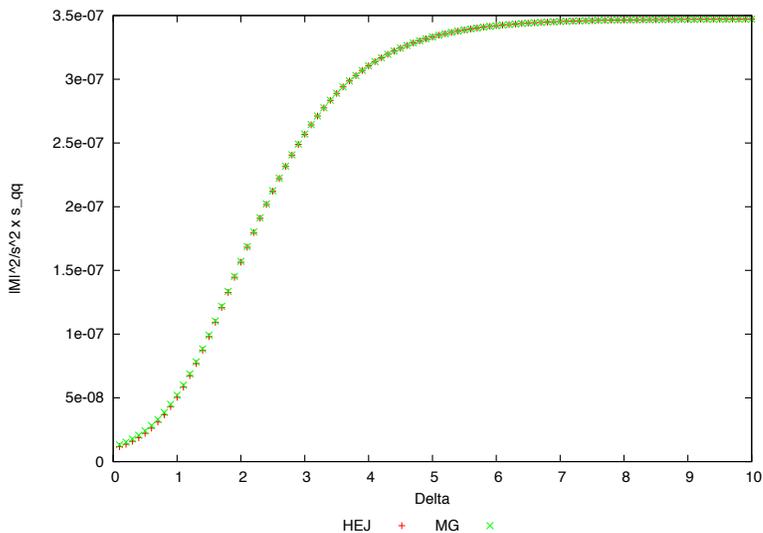

**Figure 3.5** *Effective vertex approach to the $qg \to qQ\bar{Q}$ amplitude (red) compared to the full LO (green) multiplied by the invariant mass of the quark/anti-quark pair.*

Given these plots, we are satisfied that the 'base' amplitude works as expected. To extend it, we need to be able to generalise to an arbitrary incoming state and handle extra gluon emissions. The first of these can be achieved by a simple multiplication of a colour factor at the $|M|^2$ level. Since the amplitude must contain at least one incoming gluon, there is only one extra initial state we can have:

$$|M_{gg \to gQ\bar{Q}}|^2 \sim \frac{\tilde{C}_A}{C_F} |M_{qg \to qQ\bar{Q}}|^2, \qquad (3.27)$$

where $\tilde{C}_A$ is as defined in equation 2.29. We plot this result (multiplied by $s_{Q\bar{Q}}$) along with the full leading order in figure 3.6. The difference between the two lines is minimal and the only noticeable difference is in the low $\Delta$ regime, where it should be expected to be different since the approximations valid in the High Energy Limit are less accurate here.

We then move on to discussing how extra gluon emissions are added to the amplitude. Because of the factorisation properties of the High Energy Limit, so long as we assume the extra gluon emissions are far away in rapidity from the partons already in the amplitude, we can simply insert a Lipatov vertex (defined in equation 2.40), along with a colour factor, and then divide by additional $t$-channel poles that will appear. This yields simply

$$|M_{qg \to q...Q\bar{Q}}|^2 \sim |M_{qg \to qQ\bar{Q}}|^2 \times \prod_{i=1}^{n-3} C_A \left( \frac{-V(q_i, q_{i+1}) \cdot V(q_i, q_{i+1})}{q_i^2 q_{i+1}^2} \right), \qquad (3.28)$$

where we have decided to define a division of $q_1^2 q_{n-3}^2$ within the matrix element squared[2]. Once more we plot this result against the full LO calculation in figure 3.7 and show that the agreement is still reasonable.

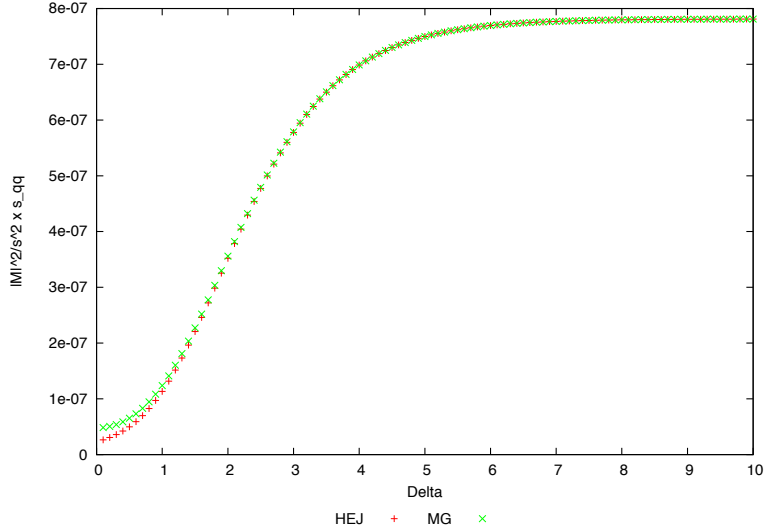

**Figure 3.6** *Effective vertex approach to the $gg \to gQ\bar{Q}$ amplitude (red) compared to the full LO (green) multiplied by the invariant mass of the quark/anti-quark pair.*

---

[2]The matrix element at tree-level has a division of $q_1^2$ and so a straight squaring of it would yield a division of $q_1^4$. However, once further emissions are added, we have the freedom to decide whether the $q_i^2$ terms should be associated with the Lipatov vertex or the base amplitude as we please, allowing for this form of the generalised matrix element. The important point is to include all $q_i^2$ divisions with the correct powers in the final expression.

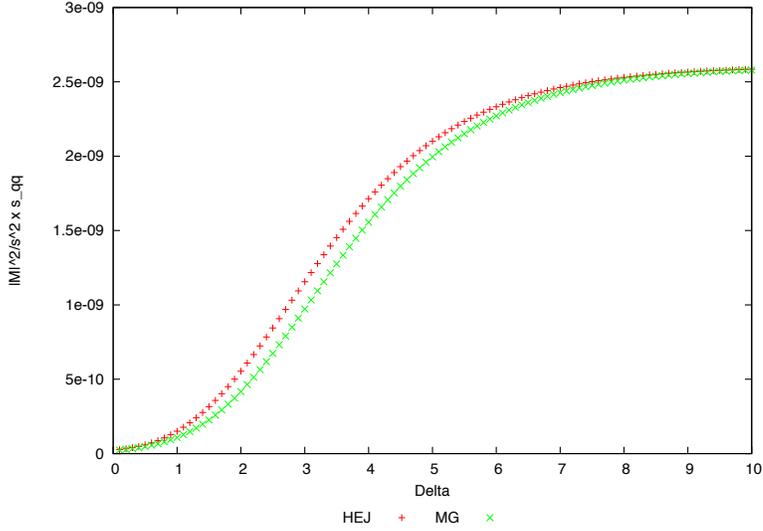

**Figure 3.7** *Effective vertex approach to the $qg \to qgQ\bar{Q}$ amplitude (red) compared to the full LO (green) multiplied by the invariant mass of the quark/anti-quark pair.*

### 3.3.3 Calculation of $qq' \to qQ\bar{Q}q'$ in the High Energy Limit

The technique for calculating the amplitude for the central process as shown in figure 3.8 is precisely the same as the one for the extremal process. In this case, we are searching for an amplitude of the form

$$M_{qq' \to qQ\bar{Q}q'} \sim \frac{\langle 1|\mu|a\rangle \, X^{\mu\nu} \, \langle 4|\nu|b\rangle}{\hat{t}_1 \hat{t}_3}. \qquad (3.29)$$

There are a total of 7 diagrams to calculate here, shown in figure 3.9. Once more, we will calculate the diagrams starting with the one in the top left and proceeding left to right. The expression for the first diagram is

$$iM_1 = \frac{-ig_s^4 T_{1q}^e T_{qa}^g T_{23}^e T_{4b}^g}{s_{23}\hat{t}_3} \left[\bar{u}_1 \gamma^\mu \frac{(\not{p}_1 + \not{p}_2 + \not{p}_3)}{(p_1+p_2+p_3)^2} \gamma^\rho u_a\right] [\bar{u}_2 \gamma_\mu v_3] [\bar{u}_4 \gamma_\rho u_b]. \qquad (3.30)$$

It is immediately clear that this diagram will not factorise into our desired form, so we expand the square bracket as a spinor chain again to attempt to make some

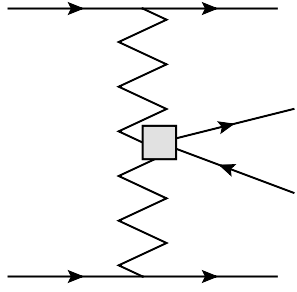

**Figure 3.8** *Effective description of $qq' \to qQ\bar{Q}q'$*

approximations:

$$\frac{1}{s_{12} + s_{13} + s_{23}} \left( \langle 1|\mu|1 \rangle \langle 1|\rho|a \rangle + \langle 1|\mu|2 \rangle \langle 2|\rho|a \rangle + \langle 1|\mu|3 \rangle \langle 3|\rho|a \rangle \right). \quad (3.31)$$

Depending on the helicities, either the second or third term in this bracket is identically zero when contracted with the quark current $\langle 2|\mu|3 \rangle$. Once more, we can use a scaling argument to eliminate other terms in this string. The first term contracted with $\langle 2|\mu|3 \rangle$ will scale as $\sqrt{s_{12}s_{13}}$ regardless of helicity choices. For the other non-zero term, the scaling will be $\sqrt{s_{23}s_{12}}$. We have no requirement that $s_{23}$ be large, but all other invariants *are* large and thus we can approximate the sub-amplitude by keeping the first term only:

$$iM_1 \approx \frac{-ig_s^4 T_{1q}^e T_{qa}^g T_{23}^e T_{4b}^g}{s_{23}\hat{t}_3(s_{12} + s_{13})} \langle 1|\rho|a \rangle \langle 4|\rho|b \rangle \times 2p_1^\mu \langle 2|\mu|3 \rangle. \quad (3.32)$$

The next graph is very similar to the previous and so the calculation proceeds in the same fashion as before. We will then skip to the final expression (having again made the relevant approximation) which is

$$iM_2 \approx \frac{ig_s^4 T_{1q}^g T_{qa}^e T_{23}^e T_{4b}^g}{s_{23}\hat{t}_3(s_{a2} + s_{a3})} \langle 1|\rho|a \rangle \langle 4|\rho|b \rangle \times 2p_a^\mu \langle 2|\mu|3 \rangle. \quad (3.33)$$

The next two graphs where the $Q\bar{Q}$ is emitted from the $p_4$ and $p_b$ legs is clearly very similar to the previous two results. Because of this, we simply state the result of the calculation here for these amplitudes which for the $p_4$ leg is

$$iM_3 \approx \frac{-ig_s^4 T_{1a}^g T_{4q}^e T_{qb}^g T_{23}^e}{\hat{t}_1 s_{23}(s_{24} + s_{34})} \langle 1|\rho|a \rangle \langle 4|\rho|b \rangle \times 2p_4^\mu \langle 2|\mu|3 \rangle, \quad (3.34)$$

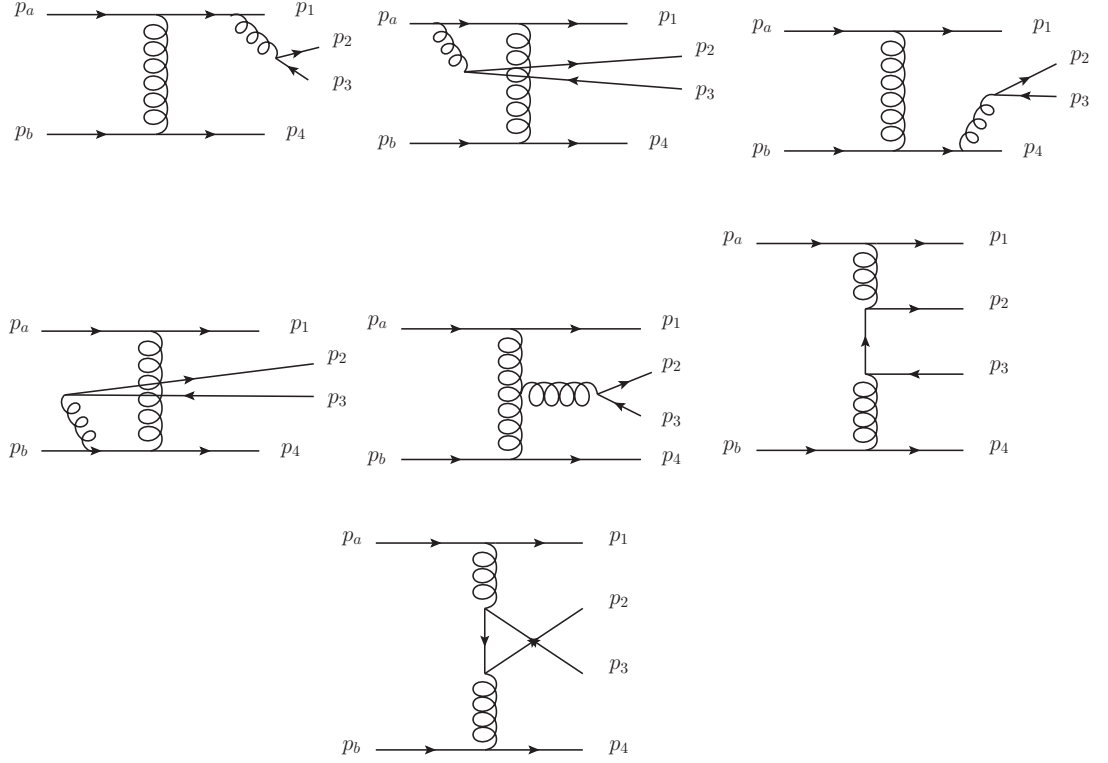

**Figure 3.9** *All LO graphs for $qq' \to qQ\bar{Q}q'$.*

and for the $p_b$ leg is

$$iM_4 \approx \frac{ig_s^4 T_{1a}^g T_{4q}^g T_{qb}^e T_{23}^e}{\hat{t}_1 s_{23}(s_{2b} + s_{3b})} \langle 1|\rho|a\rangle \langle 4|\rho|b\rangle \times 2p_b^\mu \langle 2|\mu|3\rangle . \tag{3.35}$$

The remaining graphs are already $t$-channel factorised and so we can include them exactly. The next diagram where the $Q\bar{Q}$ emission is from the $t$-channel gluon propagator has the exact expression

$$iM_5 = \frac{g_s^4 T_{1a}^g f^{geg'} T_{4b}^{g'} T_{23}^e}{\hat{t}_1 s_{23} \hat{t}_3} \left( (q_1 + p_2 + p_3)^\lambda \eta^{\nu\sigma} + (q_3 - p_2 - p_3)^\nu \eta^{\sigma\lambda} - (q_1 + q_3)^\sigma \eta^{\nu\lambda} \right)$$
$$[\bar{u}_1 \gamma_\nu u_a] [\bar{u}_4 \gamma_\lambda u_b] [\bar{u}_2 \gamma_\sigma v_3] ,$$

(3.36)

where $q_1 = p_a - p_1 = p_4 - p_b + p_3 + p_2$ and $q_3 = p_4 - p_b = p_a - p_1 - p_2 - p_3$. The final two diagrams both have quark propagators. They have the exact expressions

$$iM_6 = \frac{ig_s^4 T_{1a}^g T_{2q}^g T_{q3}^{g'} T_{4b}^{g'}}{\hat{t}_1 (p_a - p_1 - p_2)^2 \hat{t}_3} [\bar{u}_1 \gamma^\mu u_a] [\bar{u}_4 \gamma^\sigma u_b] \left[ \bar{u}_2 \gamma_\mu (\slashed{p}_a - \slashed{p}_1 - \slashed{p}_2) \gamma_\sigma v_3 \right] \quad (3.37)$$

and

$$iM_7 = \frac{-ig_s^4 T_{1a}^g T_{2q}^{g'} T_{q3}^g T_{4b}^{g'}}{\hat{t}_1 (p_a - p_1 - p_3)^2 \hat{t}_3} [\bar{u}_1 \gamma^\mu u_a] [\bar{u}_4 \gamma^\sigma u_b] \left[ \bar{u}_2 \gamma_\sigma (\slashed{p}_a - \slashed{p}_1 - \slashed{p}_3) \gamma_\mu v_3 \right] \quad (3.38)$$

respectively. Since the High Energy limit here still implies $p_a \sim p_1$ and $p_b \sim p_4$, we can approximately combine both $M_1$ with $M_2$ and $M_3$ with $M_4$. Doing the former yields

$$i(M_1 + M_2) \approx \frac{C_1 g_s^4}{s_{23} \hat{t}_3 (s_{a2} + s_{a3})} \langle 1|\rho|a \rangle \langle 4|\rho|b \rangle \times 2 p_a^\sigma \langle 2|\sigma|3 \rangle, \quad (3.39)$$

where we have defined

$$C_1 = T_{1q}^e T_{qa}^g T_{23}^e T_{4b}^g - T_{1q}^g T_{qa}^e T_{23}^e T_{4b}^g = f^{egc} T_{1a}^c T_{23}^e T_{4b}^g, \quad (3.40)$$

and a similar process on $M_3$ and $M_4$ gives

$$i(M_3 + M_4) \to \frac{-C_1 g_s^4}{s_{23} \hat{t}_1 (s_{b2} + s_{b3})} \langle 1|\rho|a \rangle \langle 4|\rho|b \rangle \times 2 p_b^\sigma \langle 2|\sigma|3 \rangle. \quad (3.41)$$

We are then in a position to combine all graphs together and derive our effective vertex $X^{\mu\nu}$:

$$X^{\mu\nu} = \frac{C_1}{s_{23}} \left( \eta^{\mu\nu} \left( 2 p_a^\sigma \left( \frac{q_1^2}{s_{a2} + s_{a3}} \right) - 2 p_b^\sigma \left( \frac{q_3^2}{s_{b2} + s_{b3}} \right) \right) + V_{3g}^{\mu\nu\sigma} \right) \langle 2|\sigma|3 \rangle$$
$$+ \frac{iC_2}{(p_a - p_1 - p_2)^2} X_{qprop}^{\mu\nu} - \frac{iC_3}{(p_a - p_1 - p_3)^2} X_{crossed}^{\mu\nu}, \quad (3.42)$$

where we have defined the following expressions:

$$V_{3g}^{\mu\nu\sigma} = (q_1 + p_2 + p_3)^\nu \eta^{\mu\sigma} + (q_3 - p_2 - p_3)^\mu \eta^{\sigma\nu} - (q_1 + q_3)^\sigma \eta^{\mu\nu} \quad (3.43\text{a})$$
$$C_2 = T_{1a}^g T_{2q}^g T_{q3}^{g'} T_{4b}^{g'} \quad (3.43\text{b})$$
$$C_3 = T_{1a}^g T_{2q}^{g'} T_{q3}^g T_{4b}^{g'} \quad (3.43\text{c})$$
$$X_{qprop}^{\mu\nu} = \bar{u}_2 \gamma^\mu (\slashed{q}_1 - \slashed{p}_2) \gamma^\nu v_3 \quad (3.43\text{d})$$
$$X_{crossed}^{\mu\nu} = \bar{u}_2 \gamma^\nu (\slashed{q}_1 - \slashed{p}_3) \gamma^\mu v_3. \quad (3.43\text{e})$$

In fact, since we assumed $p_a \sim p_1$ and $p_b \sim p_4$ in deriving this form, we can go one step further and reinstate this symmetry. Such a step is consistent with how we treat the Lipatov vertex and will only affect sub-leading terms. Our final vertex is then

$$X^{\mu\nu} = \frac{C_1}{s_{23}} \left(\eta^{\mu\nu} X^\sigma_{sym} + V_{3g}^{\mu\nu\sigma}\right) \langle 2|\sigma|3\rangle + \frac{iC_2}{(p_a - p_1 - p_2)^2} X^{\mu\nu}_{qprop} - \frac{iC_3}{(p_a - p_1 - p_3)^2} X^{\mu\nu}_{crossed},$$
(3.44)

with

$$X^\sigma_{sym} = p_a^\sigma \left(\frac{q_1^2}{s_{a2} + s_{a3}}\right) + p_1^\sigma \left(\frac{q_1^2}{s_{12} + s_{13}}\right) - p_b^\sigma \left(\frac{q_3^2}{s_{b2} + s_{b3}}\right) - p_4^\sigma \left(\frac{q_3^2}{s_{42} + s_{43}}\right).$$
(3.45)

We have intentionally used the notation of $q^2$ to make it clear that this invariant is formed from the mass of the propagator entering into the effective vertex. Once extra emissions are added, the $t_{ij}$ notation can be misleading. With this, we now have the complete expression for the effective central $Q\bar{Q}$ vertex. Furthermore, we notice again that we can rewrite our colour factors in such a way that the vertex depends only on $C_2$ and $C_3$, but we decide not to for computational ease.

### 3.3.4 Verifications of the Central $Q\bar{Q}$ Vertex

As with the previous effective vertex, we take this subsection as an opportunity to check the correctness of our calculation. The calculation is once more practically performed by splitting the effective vertex up into sub-vertices according to the colour factors and explicitly summing the contributions at the $|M|^2$ level. We choose the following parametrisation for the momenta:

$$\begin{aligned}
p_1 &= (40\cosh(\Delta), 0, 40, 40\sinh(\Delta)), \\
p_2 &= (40\cosh(\Delta/3), 40, 0, 40\sinh(\Delta/3)), \\
p_3 &= (40\cosh(-\Delta/3), 0, -40, 40\sinh(-\Delta/3)), \\
p_4 &= (40\cosh(-\Delta), -40, 0, 40\sinh(-\Delta)).
\end{aligned}$$
(3.46)

The comparison between our amplitude and the full leading order is shown in figure 3.10. We see once more the good agreement between the two across the phase space as well as the suppression at large $\Delta$, which figure 3.11 shows is again due to $s_{Q\bar{Q}}$.

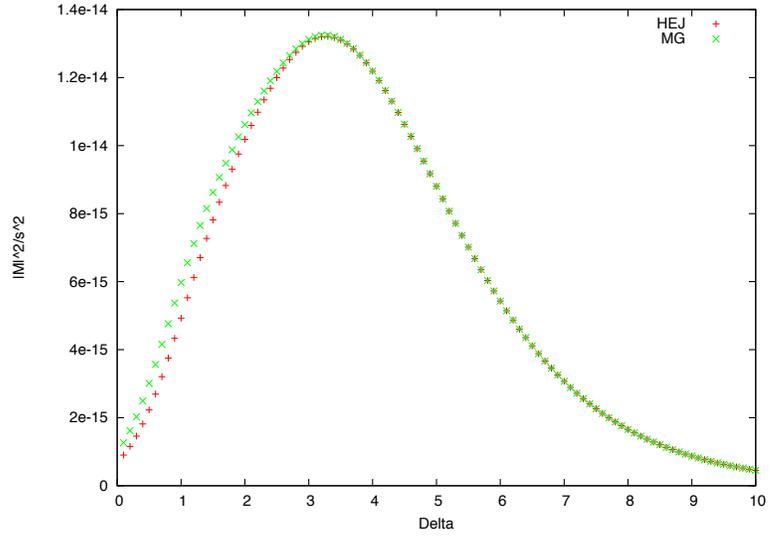

**Figure 3.10** *Effective vertex approach to the $qq' \to qQ\bar{Q}q'$ amplitude (red) compared to the full LO (green).*

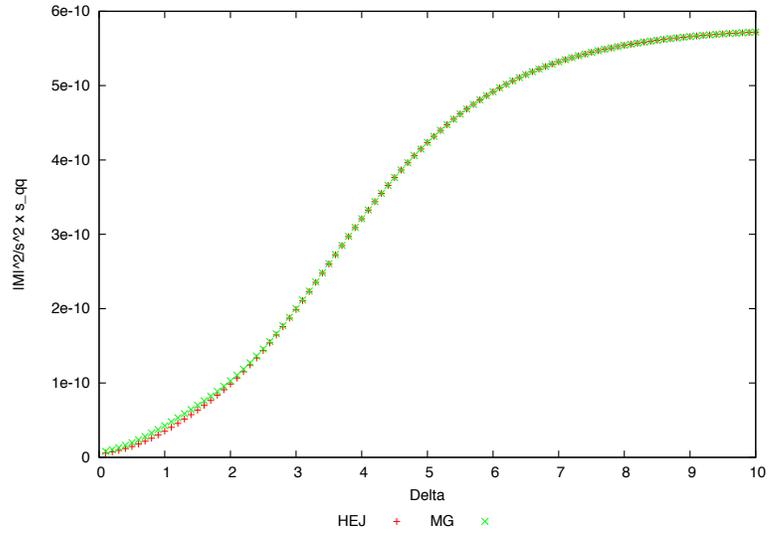

**Figure 3.11** *Effective vertex approach to the $qq' \to qQ\bar{Q}q'$ amplitude (red) compared to the full LO (green) multiplied by the invariant mass of the quark/anti-quark pair.*

To extend our result to take into account gluons in the initial state, we once more multiply by colour factors:

$$|M_{qg \to qQ\bar{Q}g}|^2 \sim \frac{\tilde{C}_A}{C_F}|M_{qq' \to qQ\bar{Q}q'}|^2,$$
$$|M_{gg \to gQ\bar{Q}g}|^2 \sim \left(\frac{\tilde{C}_A}{C_F}\right)^2 |M_{qq' \to qQ\bar{Q}q'}|^2. \quad (3.47)$$

The comparison to the full leading order result for these amplitudes (multiplied by $s_{Q\bar{Q}}$) is shown in figures 3.12 and 3.13 respectively.

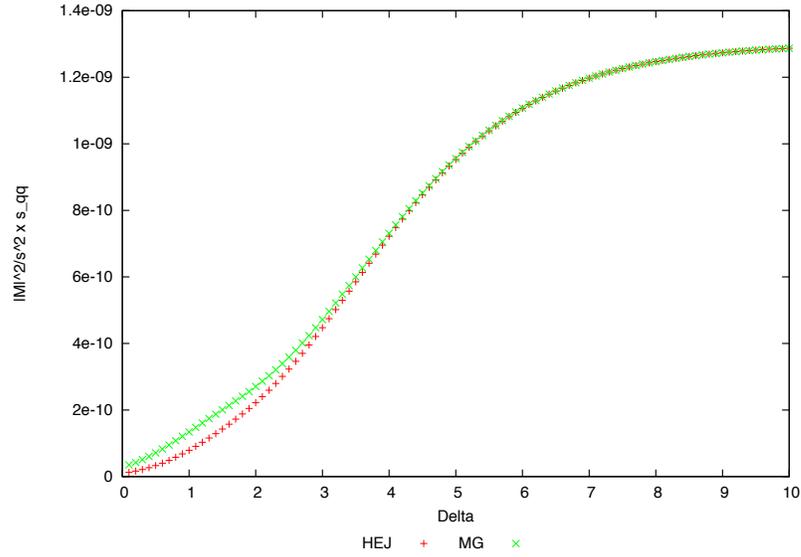

**Figure 3.12**  *Effective vertex approach to the $qg \to qQ\bar{Q}g$ amplitude (red) compared to the full LO (green) multiplied by the invariant mass of the quark/anti-quark pair.*

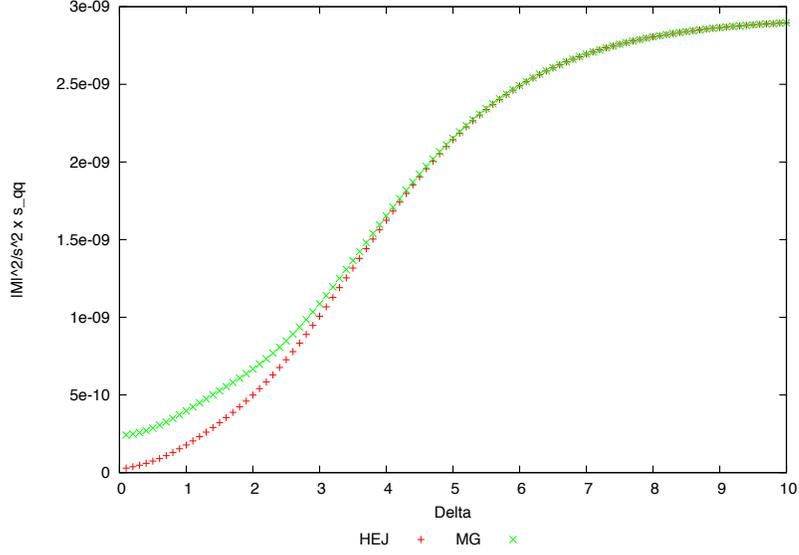

**Figure 3.13**   *Effective vertex approach to the $gg \to gQ\bar{Q}g$ amplitude (red) compared to the full LO (green) multiplied by the invariant mass of the quark/anti-quark pair.*

In this case, the addition of extra gluon emissions is slightly more complicated than in the previous case. Any extra emission can take place either before or after the effective central vertex in the rapidity chain, as shown in figure 3.14. This distinction needs to be kept track of and so our extension is

$$|M_{qq' \to q\ldots Q\bar{Q}\ldots q'}|^2 = |M_{qq' \to qQ\bar{Q}q'}|^2 \times \prod_{i=1}^{n_a} C_A \left( \frac{-V(q_i, q_{i+1}) \cdot V(q_i, q_{i+1})}{q_{i+1}^4} \right) \\ \times \prod_{j=n_a+2}^{n-2} C_A \left( \frac{-V(q_j, q_{j+1}) \cdot V(q_j, q_{j+1})}{q_j^4} \right), \quad (3.48)$$

where $n_a$ is the number of gluons more forward in rapidity than the quark/anti-quark pair and we recall there is a division of $q_1^2 q_{n-1}^2$ already inside of the base matrix element (which is then being squared). To be clear with the definitions of the $q$s, we are defining $q_1 = p_a - p_1$ and then $q_i = q_{i-1} - p_i$ for all the propagators up to the central effective vertex ($2 \leq i \leq n_a$). After this vertex, defining the $q$s down the chain like this becomes cumbersome because of the extra dependence on the quark and anti-quark momenta, so we instead use momentum conservation to start defining them from the bottom up: $q_{n-1} = p_n - p_b$ and $q_j = q_{j+1} + p_{j+1}$ for $n_a + 2 \leq j \leq n-2$. We take the simplest case of adding just one extra emission and investigate the amplitude for both options. We employ the following set of five jet momenta:

$$\begin{aligned}
p_1 &= (40\cosh(\Delta), 0, 40, 40\sinh(\Delta)), \\
p_2 &= (40\cosh(\Delta/2), -40, 0, 40\sinh(\Delta/2)), \\
p_3 &= (80\sqrt{2}, 80, -80, 0), \\
p_4 &= (40\cosh(-\Delta/2), -40, 0, 40\sinh(-\Delta/2)), \\
p_5 &= (40\cosh(-\Delta), 0, 40, 40\sinh(-\Delta)).
\end{aligned} \tag{3.49}$$

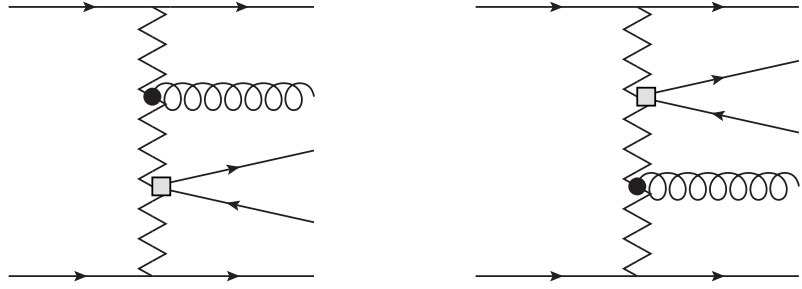

**Figure 3.14** *Extra gluon emissions can be either before (left) or after (right) the central $Q\bar{Q}$ vertex in rapidity.*

In figure 3.15 we plot the result for when the gluon is more forward in rapidity than the $Q\bar{Q}$ pair and in figure 3.16 we have the result for when it is emitted more backward in rapidity. Both of these figures show reasonable agreement with the full result and importantly, agree precisely in the high energy limit.

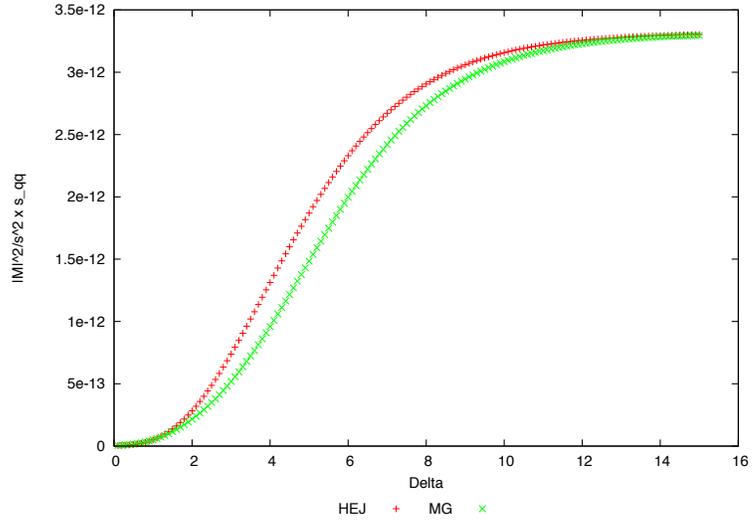

**Figure 3.15**  *Effective vertex approach to the $qq' \to qgQ\bar{Q}q'$ amplitude (red) compared to the full LO (green) multiplied by the invariant mass of the quark/anti-quark pair.*

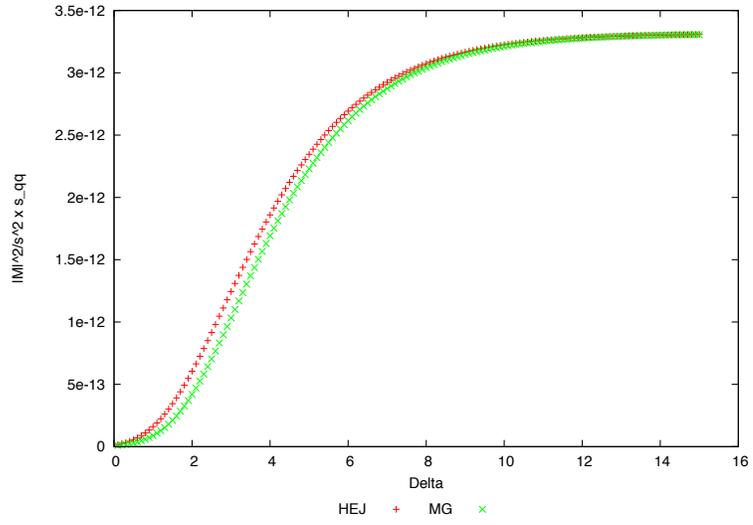

**Figure 3.16**  *Effective vertex approach to the $qq' \to qQ\bar{Q}gq'$ amplitude (red) compared to the full LO (green) multiplied by the invariant mass of the quark/anti-quark pair.*

## 3.4 All-Order Forms

For completeness, we give here the full forms of the all-order amplitudes for the $Q\bar{Q}$ processes. The resummation procedure proceeds in precisely the same way as in the FKL amplitudes (equation 2.52). We simply make the following replacement:

$$\frac{1}{\hat{t}_i} \to \frac{1}{\hat{t}_i} \exp\left[\omega_0(q_{i\perp})(y_{i+1} - y_i)\right], \tag{3.50}$$

where $\omega(q_i)$ is defined in equation 2.51. For the extremal $Q\bar{Q}$ amplitude, this prescription gives

$$\begin{aligned}|M^{HEJ}_{qg \to q\ldots Q\bar{Q}}|^2 = |M_{qg \to qQ\bar{Q}}|^2 &\times \prod_{i=1}^{n-3} C_A \left(\frac{-V(q_i, q_{i+1}) \cdot V(q_i, q_{i+1})}{\hat{t}_i \hat{t}_{i+1}}\right) \\ &\times \prod_{j=1}^{n-3} \exp\left[\omega_0(q_{j\perp})(y_{j+1} - y_j)\right].\end{aligned} \tag{3.51}$$

For the central $Q\bar{Q}$ case, the expression is

$$\begin{aligned}|M^{HEJ}_{qq' \to q\ldots Q\bar{Q}\ldots q'}|^2 = |M_{qq' \to qQ\bar{Q}q'}|^2 &\times \prod_{i=1}^{n_a} C_A \left(\frac{-V(q_i, q_{i+1}) \cdot V(q_i, q_{i+1})}{\hat{t}_{i+1}^2}\right) \\ &\times \prod_{j=n_a+2}^{n-2} C_A \left(\frac{-V(q_j, q_{j+1}) \cdot V(q_j, q_{j+1})}{\hat{t}_j^2}\right) \\ &\times \prod_{k_1=1}^{n_a+1} \exp\left[\omega_0(q_{k_1\perp})(y_{k_1+1} - y_{k_1})\right] \\ &\times \prod_{k_2=n_a+3}^{n-1} \exp\left[\omega_0(q_{k_2\perp})(y_{k_2+1} - y_{k_2})\right].\end{aligned} \tag{3.52}$$

With these forms derived, we are in a position to include them fully within the HEJ formalism. As discussed at the beginning of the chapter, these partonic subprocesses are Next-to-Leading Logarithmic in the jet process. We reiterate the fact that the derivation of these amplitudes means that HEJ is now able to include *all* subprocesses that contribute at NLL.

## 3.5 Computational Aspects

The explorer plots of the last two sections all serve to convince that the derived amplitudes behave as expected. The next step is to include them in the full HEJ program correctly. This is a very significant change to the codebase that requires the modification of many files and so a 'branch' of the codebase was created so that all work can be done independently of other projects going on in the collaboration. Once everything is completed and checked, the branch can be merged back into the main development codebase in a way that does not disrupt any other work done there since the branch was created. The points that need to be considered when adding these amplitudes in are as follows:

1. Correctly removing calls to fixed order non-FKL processes if they are now to be included in the resummation. For example, the code already contains a call to the $qg \to qQ\bar{Q}$ amplitude at leading order for matching purposes as discussed in section 2.5. We therefore need to remove calls to these particular subprocesses in that section of the code but only if the user specifies that they wish to include these processes in the resummation (although it will become default for HEJ to include them).

2. Ensuring that all possible processes are generated. Since the HEJ program is a Monte Carlo program, we must ensure that all of these extra NLL processes are 'picked'. Since we require a leading order matching, one consideration is that a new process should not be picked if the event will not cluster into at least three jets. Also, given a set number of final state partons and a central $Q\bar{Q}$ emission, we must make sure that all positions of that emission along the rapidity chain are considered. Another point is that the rapidity ordering of the $Q\bar{Q}$ pair can be either way around (quark/anti-quark or anti-quark/quark) and we must include both orderings. Failure to do any of this correctly will mean that the program will be artificially removing physics that we know should be included.

3. Checking that the leading order matching at the jet level for the NLL processes is done consistently. For example, imagine we generate a central $Q\bar{Q}$ emission at the parton level and choose partons $p_i$ and $p_j$ for the pair, where $p_i$ is more forward than and next to $p_j$ in rapidity. In order to properly implement the matching, we require that these cluster into two different jets, $p_i \to j_i$ and $p_j \to j_j$, where $j_i$ is more forward than and next to $j_j$ in rapidity. Because of how the clustering works, it is not at all guaranteed that the jets will preserve this rapidity ordering; moreover, these jets might not even still be next to each other in rapidity in the chain. All these considerations must be carefully checked by looking at the jet constituents.

It is absolutely crucial that all of these points are properly addressed and checked and such a task is necessarily time-consuming. After validating these steps by comparisons to old code with certain constraints, we can be confident that everything works as it should. In the following two sections, we move on to discuss how the inclusion of these NLL processes affects HEJ in practice.

## 3.6 Full Phase Space Analysis

One important advantage of now being able to resum these partonic subprocesses and unordered events is that it means we have a reduced reliance on fixed-order matching codes. In chapter 2, we showed how 'non-FKL' contributions are subleading in the High Energy Limit. While this is true, there are regions that (for example) the LHC can probe which are clearly not well-described by the High Energy Limit. For instance, we could imagine a region where one parton's $p_T$ is very large, creating a significant $p_T$ hierarchy that the High Energy Limit assumes not to be there. In order for HEJ to be competitive as a prediction over the entirety of the explored phase space, we need to include some description of these types of process, which we do via the inclusion of fixed-order (currently LO) matrix elements. We have shown here that some of these non-FKL processes can themselves be resummed and therefore can be removed from the fixed-order samples. In this section, we investigate the impact of this when we integrate over all of the relevant phase space. In the following figures, we show the total cross-section for inclusive 3 jets, broken down into the contribution coming from the resummation and the fixed order. It is desirable in these plots if the red line (the resummed contribution) is as close to the black line (the total contribution) as possible. For completeness, the relevant cuts applied were $p_{T,min} = 30$ GeV, $y_{max} = 4.4$, and $\sqrt{s} = 13$ TeV.

In figure 3.17, we show a 'before and after' comparison of the $m_{fb}$ distribution. It is clear that that the inclusion of these extra processes dramatically reduces the contribution of fixed-order matrix elements across the range, but especially in the low bins. The contribution of the non-FKL processes to the total cross-section in these bins is approximately 43% in the top plot, which drops to around 15% in the bottom. In the high bins, the contribution drops from around 15% to almost zero.

In figure 3.18, we show the events binned instead in the $p_T$ of the hardest jet. In it, we see that a significant proportion of the fixed order matrix elements that were very significant in the high $p_T$ region (contributing close to 60% of the cross-section there) are now included in the resummation. The net effect is to reduce the non-FKL contribution to the total cross-section down to 20%.

Finally, in figure 3.19 we show the analysis broken up in bins of the rapidity difference between the most forward and backward jets. There, we once more see a substantial improvement across the board, with the largest non-FKL percentage contribution of 30% being reduced to 12%. In the higher bins, we once more see evidence that the non-FKL contributions are tending to zero.

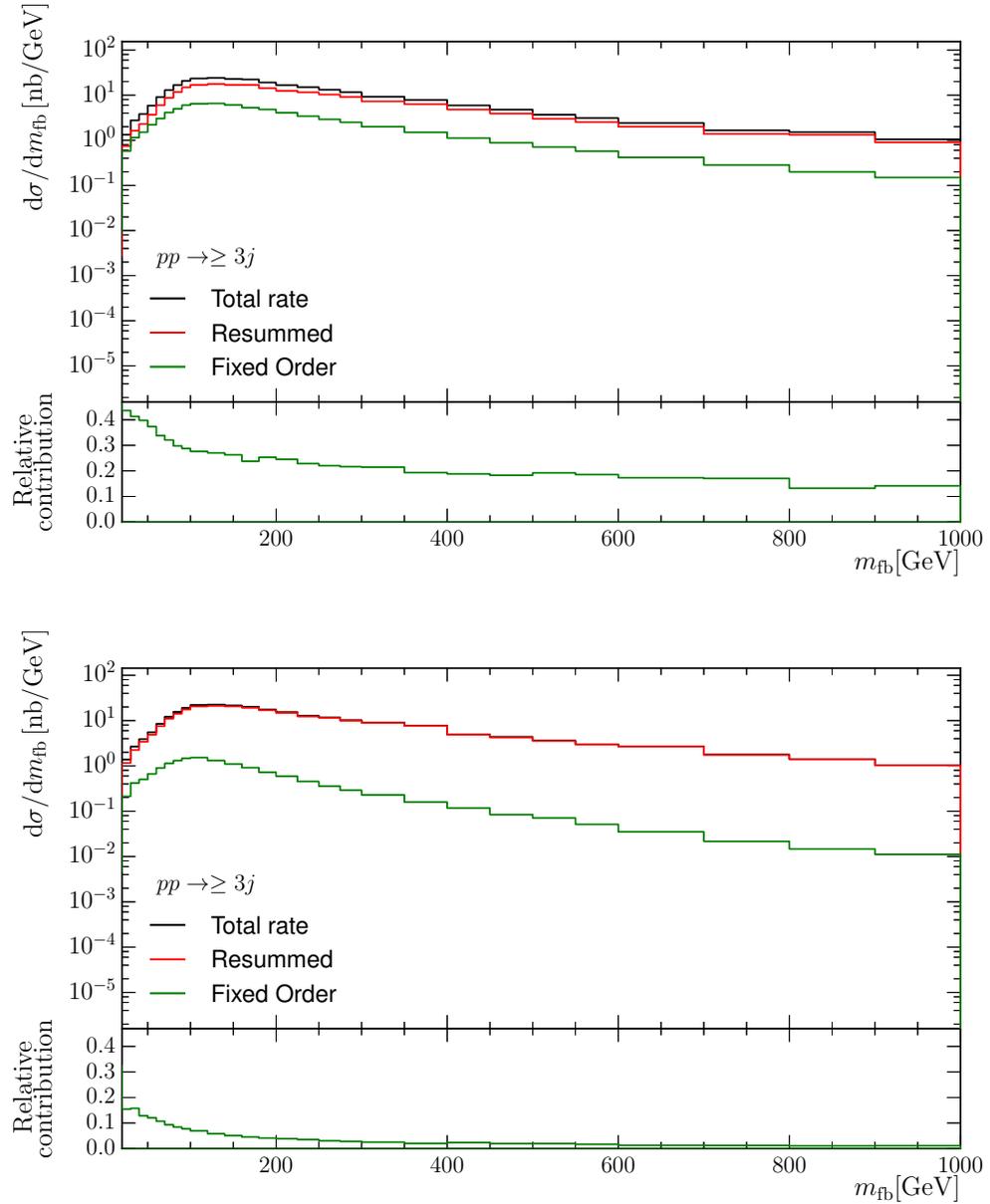

**Figure 3.17** *A breakdown of contributing parts to the jet cross-section before (top) and after (bottom) implementing the effective vertex description of the new partonic subprocesses and the unordered events, in bins of the invariant mass between the most forward and backward jets.*

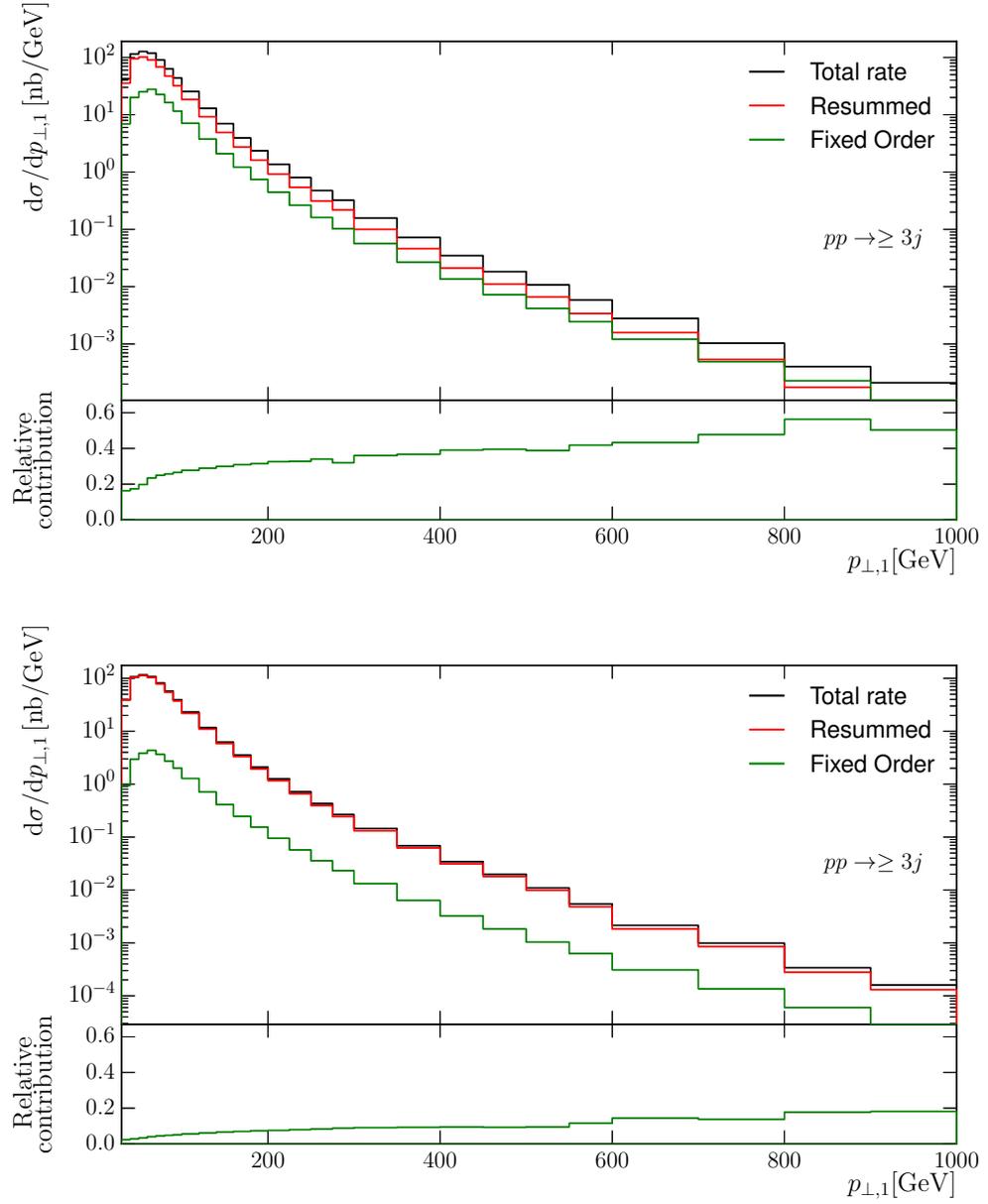

**Figure 3.18** *A breakdown of contributing parts to the jet cross-section before (top) and after (bottom) implementing the effective vertex description of the new partonic subprocesses and the unordered events, in bins of the $p_T$ of the hardest jet.*

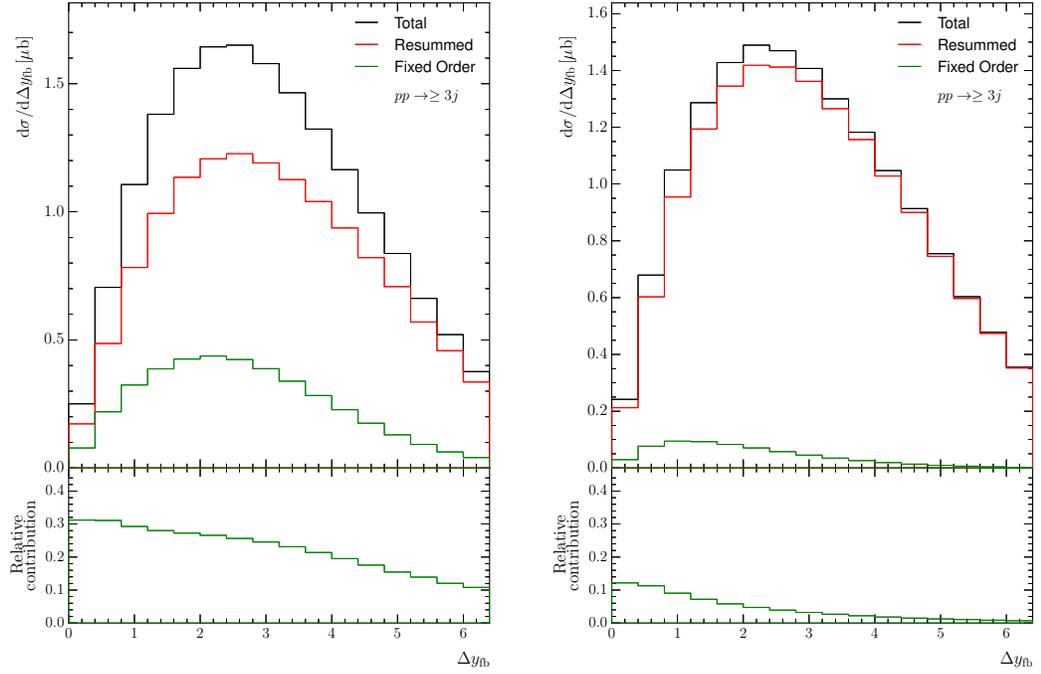

**Figure 3.19**  *A breakdown of contributing parts to the jet cross-section before (left) and after (right) implementing the effective vertex description of the new partonic subprocesses and the unordered events, in bins of the rapidity difference between the most forward and backward jets.*

## 3.7 Comparisons to Data

To conclude the chapter, we revisit the predictions for previous analyses HEJ was involved in and investigate how these new additions improve the predictions whilst comparing with real data; for example, in an ATLAS study of dijet production with a central jet veto [10]. One interesting plot is figure 6 of that paper, which is that of the average number of jets in the 'gap', defined as the rapidity region between the dijet system, which in this case is given by the two highest $p_T$ jets in a event. There are a total of four different lines, which correspond to different rapidity slices, shown in figure 3.20 which correspond to (starting at the top) $4 \leq \Delta y < 5$, $3 \leq \Delta y < 4$, $2 \leq \Delta y < 3$ and $1 \leq \Delta y < 2$. In order to distinguish them on the graph, these lines are moved up by $3, 2, 1$ and $0$ units respectively.

The $x$-axis is $\bar{p}_T$, which is the average of the dijet system's transverse momenta. For the high energy limit HEJ considers, we should not expect this to be a good variable to plot with respect to; the limit depends on all transverse scales being roughly the same order. At high average $p_T$, we can have a large $p_T$ hierarchy. This is indeed reflected at the right-hand side of the figure. It was noted that the analysis might have been better performed by considering only the resummed part of the HEJ calculation, since it is this part that gives rise to gap jets. We therefore include both the full and resummed-only lines in this figure. In the diagram on the top left, we see a considerable difference between these two predictions. By adding these new processes, we make these lines essentially indistinguishable and at the same time capture more bins on the left-hand side of the plots. It is advantageous for the two lines to come together like this, since it eradicates the need to arbitrarily choose just the resummed predictions over the full prediction. This is a practical consequence of our lesser reliance on fixed order matching in the formalism. Going back to the plot and we see that the addition of these NLL processes still does not tell the whole story; the correct description of the bins on the right-hand side still requires the addition of parton shower effects. Since HEJ and parton showers are designed to describe completely different regions of amplitudes, this is not surprising.

Indeed, in a later dijet veto ATLAS analysis [12], it was observed that the predictions from the pure partonic HEJ needed to be interfaced with a parton shower in order to get the best description of the data. Although of course there are regions where a parton shower description is always going to be important, we can investigate to see if NLL effects can push the pure partonic line closer to the data by itself. For example, consider figure 3a of [12], which plots the 'Gap Fraction', defined as

$$f(Q_0) = \frac{\sigma_{jj}(Q_0)}{\sigma_{jj}}, \qquad (3.53)$$

where $\sigma_{jj}$ is the inclusive dijet cross-section and $\sigma_{jj}(Q_0)$ is the cross-section for dijet production in the absence of additional jets with transverse momentum above the threshold $Q_0$ in the rapidity interval defined by the dijet system (once more defined by the two hardest jets in the event). Such an observable will clearly be sensitive to resummation effects, since we saw earlier that the rapidity gap plays a central role in the derivation of our amplitudes. In figure 3.21 we plot a comparison of data to four different HEJ runs, which correspond to the four different combinations of including/not including the unordered/sub-leading partonic process corrections. For the same reasons as before, we plot both the resummed and full HEJ lines for comparison. The bottom-right figure shows that the inclusion of both of these effects leads to a better description of the data everywhere, but in particular we gain a few extra bins at the left side of the figure when compared to the top-left.

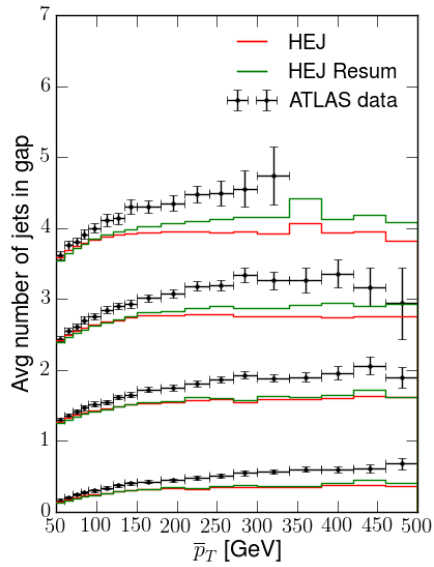
**(a)** *No NLL corrections.*

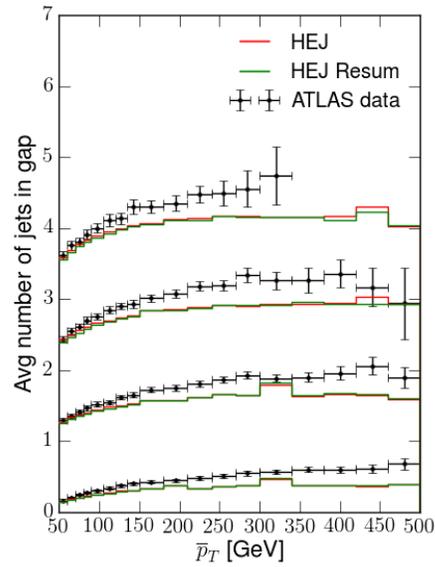
**(b)** *Unordered corrections.*

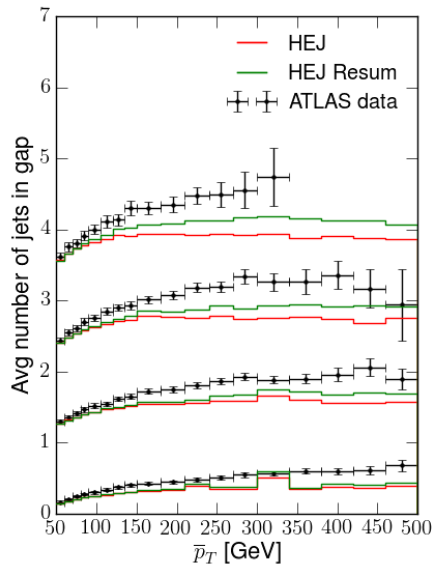
**(c)** *Sub-leading partonic processes corrections.*

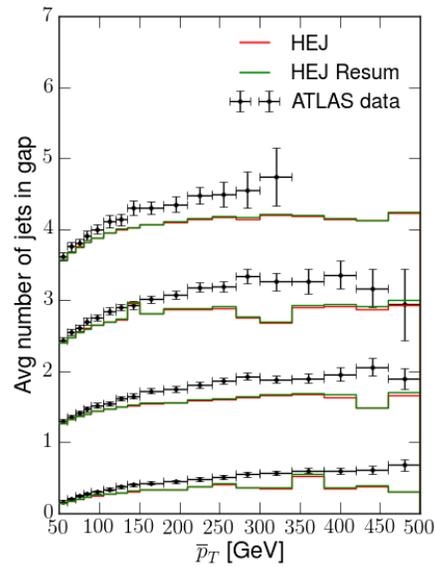
**(d)** *Both corrections.*

**Figure 3.20** *Figure 6 of [10] redone with corrections included. The red lines are the results from running the full HEJ program, the green if one only considers the resummed parts of the program and black points are the data. From the top-left: no corrections, unordered corrections, sub-leading partonic processes corrections, both corrections. In the last figure, we see how these corrections mean that there is now barely any distinction between 'resummed' and 'full' HEJ and that the agreement has improved to slightly higher values of $\bar{p}_T$.*

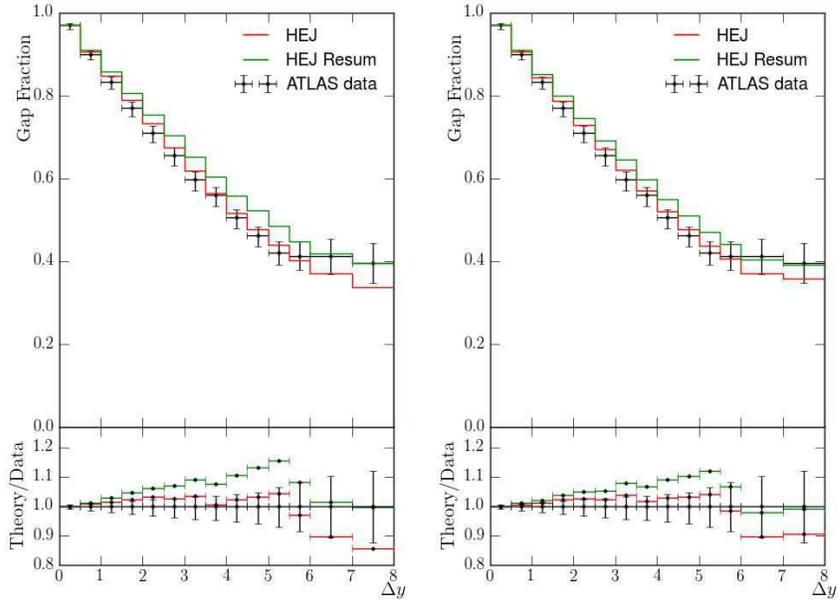

(a) *No NLL corrections.*   (b) *Unordered corrections.*

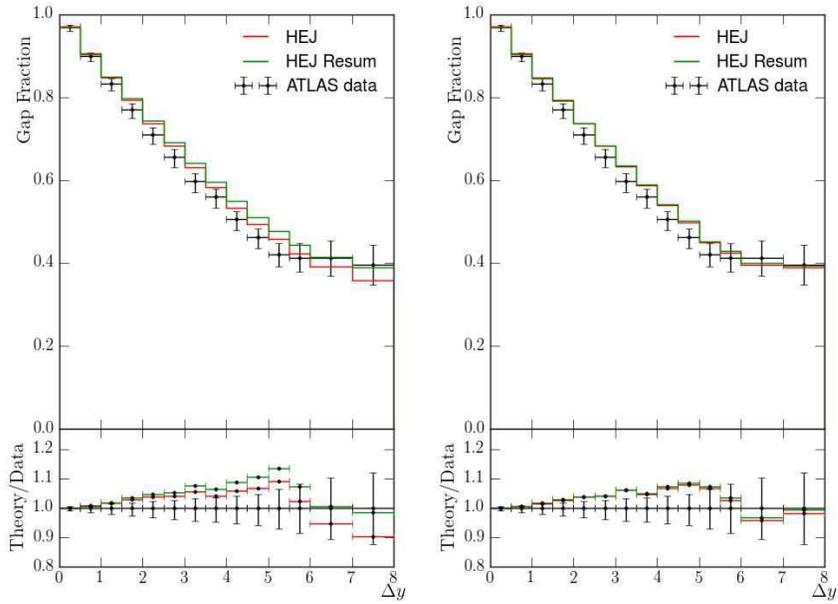

(c) *Sub-leading partonic processes corrections.*   (d) *Both corrections.*

**Figure 3.21**  *Figure 3a of [12] redone with corrections included. The red lines are the results from running the full HEJ program, the green if one only considers the resummed parts of the program and black points are the data. Further discussion in the text.*

## 3.8  Summary

In this chapter we have presented the beginnings of the work needed to extend HEJ to NLL accuracy. Specifically, we provided details of the calculation of so-called unordered amplitudes, where an FKL amplitude is modified such that an emitted gluon is taken outside of the previously extremal partons. Following on from that, we discussed how some inherently non-FKL amplitudes could also be included in our resummation procedure by virtue of the fact that the Lipatov ansatz holds at the next-to-leading level.

The derivation of these new amplitudes dominated the subject matter of the chapter and correctness checks such as gauge invariance and limit checks were presented as we went along. Once we were happy that the amplitudes were behaving in the way we expected, they were carefully and correctly incorporated into the HEJ program. Section 3.6 provided details of how this inclusion affected various distributions when integrating over phase space.

The chapter concluded with comparisons to real data. In all cases, it was seen that the addition of the corrections pushed our predictions closer towards the recorded data. We are now in a position to include these NLL processes as part of the standard HEJ package, forming part of the standard prediction for any future analyses.

# Chapter 4

# Higgs Plus Jets with Finite Quark Mass Effects

It was briefly discussed at the end of section 2.3.2 that the addition of particles not charged under QCD, such as the $W^\pm$, $Z$ and Higgs bosons, are included in HEJ by deriving a modified current that encodes the emission of the boson. In this chapter, we look more closely at how this is done for the case of the Higgs boson in various limits. We will begin with the limit where the Higgs boson is emitted far away from all other partons in rapidity and where the top quark mass is taken to infinity (the 'infinite top mass limit'). We will then allow the top quark to have a finite mass and show how the HEJ amplitude is not complicated by this. Once this is established, we will keep working with a finite top mass and allow for the Higgs boson to be emitted outside of an extremal gluon; this will constitute the author's own work. By the end of the chapter, we will have an expression that allows for the Leading Logarithmic resummation of Higgs plus at least two jets processes with full quark mass dependence, a calculation that is unique to HEJ.

## 4.1 The Infinite Top Mass Limit

If we wish to place a Higgs boson along a rapidity chain consisting of $t$-channel gluon exchanges, then it would seem we would have to deal with loops. Since the gluon is massless, the first contribution to the $gg \to H$ vertex is at the one-loop level, where we have a massive quark loop that can couple to the Higgs directly. Loops generally complicate the calculation of an amplitude, so that even the calculation to leading order for this case is far more involved than a pure QCD process. This will be demonstrated further in the later sections of this chapter. In order to get around this complication, it is usual to take the limit where the top mass tends to infinity. This allows us to only consider a single quark loop (with the top quark) and then effectively shrink it down to a point, resulting in an *effective* direct gluon-gluon-Higgs coupling in our Lagrangian[1]

$$\mathscr{L}_{QCD+ggH} = \mathscr{L}_{QCD} + \frac{A}{4}G^{\mu\nu}G_{\mu\nu}H, \tag{4.1}$$

with $A = \frac{\alpha_s}{3\pi v}$ [28] and where $G^{\mu\nu}$ is the field tensor for the gluon. This results in the following Feynman rule for the effective $ggH$ vertex [6]:

$$V_H^{\mu\nu}(q_1, q_2) = A\left(\eta^{\mu\nu} q_1 \cdot q_2 - q_1^\nu q_2^\mu\right), \tag{4.2}$$

where $q_1$ is incoming to the vertex and $q_2$ outgoing. The simplest amplitude we can imagine is $qQ \to qHQ$, which has only one Feynman diagram: it is almost the same as the diagram for $qQ \to qQ$ scattering (figure 1.3), with the only difference being the addition of the $ggH$ vertex along the $t$-channel gluon. It will be useful to absorb part of the expression for the $ggH$ vertex into the spinor chain, such that we create the object

$$S_{qQ \to qHQ}(q_1, q_2) = \langle 1|\mu|a\rangle \left(\eta^{\mu\nu} q_1 \cdot q_2 - q_1^\nu q_2^\mu\right) \langle 2|\nu|b\rangle. \tag{4.3}$$

With this, we can express the $qQ \to qHQ$ amplitude (summed and averaged over helicity and colour and exact within the infinite top mass limit) as

$$\begin{aligned}
|M_{qQ \to qHQ}|^2 &= \frac{1}{4(N_C^2 - 1)} ||S_{qQ \to qHQ}(q_1, q_2)||^2 \\
&\cdot \left(g^2 C_F \frac{1}{\hat{t}_1}\right) \\
&\cdot \left(\frac{1}{\hat{t}_1}\left(\frac{\alpha_s}{3\pi v}\right)^2 \frac{1}{\hat{t}_2}\right) \\
&\cdot \left(g^2 C_F \frac{1}{\hat{t}_2}\right).
\end{aligned} \tag{4.4}$$

---
[1]There is also a gluon-gluon-Higgs-Higgs interaction, but this thesis only concerns itself with single Higgs production, so we omit it for brevity here.

If we wish to extend the result to other incoming states and/or with extra emissions, then we must also impose the condition that the Higgs boson is produced centrally and far away in rapidity from all other partons. Assuming that, we can once more insert Lipatov vertices and colour factors as appropriate. However, if we do not make this assumption then we are in a similar situation to the case of the unordered gluon in section 3.2 – there are more terms in the amplitude that we cannot ignore. Furthermore, in contrast to the unordered case, this contribution is not suppressed in the MRK limit since the position of the Higgs boson along (or outside of) the rapidity chain does not change the colour properties of the amplitude. Such contributions must be taken into account. These are included in HEJ via the use of impact factors [28], which are strict high energy expressions that include all relevant terms in this limit.

The infinite top mass limit is remarkably successful for predictions of inclusive variables like total cross-sections. The reason for this is that the expansion of this effective coupling is a power series in $\frac{m_H^2}{4m_t^2} \sim 0.13$ and we assume the top mass is the most relevant scale in the problem. In [28], it was shown that even the addition of a large dijet invariant mass in the amplitude has little effect. However, if the transverse scales start to become larger than the top mass (in particular, the transverse scales of the gluons entering the effective vertex), there is significant deviation between the result obtained in the full and effective theories [28].

For HEJ, the infinite top mass limit was included as a first approximation, but the factorisation of amplitudes will still occur whether or not this limit is taken. Therefore, any derived vertices for Higgs production with the full quark mass included can still be implemented in a straightforward way. Furthermore, as soon as the vertex including the full quark mass is derived, we can trivially add in the effect of more than one quark loop (for example, where a bottom quark runs in the loop). Such interference has been shown to produce interesting effects at small $p_T$ scales, at least in processes involving the emission of a Higgs boson with one accompanying jet [39, 46]. For these reasons, it is important to include the effect of finite quark masses in the HEJ formalism. Finally, since the emission of extra jets is trivial in HEJ, the inclusion of a full finite-quark-mass-dependent expression for the $gg \to H$ coupling will result in a prediction for the $pp \to H+n$ jets amplitude for any $n$ without the assumption of a infinite top mass. This is a calculation that is far beyond the reach of any fixed-order approach.

## 4.2 Calculations of $H$ + Jets with Finite Quark Mass Effects

As previously discussed, there are two rapidity cases to be discussed: the first where the Higgs is emitted centrally and the second where it is emitted outside of an extremal parton. The base amplitude for the first is the $qQ \to qHQ$ amplitude, which we will discuss in the next subsection. For the case where the Higgs is emitted outside of an extremal parton, we look at the $gq \to Hgq$ amplitude, where the Higgs boson is emitted outside of the gluon. Such a process gives rise to diagrams involving box integrals which become important in this limit, making the study of amplitudes involving the Higgs emitted outside of an extremal gluon fundamentally different.

### 4.2.1 $qQ \to qHQ$

The expression for the $qQ \to qHQ$ amplitude with full finite quark mass effects is shown diagrammatically in figure 4.1. In the previous section, we showed how the amplitude was derived in the infinite top mass limit. Because the factorisation properties of the amplitude are unchanged by moving away from this limit, the extension of this amplitude to the full (finite quark mass dependent) amplitude is found by simply 'undoing' the infinite top mass limit in the vertex $V_H^{\mu\nu}$. Such an expression can be found, for example in Appendix B of [28], so long as care is taken to ensure that incoming and outgoing momenta are labelled correctly. Keeping with the convention before that $q_1$ is incoming and $q_2$ outgoing, the vertex with a finite top mass is

$$V_{H,m_t}^{\mu\nu}(q_1, q_2) = -\frac{4g_s^2 m_t^2}{v} \left[\eta^{\mu\nu} A_2(-q_1, q_2) + q_1^\nu q_2^\mu A_1(-q_1, q_2)\right], \qquad (4.5)$$

where $v$ is the Higgs vacuum expectation value ($\approx 246$ GeV) and $A_1$, $A_2$ depend on a combination of scalar integrals which appear via the integral reduction technique

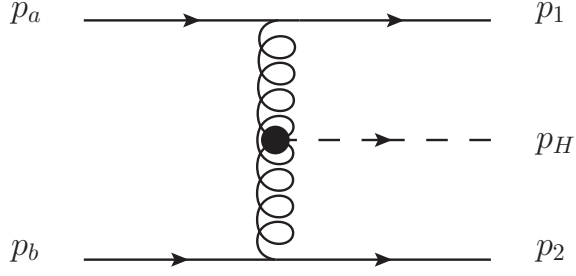

**Figure 4.1** *Diagrammatic representation of $qQ \to qHQ$ with an effective vertex for the production of the Higgs.*

[20]. We first define[2]

$$B_0(k) = \int \frac{d^4q}{(2\pi)^4} \frac{1}{(q^2 - m_t^2)((q+k)^2 - m_t^2)},$$

$$C_0(p,k) = \int \frac{d^4q}{(2\pi)^4} \frac{1}{(q^2 - m_t^2)((q+p)^2 - m_t^2)((q+p+k)^2 - m_t^2)}, \quad (4.6)$$

$$Q = -q_1 - q_2,$$

$$\Delta_3 = (q_1^2)^2 + (q_2^2)^2 + (Q^2)^2 - 2q_1^2 q_2^2 - 2q_1^2 Q^2 - 2q_2^2 Q^2,$$

which allows us to give the forms of $A_1$ and $A_2$ as

$$A_1(-q_1, q_2) = C_0(-q_1, q_2) \left[ \frac{4m_t^2}{\Delta_3}(Q^2 - q_1^2 - q_2^2) - 1 - \frac{4q_1^2 q_2^2}{\Delta_3} - \frac{12 q_1^2 q_2^2 Q^2}{\Delta_3^2}(q_1^2 + q_2^2 - Q^2) \right]$$

$$- [B_0(q_2) - B_0(Q)] \left[ \frac{2q_1^2}{\Delta_3} + \frac{12 q_1^2 q_2^2}{\Delta_3^2}(q_2^2 - q_1^2 + Q^2) \right]$$

$$- [B_0(-q_1) - B_0(Q)] \left[ \frac{2q_1^2}{\Delta_3} + \frac{12 q_1^2 q_2^2}{\Delta_3^2}(q_1^2 - q_2^2 + Q^2) \right]$$

$$- \frac{2}{\Delta_3} \frac{i}{(4\pi)^2}(q_1^2 + q_2^2 - Q^2)$$

$$A_2(-q_1, q_2) = C_0(-q_1, q_2) \left[ 2m_t^2 + \frac{1}{2}(q_1^2 + q_2^2 - Q^2) + \frac{2q_1^2 q_2^2 Q^2}{\Delta_3} \right]$$

$$+ [B_0(q_2) - B_0(Q)] \frac{1}{\Delta_3} q_2^2(q_2^2 - q_1^2 - Q^2)$$

$$+ [B_0(-q_1) - B_0(Q)] \frac{1}{\Delta_3} q_1^2(q_1^2 - q_2^2 - Q^2)$$

$$+ \frac{i}{(4\pi)^2}.$$

$$(4.7)$$

---

[2]The integral $B_0$ is actually divergent in 4 dimensions, but will always appear in combinations such that this divergence cancels in later functions.

The scalar integrals can either be evaluated numerically (via, for example, a program like LoopTools [40]) or again simply looked up,[3] so that the values of $A_1$ and $A_2$ can be determined at any point. It can be numerically checked that

$$\lim_{m_t \to \infty} 4\frac{g_s^2 m_t^2}{v} A_1(-q_1, q_2) = iA,$$
$$\lim_{m_t \to \infty} 4\frac{g_s^2 m_t^2}{v} A_2(-q_1, q_2) = -iq_1 \cdot q_2 A, \qquad (4.8)$$

and thus

$$\lim_{m_t \to \infty} V_{H,m_t}^{\mu\nu} \to -iV_H^{\mu\nu}, \qquad (4.9)$$

where the phase factor of $-i$ arises from a difference in convention and is unimportant since we are always taking the modulus squared of the amplitude. We can therefore simply insert this vertex rather than the infinite top mass vertex into our amplitude to get the result

$$|M_{qQ \to qHQ}^{HE,ggH}|^2 = \frac{1}{4(N_C^2 - 1)} ||S_{qQ \to qHQ}^{m_t}(q_1, q_2)||^2 \\
\cdot \left(g^2 C_F \frac{1}{\hat{t}_1}\right) \\
\cdot \left(\frac{1}{\hat{t}_1} \left(\frac{-4g_s^2 m_t^2}{v}\right)^2 \frac{1}{\hat{t}_2}\right) \\
\cdot \left(g^2 C_F \frac{1}{\hat{t}_2}\right) \qquad (4.10)$$

with

$$S_{qQ \to qHQ}^{m_t}(q_1, q_2) = \langle 1|\mu|a\rangle \left(\eta^{\mu\nu} A_2(-q_1, q_2) + q_1^\nu q_2^\mu A_1(-q_1, q_2)\right) \langle 2|\nu|b\rangle. \qquad (4.11)$$

To see how much of a difference this makes to the amplitude, we plot the results for the infinite and finite top mass cases in figure 4.2. The momenta used are the following:

$$p_1 = (40\sqrt{2}\cosh(\Delta), -40, 40, 40\sqrt{2}\sinh(\Delta)), \qquad (4.12\text{a})$$
$$p_H = (\sqrt{40^2 + m_H^2}, 0, -40, 0), \qquad (4.12\text{b})$$
$$p_2 = (40\cosh(-\Delta), 40, 0, 40\sinh(-\Delta)). \qquad (4.12\text{c})$$

The two results agree well until around $\Delta = 3$ when they begin to split. In the

---

[3]It is important to realise that most given results are valid only in certain kinematical regions, so care must be taken to pick the correct analytical formula.

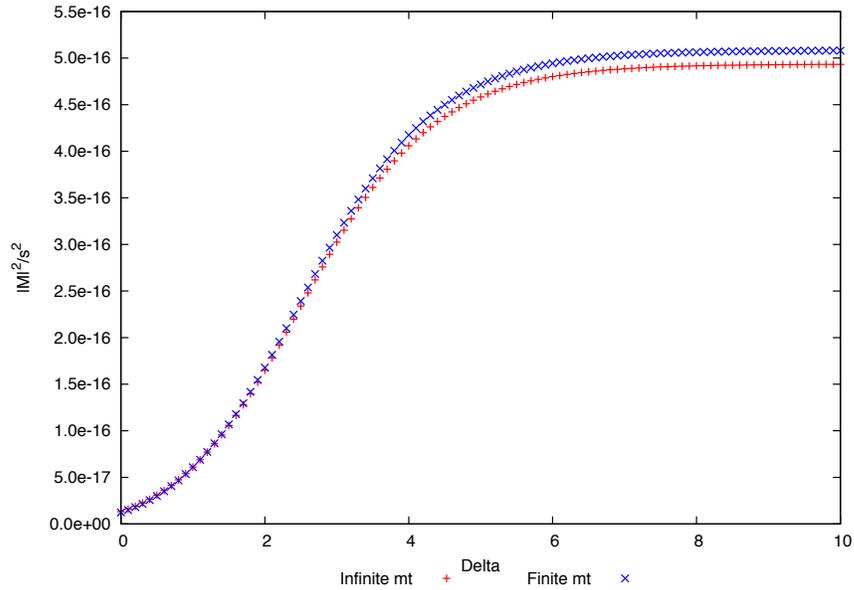

**Figure 4.2** *Behaviour of the $ud \to uHd$ amplitude in a slice of phase space where the Higgs boson is kept central in rapidity. The red line shows the result for the matrix element when the infinite top mass limit is taken and the blue line shows the result when the full top quark mass is taken into account.*

High Energy Limit, the finite top mass result is larger than the infinite top mass result by about 3%.

### 4.2.2 $gq \to Hgq$

The situation where the Higgs is emitted outside of an extremal gluon involves a much more thorough calculation. We will start by finding the general leading order expression and then use knowledge of the high energy limit considered $(y_H \sim y_1 \gg y_2)$ to again factorise the expression into the form HEJ requires, shown schematically in figure 4.3 and of the form

$$M \sim \frac{Z^\mu(p_a, p_1, p_H) \langle 2|\mu|b\rangle}{\hat{t}_2}, \qquad (4.13)$$

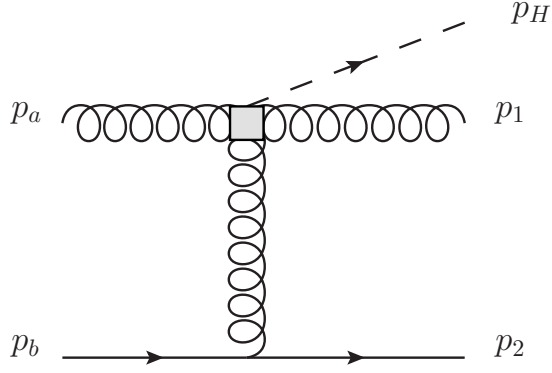

**Figure 4.3** *Diagrammatic representation of factorised $gq \to gHq$ expression.*

assuming that the polarisation vectors of the external gluons have been contracted with the effective vertex. Here, $\hat{t}_2 = p_2 - p_b$ is the $t$-channel pole we will resum (the definition of $\hat{t}_i$ used in the previous chapter no longer holds here since we have an extra momentum $p_H$ to consider). There are 20 leading order diagrams to consider in total, which is reduced to 10 after invoking Furry's Theorem [9], which states that diagrams involving an anti-quark loop can be related to diagrams with a quark loop. A selection of these diagrams is shown in figure 4.4. We will use [27] as a guide for our work. We begin by finding a suitable parametrisation for the triangle graphs that appear in the amplitudes. Using Furry's Theorem, we find that we can describe these graphs (one with a quark running around the loop and the other with an anti-quark) using only one object, $T^{\mu_1\mu_2}(q_1, q_2)$. Gauge invariance of these graphs implies $q_1^{\mu_1} T_{\mu_1\mu_2} = q_2^{\mu_2} T_{\mu_1\mu_2} = 0$, and so the generic tensor structure is

$$T^{\mu_1\mu_2}(q_1, q_2) = F_T(q_1^2, q_2^2, (q_1+q_2)^2) T_T^{\mu_1\mu_2} + F_L(q_1^2, q_2^2, (q_1+q_2)^2) T_L^{\mu_1\mu_2}, \quad (4.14)$$

where $T_T^{\mu_1\mu_2} = q_1 \cdot q_2 \eta^{\mu_1\mu_2} - q_1^{\mu_2} q_2^{\mu_1}$ and $T_L^{\mu_1\mu_2} = q_1^2 q_2^2 \eta^{\mu_1\mu_2} - q_1^2 q_2^{\mu_1} q_2^{\mu_2} - q_2^2 q_1^{\mu_1} q_1^{\mu_2} + q_1 \cdot q_2 q_1^{\mu_1} q_2^{\mu_2}$, and both $q_1$ and $q_2$ are going out from the triangle loop[4]. The full

---

[4]In our calculation, we will not conform to this convention and so we will take care with signs here.

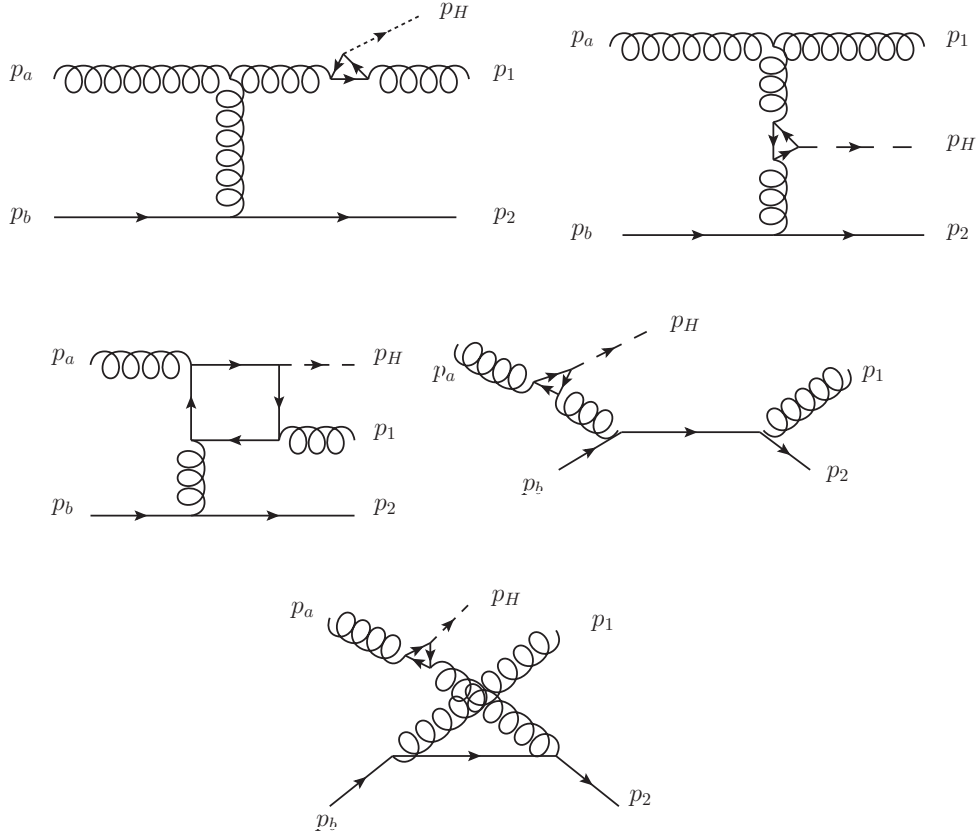

**Figure 4.4** *A selection of LO graphs for $gq \to Hgq$.*

forms for the $F_L$ and $F_T$ functions are

$$F_L(q_1^2, q_2^2, Q^2) = -\frac{1}{2\det \mathcal{Q}_2}\Bigg\{\left[2 - \frac{3q_1^2 q_2 \cdot Q}{2\det \mathcal{Q}_2}\right]\left(\tilde{B}_0(q_1) - \tilde{B}_0(Q)\right)$$
$$+ \left[2 - \frac{3q_2^2 q_1 \cdot Q}{2\det \mathcal{Q}_2}\right]\left(\tilde{B}_0(q_2) - \tilde{B}_0(Q)\right)$$
$$- \left[4m_t^2 + q_1^2 + q_2^2 + Q^2 - \frac{3q_1^2 q_2^2 Q^2}{\det \mathcal{Q}_2}\right]\tilde{C}_0(q_1, q_2) - 2\Bigg\},$$
$$F_T(q_1^2, q_2^2, Q^2) = -\frac{1}{2\det \mathcal{Q}_2}\Bigg\{Q^2\left[\tilde{B}_0(q_1) + \tilde{B}_0(q_2) - 2\tilde{B}_0(Q) - 2q_1 \cdot q_2\, \tilde{C}_0(q_1, q_2)\right]$$
$$+ \left(q_1^2 - q_2^2\right)\left(\tilde{B}_0(q_1) - \tilde{B}_0(q_2)\right)\Bigg\} - q_1 \cdot q_2\, F_L,$$

(4.15)

where $\det \mathcal{Q}_2 = q_1^2 q_2^2 - (q_1 \cdot q_2)^2$. The scalar integrals with the tilde notation are equal to the scalar integrals shown in equation 4.6 multiplied by $-16i\pi^2$. The $F_L$ and $F_T$ functions are related to the $A_1$ and $A_2$ of equation 4.7 by

$$A_1(q_1, q_2) = \frac{i}{(4\pi)^2} F_T(q_1, q_2),$$
$$A_2(q_1, q_2) = \frac{i}{(4\pi)^2} \left( F_T(q_1, q_2) q_1 \cdot q_2 + F_L(q_1, q_2) q_1^2 q_2^2 \right). \quad (4.16)$$

An interesting point to notice is that, if one of the momenta $q_1$ or $q_2$ is on-shell, we only get a contribution from the transverse ($F_T$) term in $T^{\mu\nu}$. This comes about because the presence of an on-shell momenta must mean that the vertex is going to be contracted with the relevant polarisation vector, setting the longitudinal term ($F_L$) to zero. When this happens, we will write $T_R$ in place of $T$ to remind ourselves of this. Finally, since we know that every graph contributes at the same order in $\alpha_s$, has a Yukawa coupling from the top loop and has factors from loop integrals, we conveniently define

$$F = \frac{4m_t^2}{v} \alpha_s^2, \quad (4.17)$$

and multiply $F$ into every graph as in [27]. This slightly changes the Feynman rules for QCD that we should use in that all coupling information is now factored out for convenience.

We now begin the process of finding the full LO expression. We start with the graph shown in the top left of figure 4.4, where the Higgs is emitted from the gluon leg with momentum $p_1$. The Feynman rules yield

$$A_1 = -F\varepsilon_{\mu_1}(p_a) f^{a1t} V_{3g}^{\mu_1\mu_2\mu_3}(p_a, -p_1 - p_H, -p_a + p_1 + p_H) \left( \frac{-i\eta_{\mu_2\mu_4}}{(p_1 + p_H)^2} \right)$$
$$T_R^{\mu_4\mu_5}(-p_1 - p_H, p_1) \varepsilon_{\mu_5}^*(p_1) \left( \frac{-i\eta_{\mu_3\mu_6}}{\hat{t}_2} \right) (-i) T_{2b}^t \langle 2|\mu_6|b\rangle, \quad (4.18)$$

which can be simplified to give

$$A_1 = \frac{-iF f^{a1t} T_{2b}^t}{(p_1 + p_H)^2 \hat{t}_2} \varepsilon_{\mu_1}(p_a) V_{3g}^{\mu_1\mu_2\mu_3}(p_a, -p_1 - p_H, -p_a + p_1 + p_H)$$
$$\cdot T_{R,\mu_2}^{\phantom{R,\mu_2}\mu_5}(-p_1 - p_H, p_1) \varepsilon_{\mu_5}^*(p_1) \langle 2|\mu_3|b\rangle. \quad (4.19)$$

The diagram where the Higgs is emitted from the gluon with momentum $p_a$ follows in essentially the same fashion and we just quote the final result of the calculation here, which is

$$A_2 = \frac{-iFf^{a1t}T^t_{2b}}{(p_a-p_H)^2 \hat{t}_2}\varepsilon_{\mu_1}(p_a)T^{\mu_1}_{R,\,\mu_3}(-p_a,p_a-p_H)V^{\mu_3\mu_4\mu_5}_{3g}(p_a-p_H,-p_1,-p_a+p_H+p_1)$$
$$\cdot \varepsilon^*_{\mu_4}(p_1)\langle 2|\mu_5|b\rangle\,.$$

(4.20)

We now consider the case where the Higgs is emitted from a t-channel gluon, as shown in the top right of figure 4.4. The Feynman rules give us

$$A_3 = -F\varepsilon_{\mu_1}(p_a)f^{a1t}V^{\mu_1\mu_2\mu_3}_{3g}(p_a,-p_1,p_1-p_a)\left(\frac{-i\eta_{\mu_3\mu_4}}{(p_a-p_1)^2}\right)\varepsilon^*_{\mu_2}(p_1)$$
$$\cdot T^{\mu_4\mu_5}(p_1-p_a,p_a-p_1-p_H)\left(\frac{-i\eta_{\mu_5\mu_6}}{\hat{t}_2}\right)(-i)T^t_{2b}\langle 2|\mu_6|b\rangle\,,$$

(4.21)

which simplifies to

$$A_3 = \frac{-iFf^{a1t}T^t_{2b}}{(p_a-p_1)^2\hat{t}_2}\varepsilon_{\mu_1}(p_a)V^{\mu_1\mu_2\mu_3}_{3g}(p_a,-p_1,p_1-p_a)\varepsilon^*_{\mu_2}(p_1)$$
$$\cdot T_{\mu_3}{}^{\mu_5}(p_1-p_a,p_a-p_1-p_H)\langle 2|\mu_5|b\rangle\,.$$

(4.22)

All of these three diagrams have fairly complicated forms but are automatically factorised in the form we are searching for and so require no approximation at all. The graph involving a box integral will also have this behaviour as we will see when we calculate it later on in this section. For the moment, we will discuss the other four diagrams which involve an $s$ or $u$ channel quark propagator. One such diagram is shown in the middle-right of figure 4.4. The Feynman rules for this diagram give

$$A_4 = -iFT^1_{2q}\varepsilon^*_{\mu_1}(p_1)\bar{u}_2\gamma^{\mu_1}\left(\frac{i(\slashed{p}_1+\slashed{p}_2)}{s_{12}}\right)\gamma^{\mu_2}u_b(-i)T^a_{qb}\left(\frac{-i\eta_{\mu_2\mu_3}}{(p_a-p_H)^2}\right)$$
$$\cdot T^{\mu_4\mu_3}_R(-p_a,p_a+p_H)\varepsilon_{\mu_4}(p_a),$$

(4.23)

which one can simplify to yield

$$A_4 = \frac{-FT^1_{2q}T^a_{qb}}{s_{12}(p_a-p_H)^2}\varepsilon^*_{\mu_1}(p_1)\bar{u}_2\gamma^{\mu_1}(\slashed{p}_1+\slashed{p}_2)\gamma^{\mu_2}u_b T^{\mu_4}_{R\,\mu_2}(-p_a,p_a-p_H)\varepsilon_{\mu_4}(p_a). \quad (4.24)$$

We have a similar diagram to this where there is still an $s$-channel quark but now the Higgs is emitted from the $p_1$ leg. The calculation is almost identical so we just quote the result which is

$$A_5 = \frac{-FT^1_{2q}T^a_{qb}}{s_{ab}(p_1+p_H)^2}\bar{\varepsilon}_{\mu_2}(p_a)\bar{u}_2\gamma^{\mu_1}(\slashed{p}_a+\slashed{p}_b)\gamma^{\mu_2}u_b T_R{}^{\mu_4}_{\mu_1}(-p1-p_H,p_1)\varepsilon^*_{\mu_4}(p_1). \quad (4.25)$$

Finally, we have the $u$-type diagrams. One such is shown at the bottom of figure 4.4. The Feynman rules yield

$$\begin{aligned}A_6 = &-FT^a_{2q}T^1_{qb}\bar{u}_2\gamma^{\mu_1}\left(\frac{i(\slashed{p}_b-\slashed{p}_1)}{-s_{b1}}\right)\gamma^{\mu_2}u_b\varepsilon^*_{\mu_1}(p_1)\left(\frac{-i\eta_{\mu_2\mu_3}}{(p_a-p_H)^2}\right) \\ &\cdot T^{\mu_4\mu_3}_R(-p_a,p_a+p_H)\varepsilon_{\mu_4}(p_a).\end{aligned} \quad (4.26)$$

This can be simplified to

$$A_6 = \frac{FT^a_{2q}T^1_{qb}}{s_{1b}(p_a-p_H)^2}\bar{u}_2\gamma^{\mu_1}(\slashed{p}_b-\slashed{p}_1)\gamma^{\mu}_2 u_b\varepsilon^*_{\mu_1}(p_1)T^{\mu_4}_{R\,\mu_2}(-p_a,p_a+p_H)\varepsilon_{\mu_4}(p_a). \quad (4.27)$$

The last diagram is the same except with the Higgs emitted from the gluon with momentum $p_1$ and the result is

$$A_7 = \frac{FT^a_{2q}T^1_{qb}}{s_{2a}(p_1+p_H)^2}\bar{u}_2\gamma^{\mu_1}(\slashed{p}_2-\slashed{p}_a)\gamma^{\mu}_2 u_b\varepsilon_{\mu_1}(p_a)T_R{}^{\mu_4}_{\mu_2}(-p_1-p_H,p_1)\varepsilon^*_{\mu_4}(p_1) \quad (4.28)$$

We now return to the box integrals. There are three independent contributions, related to the three different ways the two external gluons and one Higgs can be attached to the box. For ease, we use the parametrisation as described in [28] (not as in [27], though they are of course related), which writes this part of the amplitude in their notation as

$$M^{\mu\nu\rho} = \frac{2g_s^3 m_t^2}{v}if^{bac}T^c J^{\mu\nu\rho}(q_1,q_2,q). \quad (4.29)$$

The $q_i$ are related to the momenta used in this thesis by

$$q_1 = p_1, \quad (4.30\text{a})$$
$$q_2 = -p_a, \quad (4.30\text{b})$$
$$q = p_a - p_1 - p_H = p_2 - p_b, \quad (4.30\text{c})$$

and for the colour indices we map $a \to 1, b \to a, c \to t$. Finally, we write this factor as also proportional to $F$:

$$M^{\mu\nu\rho} = \left(F \times \frac{-16\pi^2}{2g_s}\right)f^{a1t}J^{\mu\nu\rho}. \quad (4.31)$$

A very important thing to notice here is that this result is taken from [28], which defines loop integrals with an overall factor of $\frac{1}{(2\pi)^4}$, whereas the loops defined in LoopTools [40] (the program we will use in the Monte Carlo integration) and [27] (the reference for all other parts of the calculation) have an overall factor of $\frac{1}{i\pi^2}$ instead. We take care of this by reweighting the LoopTools results when called here. For brevity, we will write the expression for $J$ in terms of the $q_i$ rather than the momenta of the external particles. Lifting this expression from [28] also forces us to use a particular gauge, which is the one where the polarisation vector of the gluon with momentum $p_a$ is perpendicular to $p_1$ and vice versa. Imposing this, the function $J$ is

$$J^{\mu\nu\rho} = \eta^{\mu\nu}(H_1 q_1^\rho + H_2 q_2^\rho) + \eta^{\mu\rho} H_4 q^\nu + \eta^{\nu\rho} H_5 q^\mu + H_{10} q_2^\rho q^\mu q^\nu + H_{12} q_1^\rho q^\mu q^\nu. \quad (4.32)$$

The full expressions for each of the $H$ functions are very long and are given in Appendix B. A useful study is to investigate the link between the finite and infinite top mass cases for the box diagrams. This will, for example, give us a stringent numerical check on our implementation. In the infinite top mass case, the box diagram becomes a three gluon vertex diagram multiplied by a factor, since the limit shrinks quark loops. Using this knowledge and applying it to equation 4.32 along with the factor $F$, we see that the following must hold:

$$\begin{aligned}
\lim_{m_t \to \infty} \frac{2\pi F}{\alpha_s} H_1 &= iA, \\
\lim_{m_t \to \infty} \frac{2\pi F}{\alpha_s} H_2 &= -iA, \\
\lim_{m_t \to \infty} \frac{2\pi F}{\alpha_s} H_4 &= iA, \\
\lim_{m_t \to \infty} \frac{2\pi F}{\alpha_s} H_5 &= -iA, \\
\lim_{m_t \to \infty} \frac{2\pi F}{\alpha_s} H_{10} &= 0, \\
\lim_{m_t \to \infty} \frac{2\pi F}{\alpha_s} H_{12} &= 0,
\end{aligned} \quad (4.33)$$

with $A = \frac{\alpha_s}{3\pi v}$ as before. This was tested numerically in a computer program by setting $m_t = 17400$ and seen to hold in all cases.

We can now combine all graphs together. We have three different colour structures appearing from the individual sub-amplitudes, but we use the fact that $[T^a, T^b] = if^{abc}T^c$ to see that we can reduce this down to two. For the time being, we will keep the amplitudes separated into their 'natural' colour factor and then later on use our commutator identities. Before we write the full amplitude, we introduce one more useful piece of notation. Throughout the calculation, we came across many instances where one of the polarisation vectors was contracted with a $T_R$ function from the top loop. It will be useful to define an 'effective polarisation vector', which is precisely this contraction along with the propagator invariant. In other words

$$\frac{\varepsilon_{\mu_1}(p_a)T_{R\,\mu_2}^{\mu_1}(-p_a, p_a - p_H)}{(p_a - p_H)^2} = \frac{F_T(p_a^2, (p_a - p_H)^2, p_H^2)\left(p_a \cdot p_H \varepsilon_{\mu_2}(p_a) - p_{a\,\mu_2} p_H \cdot \varepsilon(p_a)\right)}{(p_a - p_H)^2}$$
$$\equiv \varepsilon_{H,\mu_2}(p_a), \tag{4.34a}$$

$$\frac{\varepsilon_{\mu_1}^*(p_1)T_{R\,\mu_2}^{\mu_1}(-p_1 - p_H, p_1)}{(p_1 + p_H)^2} = \frac{F_T((p_1 + p_H)^2, p_1^2, p_H^2)\left(p_H \cdot \varepsilon^*(p_1) p_{1\,\mu_2} - p_1 \cdot p_H \varepsilon_{\mu_2}^*(p_1)\right)}{(p_1 + p_H)^2}$$
$$\equiv \varepsilon_{H,\mu_2}^*(p_1). \tag{4.34b}$$

Note that the idea of the effective polarisation vector has been taken from [27], but the forms look quite different because we have extra minus signs to match outgoing/incoming momentum conventions ([27] takes all momenta incoming whilst we do not). The full amplitude is then

$$M_{gq \to Hgq}^{m_t} = F\Bigg( T_{2q}^a T_{qb}^1 \bigg[ \frac{\bar{u}_2 \gamma^{\mu_1}(\slashed{p}_2 - \slashed{p}_a)\gamma^{\mu_2} u_b \, \varepsilon_{\mu_1}(p_a)\, \varepsilon_{H,\mu_2}^*(p_1)}{s_{2a}}$$
$$+ \frac{\bar{u}_2 \gamma^{\mu_1}(\slashed{p}_b - \slashed{p}_1)\gamma^{\mu_2} u_b \, \varepsilon_{H,\mu_1}(p_a)\, \varepsilon_{\mu_2}^*(p_1)}{s_{1b}} \bigg] -$$
$$T_{2q}^1 T_{qb}^a \bigg[ \frac{\bar{u}_2 \gamma^{\mu_1}(\slashed{p}_a + \slashed{p}_b)\gamma^{\mu_2} u_b \, \varepsilon_{\mu_2}(p_a)\, \varepsilon_{H,\mu_1}^*(p_1)}{s_{ab}}$$
$$+ \frac{\bar{u}_2 \gamma^{\mu_1}(\slashed{p}_1 + \slashed{p}_2)\gamma^{\mu_2} u_b \, \varepsilon_{\mu_1}^*(p_1)\, \varepsilon_{H,\mu_2}(p_a)}{s_{12}} \bigg]$$
$$- [T^a, T^1]_{2b} \frac{\langle 2|\mu_3|b\rangle}{\hat{t}_2} \bigg[ 8i\pi^2 \varepsilon_{\mu_2}(p_a)\varepsilon_{\mu_1}^*(p_1) J^{\mu_1\mu_2\mu_3}(p_1, -p_a, p_a - p_1 - p_H)$$
$$+ \varepsilon_{H,\mu_1}(p_a) V_{3g}^{\mu_1\mu_2\mu_3}(p_a - p_H, -p_1, -p_a + p_H + p_1)\, \varepsilon_{\mu_2}^*(p_1)$$
$$+ \varepsilon_{\mu_1}(p_a) V_{3g}^{\mu_1\mu_2\mu_3}(p_a, -p_1 - p_H, -p_a + p_1 + p_H)\, \varepsilon_{H,\mu_2}^*(p_1)$$
$$+ \frac{\varepsilon_{\mu_1}(p_a) V_{3g}^{\mu_1\mu_2\mu_4}(p_a, -p_1, p_1 - p_a)\, \varepsilon_{\mu_2}^*(p_1) T_{\mu_4}^{\phantom{\mu_4}\mu_3}(p_1 - p_a, p_a - p_1 - p_H)}{(p_a - p_1)^2} \bigg] \Bigg). \tag{4.35}$$

This expression has been checked term-by-term with [27] and [28] with full agreement. With the full amplitude known, we can hope to use limiting arguments to factorise the expression into the form we require. To do this, we focus on the first four terms in the amplitude, since the other terms are already in the correct form. These four terms have elements of the desired form within them. To see this, consider the numerator of the first term,

$$A_1^{num} = \bar{u}_2 \gamma^{\mu_1} (\not{p}_2 - \not{p}_a) \gamma^{\mu_2} u_b \varepsilon_{\mu_1}(p_a) \varepsilon^*_{H,\mu_2}(p_1). \tag{4.36}$$

With the completeness relation, we can rewrite the $\not{p}$ parts in terms of massless spinors, $\not{p} = |p^+\rangle \langle p^+| + |p^-\rangle \langle p^-|$. By considering the action of the projection operator $(1 \pm \gamma^5)$ it is simple to see that you only pick out one of these helicity projections in the spinor chain, which is the helicity projection corresponding to the helicity of particles $b$ and 2. Thus we can write the term as

$$A_1^{num} = (\langle 2|\mu_1|2] \langle 2|\mu_2|b] + \langle 2|\mu_1|a] \langle a|\mu_2|b]) \varepsilon_{\mu_1}(p_a) \varepsilon^*_{H,\mu_2}(p_1). \tag{4.37}$$

We recall at this point a useful parametrisation for the gluon polarisation vectors, as detailed in [31]:

$$\varepsilon^\pm_\mu(k, q) = \pm \frac{\langle q^\mp |\mu| k^\mp \rangle}{\sqrt{2} \langle q^\mp | k^\pm \rangle}, \tag{4.38}$$

where $k$ is the momentum of the gluon and $q$ is an arbitrary massless reference momentum which reflects our gauge freedom. This notation is useful because it allows us to apply some of the identities established in section 1.6 to perform dot products. As previously discussed, the parametrisation of the box function was taken from [28]. This parametrisation is only valid in a certain gauge choice (the one where both external gluons are orthogonal to both $p_a$ and $p_b$) that we must therefore conform to, yielding the following forms of the polarisation vectors:

$$\begin{aligned} \varepsilon^\pm_\mu(p_a) &= \pm \frac{\langle 1^\mp |\mu| a^\mp \rangle}{\sqrt{2} \langle 1^\mp | a^\pm \rangle}, \\ \varepsilon^\pm_\mu(p_1) &= \pm \frac{\langle a^\mp |\mu| 1^\mp \rangle}{\sqrt{2} \langle 1^\mp | a^\pm \rangle}. \end{aligned} \tag{4.39}$$

Using this, we can perform the $\mu_1$ contraction in equation 4.37 and see that if the spinor chain and the polarisation vector $\varepsilon(p_a)$ have the same helicity, the second term is identically zero – this is true given any gauge choice. When they have opposite helicity, then the dot product goes like $\langle a1 \rangle [a2]$. The first term will instead go like $\langle 21 \rangle [a2]$. The magnitudes of the square and angled brackets are square roots of invariants, so the ratio of these terms is like $\sqrt{s_{a1}/s_{12}}$. Since we are considering the limit where $s_{12}$ is large and $s_{a1}$ not necessarily so, we can then neglect the second term. This will also apply to the $\slashed{p}_a$ part of the third term of the full amplitude; a similar argument can be made for the $\slashed{p}_1$ parts of the second and fourth terms. Thus we can approximate the first four terms as

$$\tilde{A} \equiv iF\left(T_{2q}^a T_{qb}^1 \left[\frac{\bar{u}_2 \gamma^{\mu_1} \slashed{p}_2 \gamma^{\mu_2} u_b \varepsilon_{\mu_1}(p_a) \varepsilon^*_{H,\mu_2}(p_1)}{s_{2a}} + \frac{\bar{u}_2 \gamma^{\mu_1} \slashed{p}_b \gamma^{\mu_2} u_b \varepsilon_{H,\mu_1}(p_a) \varepsilon^*_{\mu_2}(p_1)}{s_{1b}}\right] \right.$$
$$\left. - T_{2q}^1 T_{qb}^a \left[\frac{\bar{u}_2 \gamma^{\mu_1} \slashed{p}_b \gamma^{\mu_2} u_b \varepsilon_{\mu_2}(p_a) \varepsilon^*_{H,\mu_1}(p_1)}{s_{ab}} + \frac{\bar{u}_2 \gamma^{\mu_1} \slashed{p}_2 \gamma^{\mu_2} u_b \varepsilon^*_{\mu_1}(p_1) \varepsilon_{H,\mu_2}(p_a)}{s_{12}}\right]\right).$$
(4.40)

Because we will be performing contractions and evaluating spinor brackets with the polarisation vectors, we will rewrite them in a different (though of course equivalent) form that conforms with our spinor definitions. The formula we quoted from [31] is designed for the case where all momenta are taken as outgoing, so when it comes to writing out these contractions explicitly (not just schematically like we did for the scaling argument) we cannot be confident that the convention is the same. We therefore employ polarisation vectors for explicit calculations that are analogous to those used in [7], which works within the same conventions for the spinors as we have here. The explicit forms of the vectors are:

$$\varepsilon(p_a)^+ = \frac{\langle a^-|\mu|1^-\rangle}{\sqrt{2}[a1]} = \frac{\langle 1^+|\mu|a^+\rangle}{\sqrt{2}[a1]}, \tag{4.41a}$$

$$(\varepsilon(p_1)^+)^* = -\frac{\langle a^-|\mu|1^-\rangle}{\sqrt{2}\langle 1a\rangle} = -\frac{\langle 1^+|\mu|a^+\rangle}{\sqrt{2}\langle 1a\rangle}, \tag{4.41b}$$

$$\varepsilon(p_a)^- = \frac{\langle 1^-|\mu|a^-\rangle}{\sqrt{2}\langle 1a\rangle} = \frac{\langle a^+|\mu|1^+\rangle}{\sqrt{2}\langle 1a\rangle}, \tag{4.41c}$$

$$(\varepsilon(p_1)^-)^* = -\frac{\langle 1^-|\mu|a^-\rangle}{\sqrt{2}[a1]} = -\frac{\langle a^+|\mu|1^+\rangle}{\sqrt{2}[a1]}. \tag{4.41d}$$

We now expand our expression for $\tilde{A}$ using the completeness relation to give

$$\tilde{A} \equiv F\left(T_{2q}^a T_{qb}^1 \left[\frac{2p_2^{\mu_1}\langle 2|\mu_2|b\rangle \varepsilon_{\mu_1}(p_a)\varepsilon^*_{H,\mu_2}(p_1)}{s_{2a}} + \frac{\langle 2|\mu_1|b\rangle 2p_b^{\mu_2}\varepsilon_{H,\mu_1}(p_a)\varepsilon^*_{\mu_2}(p_1)}{s_{1b}}\right] - \right.$$
$$\left. T_{2q}^1 T_{qb}^a \left[\frac{\langle 2|\mu_1|b\rangle 2p_b^{\mu_2}\varepsilon_{\mu_2}(p_a)\varepsilon^*_{H,\mu_1}(p_1)}{s_{ab}} + \frac{2p_2^{\mu_1}\langle 2|\mu_2|b\rangle \varepsilon^*_{\mu_1}(p_1)\varepsilon_{H,\mu_2}(p_a)}{s_{12}}\right]\right).$$
(4.42)

Note that we have not made any choice for the helicity of the quark line $\langle 2|\mu|b\rangle$ nor will we need to; we aim to factor out this string from our expression, and the terms $2p_b$ and $2p_2$ will appear regardless of the helicity choice. We will now explicitly calculate the contractions with the polarisation vectors using the same spinor convention as outlined in section 1.6. Since our aim is to remove all $p_b$ and $p_2$ dependence from our effective vertex, it will be useful to consider which spinor bracket combinations are independent of these. Two useful results are (we will assume $p_a$ is moving in the + direction for now, but generalise later)

$$\frac{[1b]}{[ba]} = -\sqrt{\frac{p_1^+}{p_a^+}}, \tag{4.43a}$$

$$\frac{\langle ba\rangle}{\langle 1b\rangle} = -\sqrt{\frac{p_a^+}{p_1^+}}. \tag{4.43b}$$

These are exact. However, if we consider the limit $p_2^- \sim p_b^-$ that is still valid here, then we have additional, approximate results we can use. For example

$$\langle 12\rangle = \sqrt{p_2^+ p_1^-}\, e^{i\phi_1} - \sqrt{p_2^- p_1^+}\, e^{i\phi_2} \approx -\sqrt{p_2^- p_1^+}\, e^{i\phi_2}, \tag{4.44}$$

where we use the fact that both $p_2^+$ and $p_1^-$ are suppressed in comparison to $p_2^-$ and $p_1^+$. Using this, we also use the results (only valid the High Energy Limit)

$$\frac{[12]}{[a2]} = \sqrt{\frac{p_1^+}{p_a^+}}, \tag{4.45a}$$

$$\frac{\langle a2\rangle}{\langle 12\rangle} = \sqrt{\frac{p_a^+}{p_1^+}}. \tag{4.45b}$$

Let us now return to our expression for $\tilde{A}$. We calculate the dot product between the pure momentum term (either $p_b$ or $p_2$) and the polarisation vector. There are two cases we need to consider: firstly, when the helicity of the gluons with momentum $p_a$ and $p_1$ are the same (helicity-conserving); and secondly, when they differ (helicity non-conserving). Though there are of course four total choices for the helicities, we need only consider two, being able to get the other two by parity relations. We start with the helicity-conserving case and choose gluons $a, 1$ to both have positive helicity. The first term is

$$\frac{2p_2^{\mu_1} \langle 2|\mu_2|b] \, \varepsilon_{\mu_1}^+(p_a)\varepsilon_{H,\mu_2}^{+,*}(p_1)}{s_{2a}} = \frac{2 \langle 2a \rangle [12] \langle 2|\mu_2|b] \, \varepsilon_{H,\mu_2}^{+,*}(p_1)}{\sqrt{2}[a1]\langle 2a \rangle [a2]}$$
$$= \frac{\sqrt{2} \langle 2|\mu_2|b] \, \varepsilon_{H,\mu_2}^{+,*}(p_1)}{[a1]} \sqrt{\frac{p_1^+}{p_a^+}}, \qquad (4.46)$$

and therefore (once we factor out the spinor current) is completely independent of both $p_b$ and $p_2$. Similar results occur with the other three terms and, in fact, terms 1 and 3, and 2 and 4, will become equal. At that point, we can rewrite $\tilde{A}$ as proportional to the colour commutator $[T^a, T^1]$. The result is

$$\tilde{A}_{++} = \sqrt{2}F[T^a, T^1]\langle 2|\mu|b] \left( \sqrt{\frac{p_1^+}{p_a^+}} \frac{\varepsilon_{H,\mu}^{+,*}(p_1)}{[a1]} - \sqrt{\frac{p_a^+}{p_1^+}} \frac{\varepsilon_{H,\mu}^+(p_a)}{\langle 1a \rangle} \right). \qquad (4.47)$$

For the helicity non-conserving case, we choose the helicity of the gluon with momentum $p_a$ to be positive and the other gluon to have negative helicity, yielding similar results:

$$\tilde{A}_{+-} = \sqrt{2}F[T^a, T^1]\frac{\langle 2|\mu|b]}{[a1]} \left( \sqrt{\frac{p_1^+}{p_a^+}} \varepsilon_{H,\mu}^{-,*}(p_1) - \sqrt{\frac{p_a^+}{p_1^+}} \varepsilon_{H,\mu}^+(p_a) \right). \qquad (4.48)$$

If instead we chose $p_a$ to be moving in the - direction, we find by direct calculation that we need to multiply $\tilde{A}_{++}$ and the first term of $\tilde{A}_{+-}$ by $-\frac{p_{1,\perp}^*}{|p_{1,\perp}|}$, and the second term of $\tilde{A}_{+-}$ by $-\frac{p_{1,\perp}}{|p_{1,\perp}|}$.

We have now achieved our goal of creating a factorised matrix element. Before we write down the explicit expressions, we consider if the form can be simplified. For example, in the helicity conserving case, our gauge actually removes the contribution from one diagram. The graph shown in the top right of figure 4.4 has a part which is a three-gluon vertex contracted with symmetric polarisation vectors that depend only on the momenta along the top line. It is easy to show by direct calculation that this yields a result of zero. Because we are going to implement this in a numerical integration program, it is important to find these contributions where the analytical result is zero. If we did not, depending on the accuracy of the calculation, a computer might give you a small (but vitally non-zero) answer that could lead to instabilities.

We now have everything we need to state the result of $gq \to Hgq$ in the limit where the Higgs is emitted outside of the gluon in rapidity space. For our High

Energy expressions, we will present a form that will conform with equation 4.13 by performing some of the index contractions. After some manipulation, we find for the helicity conserving amplitude (we will take $p_a$ to be moving in the + direction, but recall that we know how to immediately get to the situation where it is going in the - direction)

$$\begin{aligned}
A_{++} =& F[T^a,T^1]\frac{\langle 2|\mu|b\rangle}{\hat{t}_2}\Bigg[\sqrt{\frac{2p_1^+}{p_a^+}}\frac{\varepsilon_{H,\mu}^{+,*}(p_1)\hat{t}_2}{[a1]} - \sqrt{\frac{2p_a^+}{p_1^+}}\frac{\varepsilon_{H,\mu}^{+}(p_a)\hat{t}_2}{\langle 1a\rangle} \\
& + \frac{\langle 1^+|H|a^+\rangle}{\sqrt{2}\langle 1a\rangle}\varepsilon_{H,\mu}^+(p_a) + \frac{\langle 1^+|H|a^+\rangle}{[a1]}\varepsilon_{H,\mu}^{+,*}(p_1) \\
& - \frac{\sqrt{2}F_{Ta}p_a\cdot p_1\langle 1^+|H|a^+\rangle}{[a1]}\varepsilon_\mu^{+,*}(p_1) - \frac{\sqrt{2}F_{T1}p_a\cdot p_1\langle 1^+|H|a^+\rangle}{\langle 1a\rangle}\varepsilon_\mu^+(p_a) \\
& - \frac{\langle 1^+|H|a^+\rangle}{\sqrt{2}[a1]}\varepsilon_\mu^{+,*}(p_1)RH_4 + \frac{\langle 1^+|H|a^+\rangle}{\sqrt{2}\langle 1a\rangle}\varepsilon_\mu^+(p_a)RH_5 \\
& + \frac{\langle 1^+|H|a^+\rangle^2}{2\langle 1a\rangle[a1]}\{p_{a,\mu}RH_{10} - p_{1,\mu}RH_{12}\}\Bigg]
\end{aligned}$$

(4.49)

where $F_{Ta} = \frac{F_T(0,(p_a-p_H)^2,m_H^2)}{(p_a-p_H)^2}$, $F_{T1} = \frac{F_T((p_1+p_H)^2,0,m_H^2)}{(p_1+p_H)^2}$ and $R = 8i\pi^2$. The part in the square brackets can be interpreted (up to some overall constant) as the effective vertex we were searching for, in the helicity conserving case. What remains is the helicity non-conserving case, which is given by

$$
\begin{aligned}
A_{+-} = F[T^a, T^1] \frac{\langle 2|\mu|b\rangle}{\hat{t}_2} &\Bigg[ \sqrt{\frac{2p_1^+}{p_a^+}} \frac{\varepsilon_{H,\mu}^{+,*}(p_1)\hat{t}_2}{[a1]} - \sqrt{\frac{2p_a^+}{p_1^+}} \frac{\varepsilon_{H,\mu}^{+}(p_a)\hat{t}_2}{[a1]} \\
&+ \frac{\langle 1^+|H|a^+\rangle}{\sqrt{2}[a1]} \varepsilon_{H,\mu}^{+}(p_a) + \frac{\langle 1^+|H|a^+\rangle}{[a1]} \varepsilon_{H,\mu}^{+,*}(p_1) \\
&- \frac{\sqrt{2} F_{Ta} p_a \cdot p_1 \langle 1^+|H|a^+\rangle}{[a1]} \varepsilon_\mu^{+,*}(p_1) - \frac{\sqrt{2} F_{T1} p_a \cdot p_1 \langle 1^+|H|a^+\rangle}{[a1]} \varepsilon_\mu^{+}(p_a) \\
&- \frac{\langle 1^+|H|a^+\rangle}{\sqrt{2}[a1]} \varepsilon_\mu^{+,*}(p_1) RH_4 + \frac{\langle 1^+|H|a^+\rangle}{\sqrt{2}[a1]} \varepsilon_\mu^{+}(p_a) RH_5 \\
&+ \frac{\langle 1^+|H|a^+\rangle^2}{2[a1]^2} \{p_{a,\mu} RH_{10} - p_{1,\mu} RH_{12}\} \\
&+ e^{i\phi} RH_1 p_{1,\mu} - e^{i\phi} RH_2 p_{a,\mu} + 2 e^{i\phi} F_{T1} p_1 \cdot p_H p_{a,\mu} - 2 e^{i\phi} F_{Ta} p_a \cdot p_H p_{1,\mu} \\
&- e^{i\phi} (p_a + p_1)_\mu F_\alpha \frac{(p_1 - p_a) \cdot (p_a - p_1 - p_H)}{(p_a - p_1)^2} \\
&+ e^{i\phi} (p_a - p_1 - p_H) \cdot (p_a + p_1) \frac{(p_1 - p_a)_\mu}{(p_a - p_1)^2} F_\alpha \\
&- e^{i\phi} F_\beta (p_a - p_1 - p_H)^2 (p_a + p_1)_\mu \Bigg]
\end{aligned}
$$

(4.50)

where $F_\alpha = F_T(p_1 - p_a, p_a - p_1 - p_H, p_H)$, $F_\beta = F_L(p_1 - p_a, p_a - p_1 - p_H, p_H)$ and $\varepsilon^{-,*}(p_1) \cdot \varepsilon^{+}(p_a) = e^{i\phi}$ is a phase factor.

With both the helicity conserving and non-conserving vertices found, we are able to provide a High Energy approximation to the whole amplitude by manually performing the colour/helicity sum/average. The colour sum is simple, since the approximation is proportional only to $f^{a1t}T^t$ and, using the normalisation $\text{tr}(T^a T^b) = \frac{1}{2}\delta^{ab}$, we find that the sum yields an answer of 12. Since we have a quark and gluon incoming, the average factor is $\frac{1}{3\times 8} = \frac{1}{24}$. For the helicities, there are are total of eight combinations, of which we need only work out four because the other four are related by parity, and we average by a factor of 4. This gives the full approximation (where the subscripts refer to the helicities of particles with momenta $p_a, p_1, p_2 \& p_b$ respectively and $P$ is the parity operation)

$$
|M_{gq \to Hgq}^{m_t, HE}|^2 = \frac{1}{8} \Big( |A_{++,-}|^2 + |A_{++,+}|^2 + |A_{+-,-}|^2 + |A_{+-,+}|^2 + \\
|P(A_{++,-})|^2 + |P(A_{++,+})|^2 + |P(A_{+-,-})|^2 + |P(A_{+-,+})|^2 \Big).
$$

(4.51)

### 4.2.3 Checks and Verifications of Amplitudes

At this point, it is sensible to run some checks on the amplitude to have confidence in its correctness. We first check a phase space configuration where the Higgs boson is kept central in rapidity and the two extremal jets are allowed to move further and further apart. The precise momentum configuration used is the same as the one used at the end of section 4.2.1:

$$p_1 = (40\sqrt{2}\cosh(\Delta), -40, 40, 40\sqrt{2}\sinh(\Delta)), \quad (4.52a)$$

$$p_H = (\sqrt{40^2 + m_H^2}, 0, -40, 0), \quad (4.52b)$$

$$p_2 = (40\cosh(-\Delta), 40, 0, 40\sinh(-\Delta)). \quad (4.52c)$$

This configuration is useful to check against for two reasons. Firstly, it is a more restrictive limit than the one we considered to derive our amplitude (here, $s_{1H}$ also must be large) and so our calculation should correctly describe this situation too. Secondly, High Energy theory tells us that this process should behave like the $qQ \to qHQ$ case where the Higgs is produced far from the parton rapidities with a reweighting due to the quark being replaced by a gluon. Figure 4.5 shows both of these points clearly, where we also plot the full LO line given by MadGraph, which recently gained the ability to include the full finite top mass effects for suitably low-multiplicity matrix elements [1].

Our second phase space parametrisation fixes the Higgs boson to always be close to the gluon in rapidity whilst the two jets move further apart in rapidity. Explicitly

$$p_1 = (40\sqrt{2}\cosh(\Delta), -40, 40, 40\sqrt{2}\sinh(\Delta)), \quad (4.53a)$$

$$p_H = (\sqrt{40^2 + m_H^2}\cosh(\Delta+0.5), 0, -40, \sqrt{40^2 + m_H^2}\sinh(\Delta+0.5)), \quad (4.53b)$$

$$p_2 = (40\cosh(-\Delta), 40, 0, 40\sinh(-\Delta)). \quad (4.53c)$$

In figure 4.6, we plot this new effective vertex approach against the full LO result with this set of momenta. We also include the result of the reweighted $qQ \to qHQ$ amplitude to show that this is not an appropriate result in this limit. In both cases, we see clear agreement between our result and the full LO result from MadGraph in the High Energy (high $\Delta$) Limit and relatively small deviations below this. We now move on to check the effects of having a finite quark mass as opposed to an infinite one. By simply putting a high value for the top mass in our amplitude (numerically, we see that $m_t = 17400$ is a good choice – higher values are unstable) we can generate results that correspond to the effective theory where the top mass is treated as an infinite parameter. Additionally, we can very easily add the interference via bottom loops to our result (when working out the amplitudes for helicity configurations, simply add the same amplitude with the bottom mass before squaring it) so we can also see how large an effect this has. We will begin by looking at these three cases with a central Higgs (the first set of momenta), which is plotted in figure 4.7. We see that there are clear differences between the cases. The finite top mass case is always greater than the infinite top mass case – in the High Energy Limit, this difference is around 3%. The addition of the bottom quark increases this difference by a further 3%. The same comparison for a Higgs close to the gluon (the second set of momenta) is plotted in figure 4.8. In that case, we have slightly different behaviour - the infinite top mass case is still lower (approximately 5% lower than the finite top mass line), but now the addition of the bottom loop (slightly) decreases the ME from the case where only the top quark is considered.

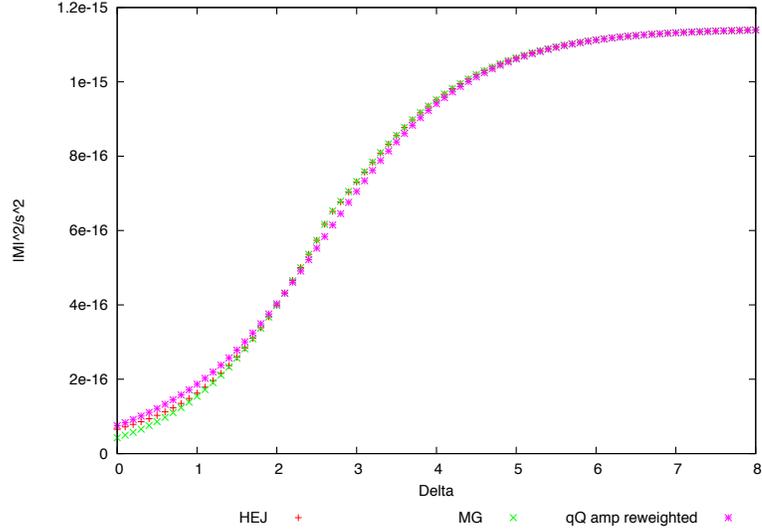

**Figure 4.5** *Comparison between the HEJ effective vertex (red), the full LO result (green) and the result of the $qQ \to qHQ$ LO calculation reweighted by a colour factor for a central $H$ in $gq \to Hgq$ (purple) with full top mass dependence.*

The addition of extra emissions is again achieved by multiplying Lipatov vertices

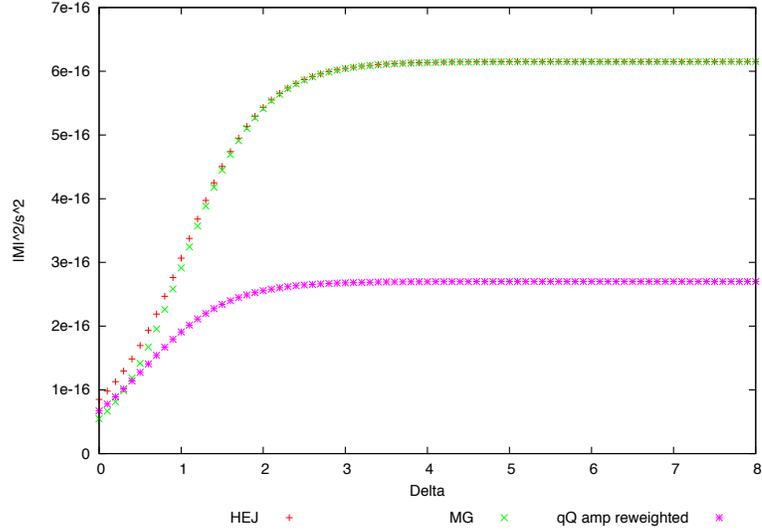

**Figure 4.6** *Comparison between the HEJ effective vertex (red), the full LO result (green) and the result of the $qQ \to qHQ$ LO calculation reweighted by a colour factor (purple) for an outside $H$ in $gq \to Hgq$ with full top mass dependence.*

into the amplitude we have derived. Some care is required to ensure the correct $t$-channel momenta are taken, resulting in a general $gq \to Hg...q$ amplitude (where the ... represent an arbitrary number of gluons) which is simply

$$|M^{HE,m_t}_{gq \to Hg...q}|^2 = |M^{HE,m_t}_{gq \to Hgq}|^2 \times \prod_{i=1}^{n-2} \frac{-g_s^2 C_A V^\mu(q_i, q_{i+1}) V_\mu(q_i, q_{i+1})}{q_i^2 q_{i+1}^2}, \qquad (4.54)$$

where $n$ is the number of final state jets and the $q_i$ are the $t$-channel momenta entering the Lipatov vertices – we define $q_1 = p_a - p_1 - p_H$ and $q_i = q_{i-1} - p_i$ for $i > 1$. The matrix element squared has an implicit division of $q_1^2 q_{n-1}^2$. As a check, we will generate more explorer plots for the case of one extra emission. We will choose the process $gu \to Hggu$ with a few different choices for the rapidity of the extra gluon and the Higgs to ensure we are calculating correctly. Our first configuration is used for when the Higgs is being emitted close to the extremal

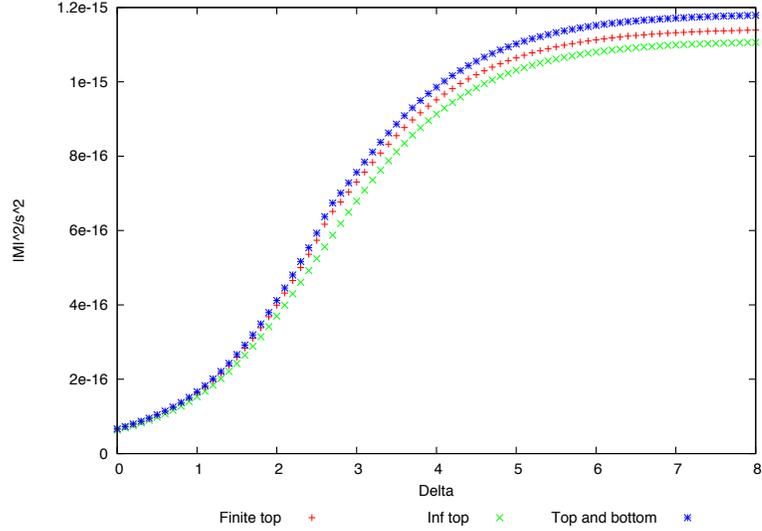

**Figure 4.7** *Comparison of infinite top (green), finite top (red) and finite top plus finite bottom (blue) HEJ matrix elements with a central Higgs.*

gluon

$$p_1 = (40\cosh(\Delta), -40, 0, 40\sinh(\Delta)), \tag{4.55a}$$

$$p_H = (\sqrt{40^2 + m_H^2}\cosh(\Delta + 0.5), 0, -40, \sqrt{40^2 + m_H^2}\sinh(\Delta + 0.5)), \tag{4.55b}$$

$$p_2 = (40\cosh(-\Delta/3), 0, 40, 0), \tag{4.55c}$$

$$p_3 = (40\cosh(-\Delta), 40, 0, 40\sinh(-\Delta)). \tag{4.55d}$$

Unfortunately, for these amplitudes the finite $m_t$ result available in MadGraph is both slow and numerically unstable at high $\Delta$, as shown in figure 4.9. We see that the two results begin to agree with each other in the High Energy Limit as expected, but then that the MadGraph result starts to give 'NaN' as a result – these correspond to missing points on the line. We will then instead set $m_t$ to 17400 and compare to the full LO effective theory matrix element (available in earlier versions of MadGraph) for the remainder of this subsection. We show $|M|^2/\hat{s}^2$ as a function of $\Delta$ in figure 4.10 for this momentum configuration in the infinite top mass limit. The agreement between our result and MadGraph's finite top mass LO calculation before this point serves as evidence that the finite quark mass element of the amplitude is correct. The agreement in figure 4.10 is good across the phase space and almost identical above $\Delta = 6$. Another slice of phase space we can take is one where all particles gradually move apart in rapidity:

$$p_1 = (40\cosh(\Delta), -40, 0, 40\sinh(\Delta)), \tag{4.56a}$$

$$p_H = (\sqrt{40^2 + m_H^2}\cosh(\Delta/3), 0, -40, \sqrt{40^2 + m_H^2}\sinh(\Delta/3)), \tag{4.56b}$$

$$p_2 = (40\cosh(-\Delta/3), 0, 40, 40\sinh(-\Delta/3)), \tag{4.56c}$$

$$p_3 = (40\cosh(-\Delta), 40, 0, 40\sinh(-\Delta)). \tag{4.56d}$$

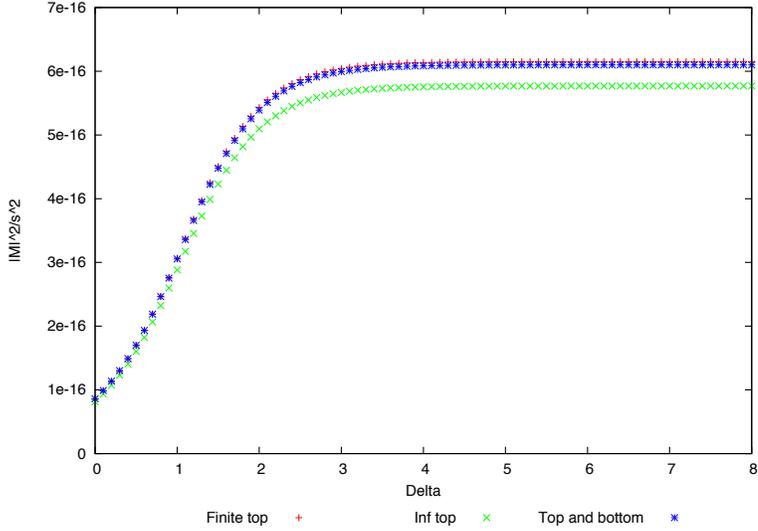

**Figure 4.8** *Comparison of infinite top (green), finite top (red) and finite top plus finite bottom (blue) HEJ matrix elements with an outside Higgs.*

The explorer plot for this configuration is shown in figure 4.11. Again, the agreement is good between the two lines across the full range. The other configuration to check is the one where the Higgs boson is more behind the extra emission. This contribution cannot be described by this matrix element for the $gq$ incoming state, but it can describe the process with the $gg$ incoming state. In that case, we can send $p_z \to -p_z$ in $p_2$ and $p_H$ in our momenta sets to probe this contribution. In figures 4.12, 4.13, 4.14 and 4.15, we show $gg \to Hggg$ with the Higgs being more forward, central but more forward than the extra emission, central but more backward than the extra emission and more backward respectively. All plots show agreement in the large $\Delta$ region, as expected, and track the LO result fairly well over the entire range. For small values of $\Delta$, the deviation of our result from the full LO result is quite pronounced. This is typical of the $gg$ incoming state for all HEJ matrix elements [6] since there are contributions from diagrams that are strongly suppressed in the high $\Delta$ limit (and therefore dropped in the derivation of our amplitudes) but are quite important at low $\Delta$. Confident in our forms for the matrix elements, we can move on to implementing them within the HEJ program and investigating their behaviour over the whole range of integrated phase space.

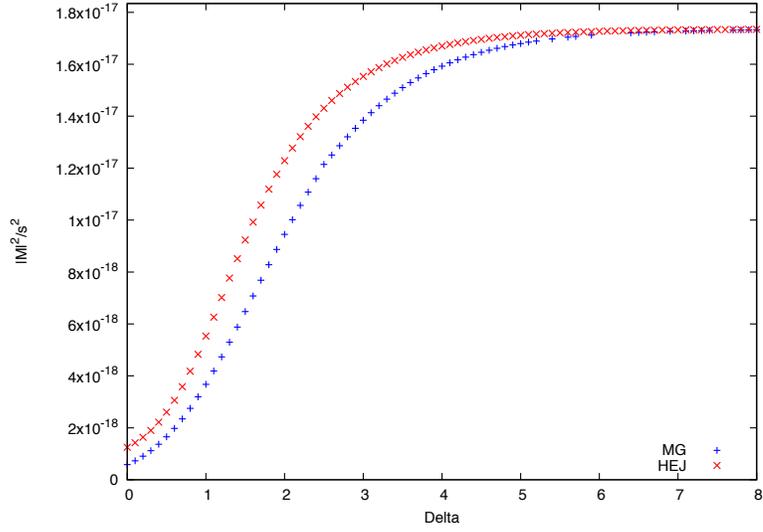

**Figure 4.9**  *An attempt to plot the $gu \to Hggu$ amplitude for the HEJ result (red) and the full LO result (blue) with the full finite top mass effects included. The blue line is missing some points in the high $\Delta$ region where the calculation evaluated to 'NaN'. The distribution is cut off at $\Delta = 8$ since the LO result always gives 'NaN' over this point. Despite this instability, it is still clear that the two results are in agreement in the High Energy Limit.*

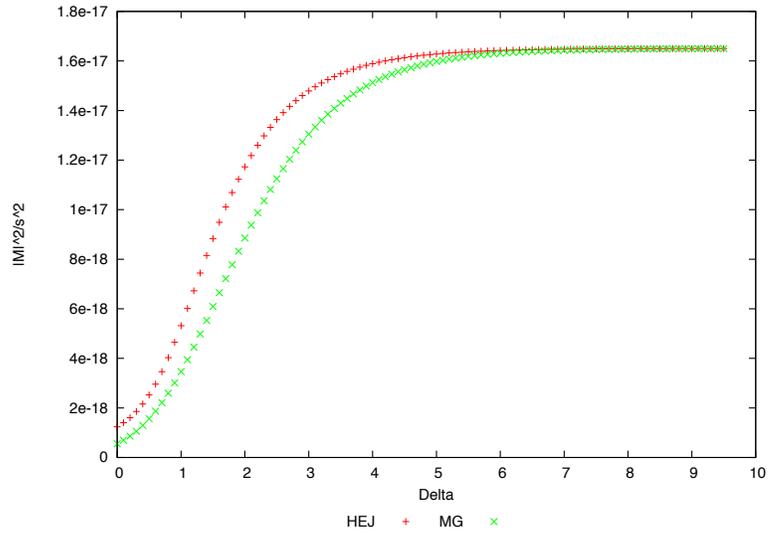

**Figure 4.10** *Comparison between the HEJ effective vertex (red) and the full LO result (green) of the $gu \to Hggu$ amplitude for a forward $H$ with an infinite top mass.*

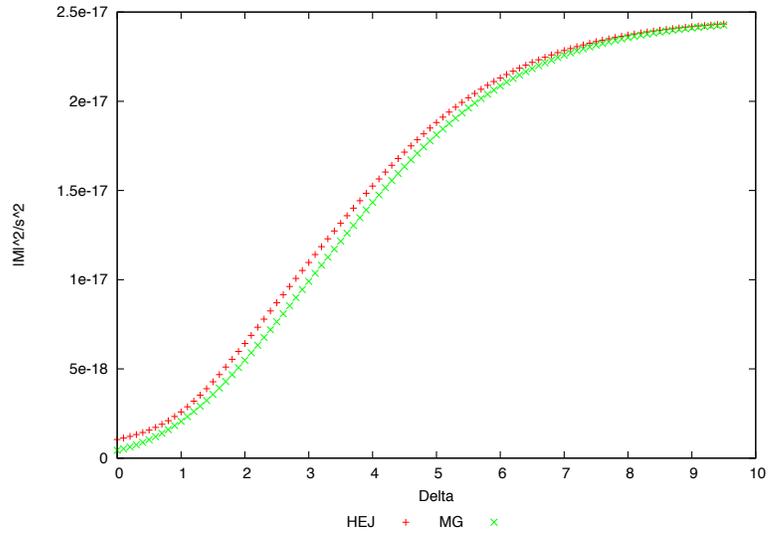

**Figure 4.11** *Comparison between the HEJ effective vertex (red) and the full LO result (green) of the $gu \to gHgu$ amplitude for a central $H$ with an infinite top mass.*

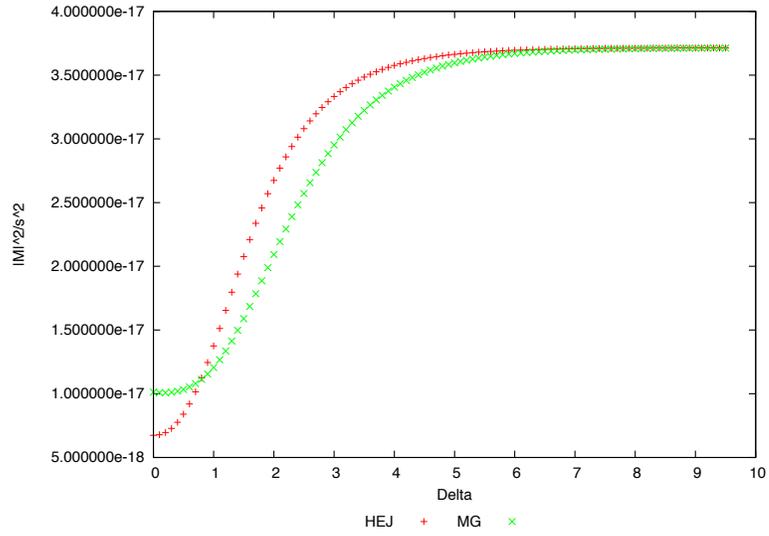

**Figure 4.12** *Comparison between the HEJ effective vertex (red) and the full LO result (green) of the $gg \to Hggg$ amplitude for a forward $H$ with an infinite top mass.*

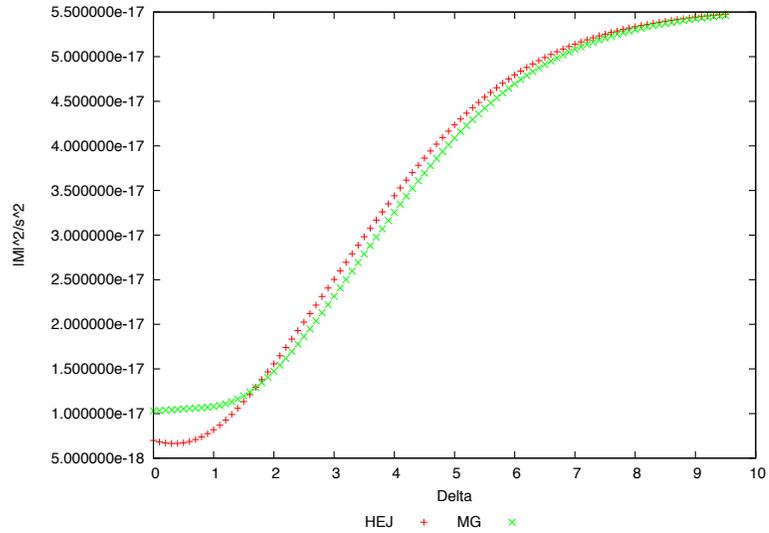

**Figure 4.13** *Comparison between the HEJ effective vertex (red) and the full LO result (green) of the $gg \to gHgg$ amplitude for a central $H$ next to the extremal forward parton with an infinite top mass.*

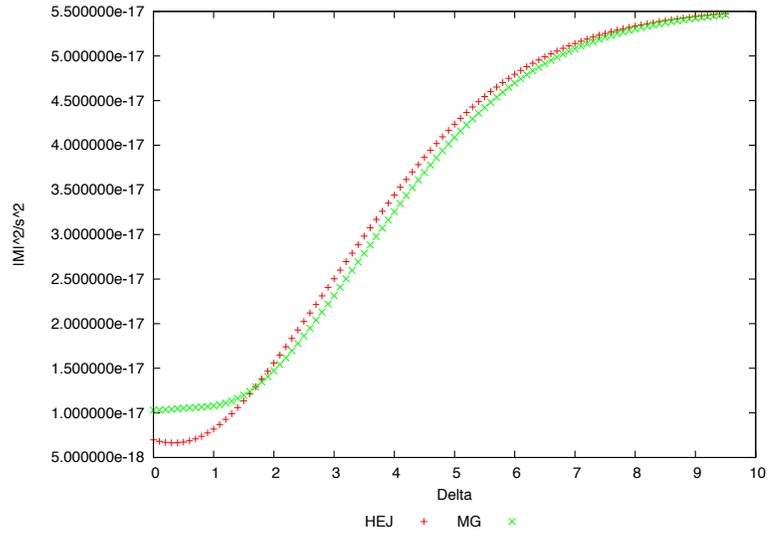

**Figure 4.14**  *Comparison between the HEJ effective vertex (red) and the full LO result (green) of the $gg \to ggHg$ amplitude for a central $H$ next to the extremal backward parton with an infinite top mass.*

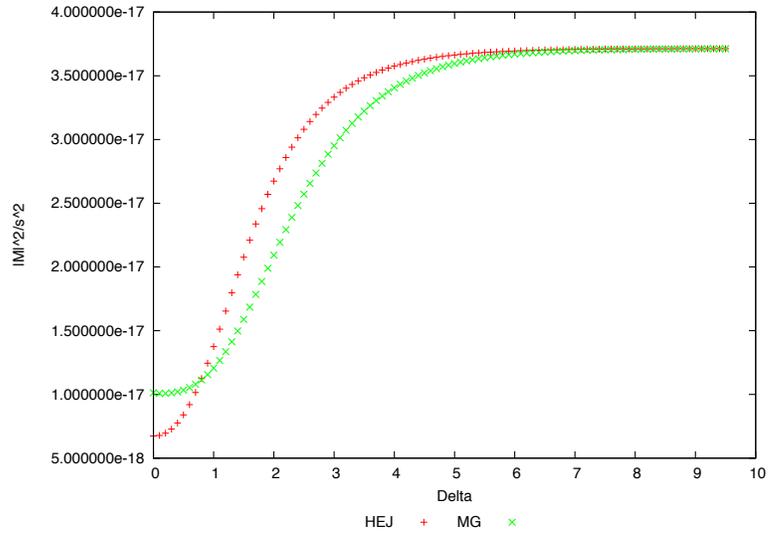

**Figure 4.15**  *Comparison between the HEJ effective vertex (red) and the full LO result (green) of the $gg \to gggH$ amplitude for a backward $H$ with an infinite top mass.*

## 4.3 Final Extensions and the All-Order Amplitude

We are almost ready to present a form for the all-order amplitude. Before we do, we first must revisit one point we made earlier: that the analytical form of the matrix element written here assumed that the gluon was the forward-moving particle. In any real analysis, we will have the situation where the gluon is travelling either backwards or forwards and the mathematical difference is the multiplication of a phase factor in some of the amplitude's terms. There is clearly a symmetry between the two cases and so any integration over phase space should yield the result that the contribution from this process with a forward-moving gluon should equal the contribution from the same process with a backward-moving gluon. We will show this to be the case in the next section.

With that taken into account, we are ready to discuss the all-order expression. As with all other HEJ amplitudes, the generalisation to all orders is simple. In the previous section, we derived an expression for the $gq \to Hgq$ amplitude in a $t$-channel factorised form and then showed how extra gluon emissions are taken into account. The all-order resummation is once more performed by making the following replacement for all $t$-channel gluon propagators:

$$\frac{1}{\hat{t}_i} \to \frac{1}{\hat{t}_i} \exp\left[\omega_0(q_{i\perp})(y_{i+1} - y_i)\right], \tag{4.57}$$

where $\omega(q_i)$ is defined in equation 2.51.

This gives the form of the all-order amplitude to be (where we emphasise that the rapidity ordering can be either backwards-to-forwards or forwards-to-backwards with the correct phase multiplication of terms in the base amplitude)

$$\begin{aligned}|M_{gf_2 \to Hg\ldots f_2}^{HEJ,m_t}|^2 =\ & \frac{1}{4(N_C^2 - 1)} |M_{gq \to Hgq}^{HE,m_t}|^2 \cdot \frac{C_{f_2}}{C_F} \\ & \cdot \prod_{i=1}^{n-2} \frac{-g^2 C_A V(q_i, q_{i+1}) \cdot V(q_i, q_{i+1})}{\hat{t}_i \hat{t}_{i+1}} \\ & \cdot \prod_{j=1}^{n-1} \exp\left[\omega_0(q_{j\perp})(y_{j+1} - y_j)\right], \end{aligned} \tag{4.58}$$

where $C_{f_2} = \tilde{C}_A$ if $f_2$ is a gluon or $C_F$ if it is a quark and the notation '...' signifies the emission of $n - 2$ gluons with $n$ being the total number of colour-charged particles in the final state. We define $q_1 = p_a - p_1 - p_H$ and $q_i = q_{i-1} - p_i$ after that and remind ourselves that there is already a division by $\hat{t}_1 \hat{t}_{n-1}$ defined within the squared base amplitude, so that the powers and numbers of $t$-channel propagators in this equation is correct.

In the next section, we describe the considerations needed when incorporating this amplitude in the HEJ program.

## 4.4 Computational Aspects

The addition of this new amplitude into the HEJ program is a much simpler task than it was for the NLL processes. Since we already have the amplitude in terms of the impact factors for the infinite top mass cases, the implementation is a case of adding the option to run with this finite top mass element instead. However, in order to calculate the scalar integrals in this amplitude, a HEJ interface to LoopTools is required. This is achieved in the program by setting a special instruction in the makefile; this way, if a user is not interested in using HEJ to generate for these types of events, they do not need LoopTools on their system. On the other hand, with the setting of a few library paths and setting the value for the special flag, LoopTools is quickly and simply added to the program such that running these new amplitudes can be performed 'out of the box'.

Another consideration is how to implement the matching for the finite top mass case. Given the length of time it would take to evaluate the matrix element with the full top mass included (especially in the three jet case), it would not be feasible for realistic analyses. Instead, since the infinite top mass elements are much quicker to evaluate, the matching is implemented by applying the infinite top mass limit to the HEJ amplitude and dividing that by the full LO result in the effective theory. This yields an acceptable compromise as the correction of the Born approximation should be independent of the $m_t$ effects, but it would clearly be ideal to match to the full result if the amount of time needed to do so was significantly reduced. This latter point is one motivation behind the development of the 'inverse HEJ' technique. Since there are many resummation processes that can map to one jet process, the idea is to instead generate the jet level matrix element and work backwards from that to generate many resummation points to evaluate: hence, 'inverse'. This will drastically reduce the amount of calls made to the full finite top mass LO matrix element and thus allow us to include this matching.

## 4.5 Results

Our first integrated distributions will be for the $gu \to gu + H$ process, where we restrict ourselves to the case where the gluon is the forward-moving particle and the $u$ quark is the backward-moving particle. The Higgs can then be more forward than the gluon ($Hgu$), more backward than the quark ($guH$) or in between the two particles in rapidity ($gHu$). For the forward case, we will use the matrix element derived in section 4.2 to describe the process. For the central case, we will use the reweighted $qQ$ amplitude from section 4.1. Although our new matrix element will get this case correct, the reason we do not use it here is in anticipation of our resummation technique. Because we resum $t$-channel gluons, it is important to have a system where we can always unambiguously define what those gluons will be (or, in other words, we must have a consistent 'resummation region', which for us will be the rapidity space between the gluon and quark). With our new amplitude, we can only resum the gluon connecting the effective vertex to the quark line, but for the reweighted $qQ$ case, we can resum the gluon from the top quark line to the triangle loop in the middle of the diagram and then from the triangle loop down to the bottom quark line. Choosing which matrix element we use based on the rapidity of the Higgs makes it clear which resummation we will be doing. Finally, since we have not discussed the case where the Higgs is behind the outgoing quark, we will set this process to zero for the time being (though in the full HEJ program, this is of course properly incorporated with the appropriate impact factor).

An interesting plot to look at is the Higgs $p_T$ spectrum. The infinite top mass limit should not do well at high values of Higgs $p_T$, as discussed in [28], because it breaks the idea of the top quark mass being the largest relevant scale. This distribution is shown in figure 4.16. We see the expected behaviour: at large Higgs $p_T$, the results obtained with the effective theory and the theory with finite quark mass effects are very different. The difference when also including a bottom quark is minimal, leading to slight difference in the low $p_T$ bins which is not large enough to show up on the logarithmic plot.

Another interesting distribution to look at is the rapidity difference between the gluon and the quark. One could expect that the presence of a large rapidity gap (and so a large dijet invariant mass) might break the infinite top mass limit, but [28] showed this not to be the case. The distribution shown in figure 4.17 also shows little difference and hence agrees with their conclusion. The difference

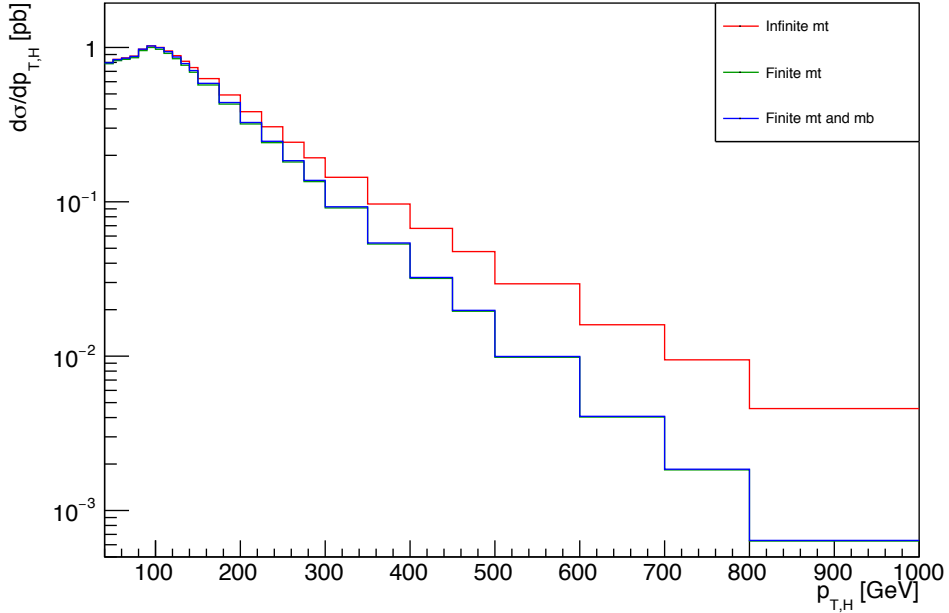

**Figure 4.16**  *Comparison of infinite top mass, finite top mass and finite top plus finite bottom mass cross-sections in $gu \to guH$, binned in Higgs $p_T$*

when including a bottom quark is clearer in this distribution, where we see a slight (but definite) increase of the cross-section in the low $\Delta y$ bins. For this analysis, the integrated cross-section is larger by about 2% because of this inclusion.

We should also remember that, given kinematical constraints on the energy of the collider, the high $\Delta y$ region must come with relatively low transverse scales, so it further lends support to the idea that the transverse scales are the defining ones in terms of how accurate the effective theory is. We can do the same analysis with a $gg$ incoming state, which in the high energy regime is just the $qg$ amplitude reweighted by a colour factor. In this case, the Higgs can be either forward of a forward-moving gluon, backward of a backward-moving one or in-between. We can describe all of these configurations: the first and second with our new amplitude and the latter with a reweighted $qQ$ amplitude.

As mentioned in section 4.3, we must have that the cross-section contribution from the forward-moving gluon is equal to the contribution from the backward-moving gluon. We show that this is the case in figure 4.18, where the green and blue lines are so close together as to be essentially indistinguishable. The results

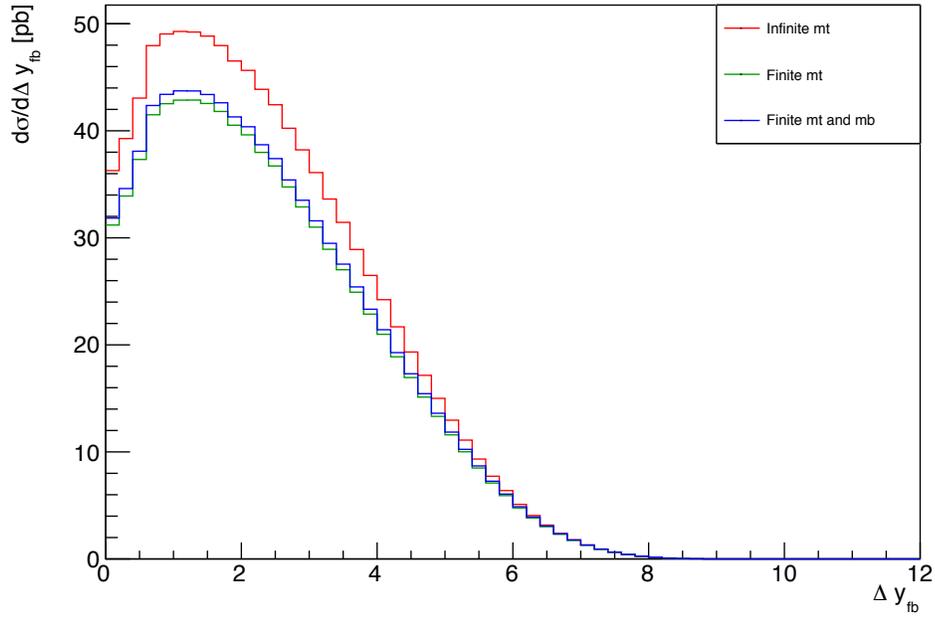

**Figure 4.17**  *Comparison of infinite top mass, finite top mass and finite top plus finite bottom mass cross-sections in $gu \to guH$, binned in rapidity difference between the gluon and up quark*

for the Higgs $p_T$ and $\Delta y_{12}$ distributions are extremely similar to the $gq$ case and we present them here for completeness in figures 4.19 and 4.20. The most obvious difference is that the $\Delta y$ distribution peaks more to the left of the plot – an effect of the gluon parton distribution function.

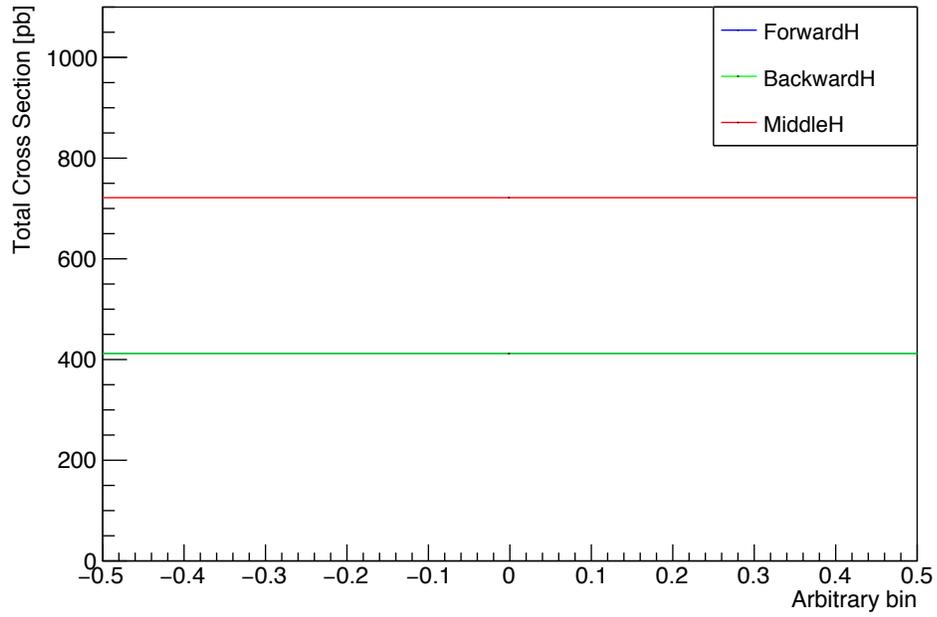

**Figure 4.18** *Cross-section breakdown in $gg \to ggH$ into forward Higgs production, backward Higgs production and central Higgs with infinite top mass.*

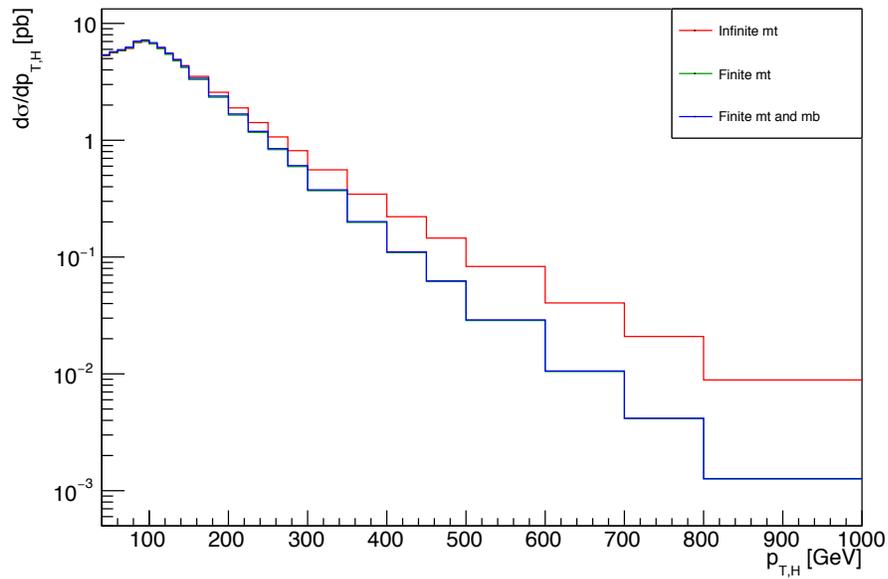

**Figure 4.19** *Comparison of infinite top mass, finite top mass and finite top plus finite bottom mass cross-sections in $gg \to ggH$, binned in Higgs $p_T$.*

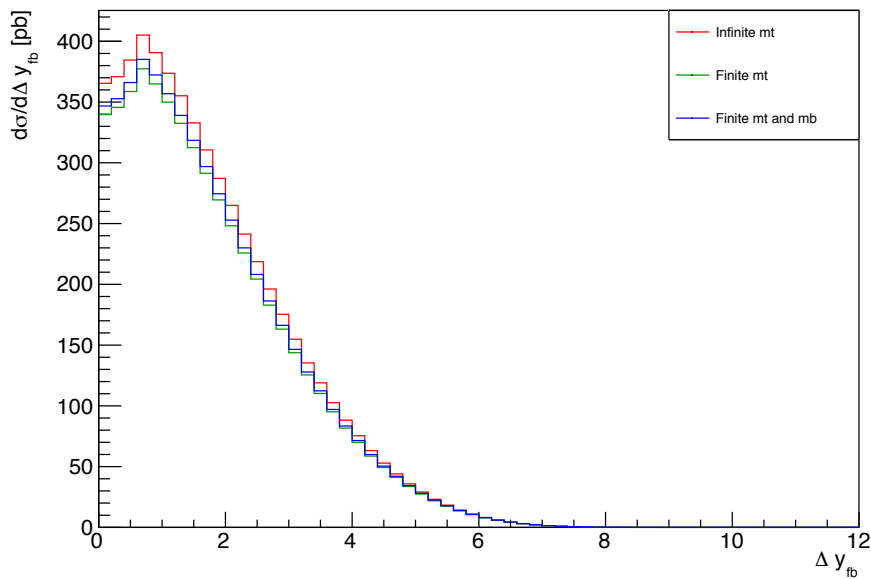

**Figure 4.20** *Comparison of infinite top mass, finite top mass and finite top plus finite bottom mass cross-sections in $gg \to ggH$, binned in rapidity difference between the gluon and up quark.*

We also investigate the effect of adding a third jet so as to perform a $gg \to gggH$ analysis. An interesting plot to show is, again, the Higgs $p_T$ as shown in figure 4.21. We see a significant difference between the infinite top mass results and the finite top mass results in all bins – strikingly, at low Higgs $p_T$. This would seem to contradict our prediction that low transverse scales imply that the effective theory is valid. The problem is that, in a three jet event, you can manufacture a situation whereby there is a large hierarchy between the transverse scales that enter the Higgs vertex. If instead we plot the cross-section as a function of the largest transverse scale that enters into this vertex, we should then once more see the agreement in the low $p_T$ end. Figure 4.22 shows this clearly.

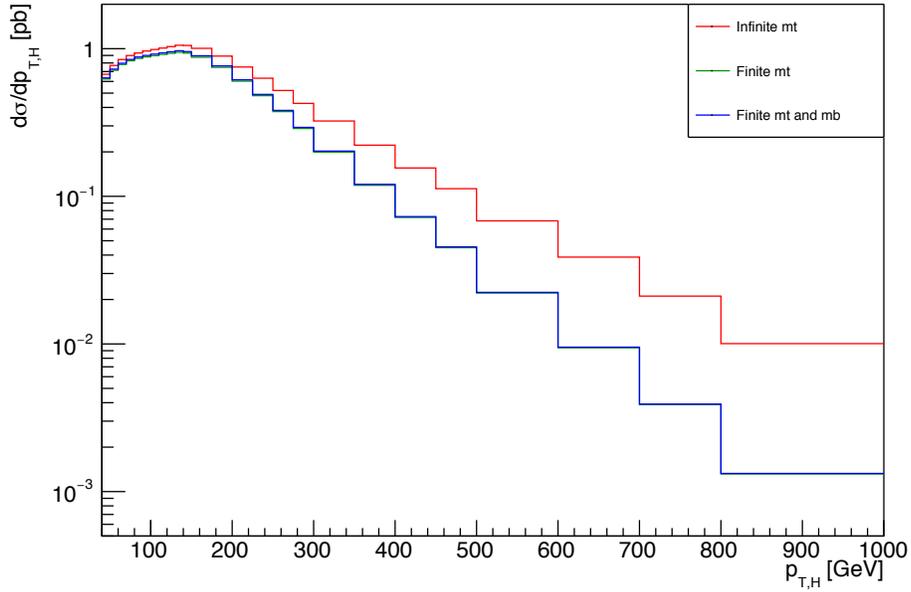

**Figure 4.21** *Comparison of infinite top mass, finite top mass and finite top plus finite bottom mass cross-sections in $gg \to gggH$, binned in Higgs $p_T$.*

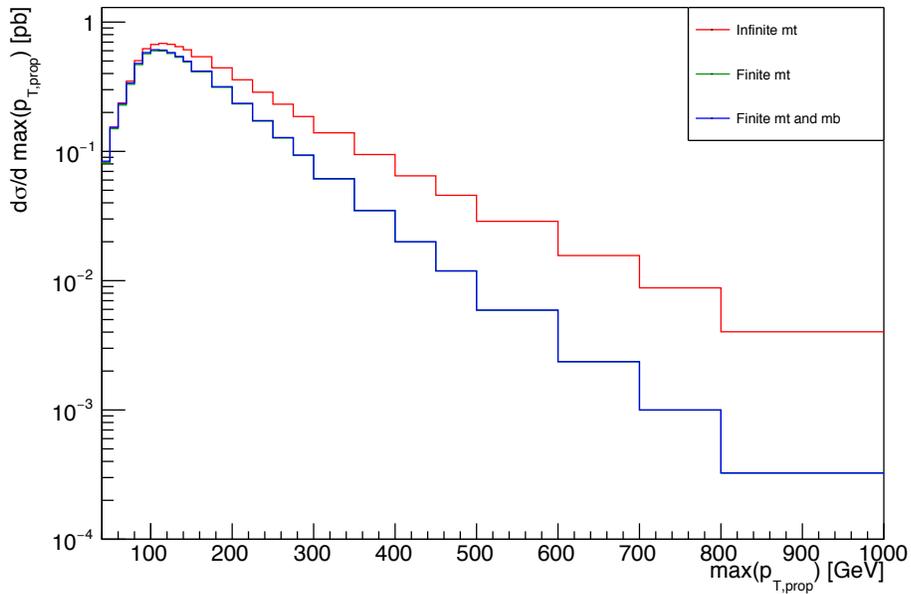

**Figure 4.22** *Comparison of infinite top mass, finite top mass and finite top plus finite bottom mass cross-sections in $gg \to gggH$, binned in the maximum $p_T$ of a gluon entering into the Higgs vertex.*

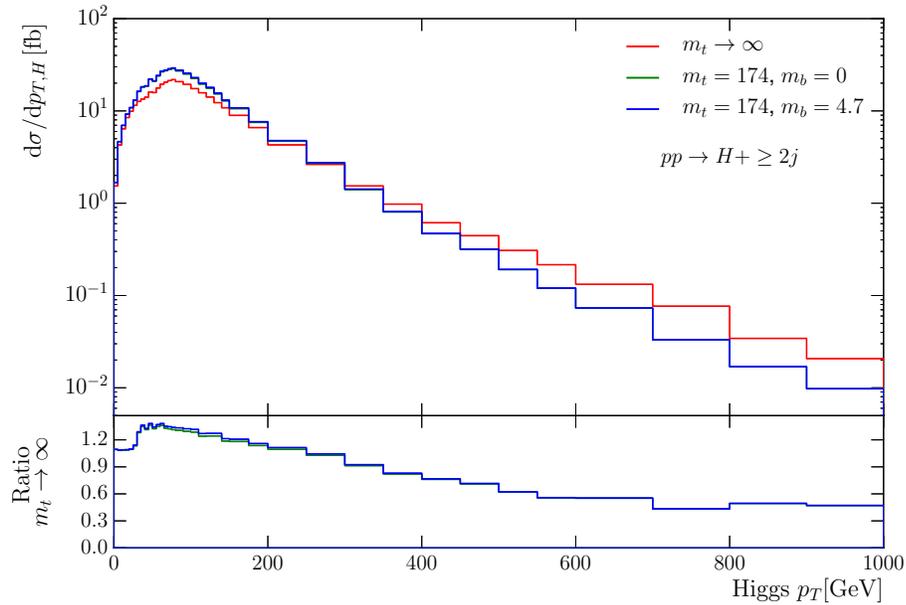

**Figure 4.23** *Comparison of infinite top mass, finite top mass and finite top plus finite bottom mass cross-sections in $pp \to H+ \geq 2j$, binned in the $p_T$ of the Higgs.*

Unfortunately, there have so far not been many analyses of Higgs plus jets physics from the LHC and the ones that do exist focus on weak boson fusion rather than gluon fusion [15]. For this reason, we are not yet able to compare these predictions to real data. We are, however, now in a prime position to provide predictions for any such data when it arrives. We conclude this section instead with an inclusive Higgs plus dijets prediction and again point out the difference between the full $m_t$ and infinite $m_t$ approaches. The $p_T$ spectra of the Higgs for this case is plotted in figure 4.23, along with the ratio of the finite $m_t$ and finite $m_t + m_b$ lines to the infinite $m_t$ line. We see clearly that the distribution obtained with the effective theory is significantly different. The same is true if we look at the $\Delta y$ distribution, shown in figure 4.24.

In all of these distributions, we see that the addition of a bottom loop has a small overall effect. This would suggest that the Higgs + 1 jet studies of [46] and [39], where a sizeable difference was seen, is an effect of having only one jet accompanying the Higgs.

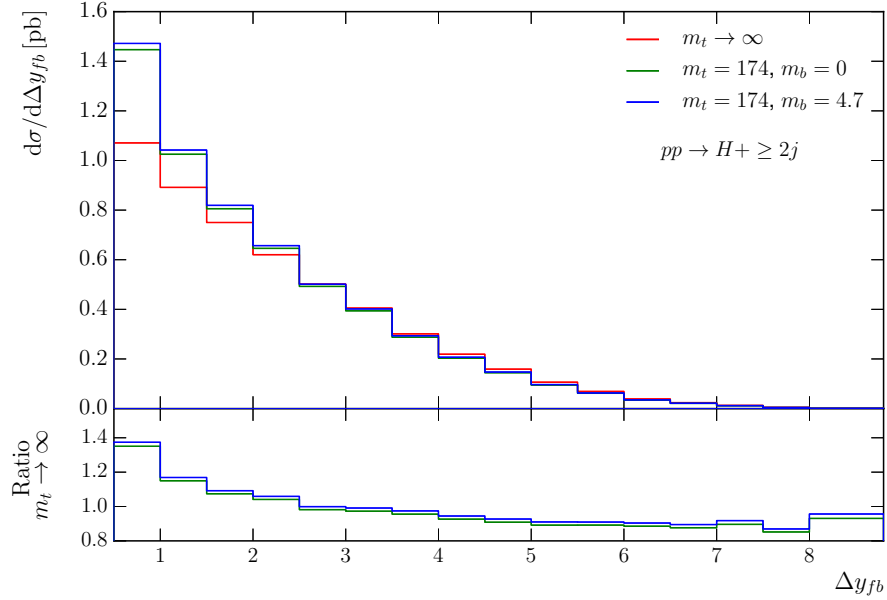

**Figure 4.24** *Comparison of infinite top mass, finite top mass and finite top plus finite bottom mass cross-sections in $pp \to H+ \geq 2j$, binned in the rapidity difference between the most forward/backward jets.*

## 4.6 Summary

In this chapter, we have discussed how HEJ can be extended to describe processes involving the production of a Higgs boson along with two or more jets. We discussed how, due to the added complexity of calculating loop diagrams, many fixed order approaches to the description of such amplitudes employ the infinite top mass limit. Using this as a first approximation, we showed how the HEJ amplitude looks in this limit by a simple redefinition of the current object.

From here, we argued that the employment of the infinite top mass limit is not required in HEJ: the High Energy and infinite top mass limits commute. Taking the simple example of $qQ \to qHQ$, we showed how simple it is for HEJ to keep the full finite quark mass effects in its amplitudes. By considering the limit where a Higgs boson is produced outside of a gluon, the bulk of the chapter describes the more involved calculation of the HEJ amplitude for this case, again keeping full finite quark mass effects.

Analysis of this amplitude over the LHC phase space confirmed previously known results: namely, that the presence of a large dijet invariant mass does little to invalidate the infinite top mass limit. Instead, the relevant variable is the transverse scales that feature in the quark loop, where we saw the striking result that at high values of the Higgs $p_T$, the difference between the effective and full theories is a factor of two.

The main result of this work is that HEJ is now able to provide predictions for Higgs plus $\geq$ 2 jet processes with full top and bottom mass effects included. Although leading order results are known for the full theory with small multiplicities of jets [29, 47], HEJ's ability to resum the amplitude to all orders in $\alpha_s$ whilst keeping the full dependence on the quark masses is unique.

# Chapter 5

# Conclusions

In this thesis we have studied the perturbative approach to the solution of QCD scattering in the specific context of the Large Hadron Collider and discussed the limitations thereof. In chapter 1, we introduced the base theoretical knowledge needed to understand the technique, which included the discussion of gauge invariance and a general overview of the path integral formulation of Quantum Field Theory. This was used to summarise that calculations could be done by simply following a set of Feynman Rules and we explicitly calculated the $qQ \to qQ$ amplitude at leading order in $\alpha_s$ using these rules. We saw how this amplitude is converted into the physical cross-section which can be measured at the collider and then re-did the calculation with the spinor helicity formalism to show how it makes the amplitude calculations neater and quicker.

In chapter 2, we explicitly showed how a truncation in the perturbative series is problematic when high energy particles are involved in the scattering. In particular, it was proved that terms that go like $\alpha_s^n \log^{n-1}\left(\frac{s}{-t}\right)$ appear at higher orders and, given their size, should not be neglected. This naturally led on to the discussion of Regge Theory and the formalism of High Energy Jets in general. The formalism exploits elements of the High Energy Limit in order to write amplitudes as simple current contractions over $t$-channel poles. By considering real and virtual corrections to processes in that limit, we ended up with a matrix element that can resum the problematic High Energy Logarithms to all orders in perturbation theory. In order to be of physical relevance, this matrix element is integrated over the relevant phase space via a Monte Carlo technique and the computational considerations of doing this are discussed. The chapter ends with a selection of distributions from real LHC analyses, where we show the capture of the High Energy Logarithms is already phenomenologically important and will only increase in importance as the centre-of-mass energy increases.

In chapter 3, we discussed how the formalism can be extended to capture some of the Next-to-Leading Logarithmic contributions to jet processes, which will behave like $\alpha_s^n \log^{n-2}\left(\frac{s}{-t}\right)$. The addition of 'unordered' gluon emissions allowed us to capture sub-leading contributions from already included partonic channels. The author's own work focused on the inclusion of matrix elements for entirely new partonic configurations which, although formally sub-leading in the jet process, are Leading Logarithmic in the particular subprocess. As such, the applicability of HEJ has increased and many checks of these new elements are presented along with a discussion of the computational challenges faced in including them within the Monte Carlo program. We concluded the chapter by investigating the effect of the NLL contributions on real data and showed that they provided a significant improvement to our predictions.

In chapter 4, we introduced how HEJ is also able to describe jet events accompanied by the emission of a Higgs boson. The effective theory where the mass of the top quark is taken to be infinite is discussed and its limitations laid out. The factorisation property that gives HEJ its resummation power does not rely on taking the infinite top mass limit and so amplitudes with full quark mass dependence were derived within the formalism. As a result, HEJ is unique in its ability to provide a prediction for such processes with both High Energy resummation and finite quark mass effects taken into account. By the presentation of a set of distributions, we saw how the effect of finite quark mass loops can lead to significantly different results – most drastically, in the behaviour of the tails of the Higgs transverse momentum spectrum.

In conclusion, the effect of large logarithmic contributions on QCD amplitudes is seen to be large and we must take them seriously in order to provide accurate Standard Model predictions. As the energy of colliders increase, these effects will only become more significant. It is therefore in the interests of both the HEJ collaboration and the phenomonology community in general to ensure an accurate inclusion of them. For HEJ, one aspect of this is to include some sub-leading partonic configurations in the resummation, since this reduces our dependence on leading order matching techniques. It remains a long term goal of the collaboration to include next-to-leading order matrix elements in the matching routine but the process is complicated by ensuring that no 'double counting' occurs. Furthermore, the contributions of collinear terms are necessarily ignored by the approximations that underlie HEJ but must be considered for a complete description. This is currently done via an interface to the parton shower ARIADNE and there are plans to extend this to use the state-of-the-art parton shower of Pythia [52]. As well as the discussion of pure jet final states, HEJ also provides predictions for final states involving an electroweak boson. We now have the capability to move away from the currently implemented infinite top mass limit when discussing the emission of a Higgs boson along with jets. In order to detect any deviations in the discovered Higgs boson from the Standard Model predictions, it is vitally important that all known effects are correctly accounted for and this new capability of HEJ will help to a great degree towards that goal.

# Appendix A

# Relating the Unordered Vertex to the Extremal $Q\bar{Q}$ Vertex

A fundamental symmetry of QFTs is crossing symmetry which relates amplitudes involving outgoing particles with amplitudes with incoming anti-particles and vice versa. For example, given a theory that involves a scalar particle $\phi$ and its anti-particle $\bar{\phi}$, we have that

$$M(\phi(p) + \ldots \to \ldots) = M(\ldots \to \ldots + \bar{\phi}(-p)). \tag{A.1}$$

With this argument, we might expect that the unordered amplitude $qQ \to qQg$ is related to our result for $qg \to qQ\bar{Q}$, since we get there by moving the gluon from the final to the initial state and the quark from the initial to the final (thereby also converting it to an anti-quark). To put it another way

$$M(q(p_a)+Q(p_b) \to q(p_1)+Q(p_2)+g(p_g)) = M(q(p_a)+g(-p_g) \to q(p_1)+Q(p_2)+\bar{Q}(-p_b)). \tag{A.2}$$

This equation is certainly true for the full amplitude. However, it is not so clear that this will hold for our equations involving impact factors, since they explicitly considered limiting arguments on the full amplitudes which may not carry over when this crossing symmetry is applied. Here we explicitly check to see whether the symmetry holds for our amplitudes, which is equivalent to showing that the unordered vertex transforms into the extremal $q\bar{q}$ vertex under this symmetry. Although it can be done either way around, we decide to start from the unordered vertex and aim for the $q\bar{q}$ vertex.

We remind ourselves from section 3.3.1 that the final form of the extremal $q\bar{q}$ vertex was

$$Q^{\mu\nu} = -\frac{C_1}{t_{3b}}\left(\bar{u}_2\gamma^\mu(\slashed{p}_3 - \slashed{p}_b)\gamma^\nu v_3\right) + \frac{C_2}{t_{2b}}\left(\bar{u}_2\gamma^\nu(\slashed{p}_2 - \slashed{p}_b)\gamma^\mu v_3\right)$$
$$+ i\frac{C_t}{s_{23}}\left[((2p_2 + 2p_3)^\nu\eta^{\mu\rho} - 2p_b^\mu\eta^{\nu\rho} + 2p_b^\rho\eta^{\nu\mu})\langle 2|\rho|3\rangle + \frac{2p_a^\nu q_1^2}{s_{ab}}\langle 2|\mu|3\rangle\right],$$
(A.3)

with $q_1 = p_a - p_1 = p_3 - p_b + p_2$ and where we have reinstated the term we removed when we picked a convenient gauge for generality. The colour factors here are

$$C_1 = T_{1a}^g T_{q3}^b T_{2q}^g,$$
$$C_2 = T_{1a}^g T_{2q}^b T_{q3}^g, \quad (A.4)$$
$$C_t = f^{gbc} T_{1a}^c T_{23}^g.$$

The unordered effective vertex from section 3.2 has the form

$$j_{uno}^{\mu\nu} = -i\frac{\tilde{C}_1}{s_{2g}}\left(\langle 2|\nu|g\rangle\langle g|\mu|b\rangle + 2p_2^\nu\langle 2|\mu|b\rangle\right) + i\frac{\tilde{C}_2}{s_{bg}}\left(2p_b^\nu\langle 2|\mu|b\rangle - \langle 2|\mu|g\rangle\langle g|\nu|b\rangle\right)$$
$$+ \frac{\tilde{C}_t}{s_{b2}}\left[\langle 2|\rho|b\rangle\left(2p_g^\rho\eta^{\mu\nu} - 2p_g^\mu\eta^{\nu\rho} - (q_1 + q_2)^\nu\eta^{\mu\rho}\right) + \frac{q_1^2}{2}\langle 2|\mu|b\rangle\left(\frac{p_1^\nu}{p_g\cdot p_1} + \frac{p_a^\nu}{p_g\cdot p_a}\right)\right],$$
(A.5)

where we have $q_1 = p_a - p_1 = p_2 - p_b + p_g$, $q_2 = p_2 - p_b = p_a - p_1 - p_g$ and the colour factors are

$$\tilde{C}_1 = T_{2i}^g T_{ib}^d d_{1a}^d$$
$$\tilde{C}_2 = T_{2i}^d T_{ib}^g t_{1a}^d \quad (A.6)$$
$$\tilde{C}_t = f^{gde} T_{2b}^e T_{1a}^d.$$

Before applying crossing symmetry, let us rewrite the unordered vertex in a different manner. Firstly, we notice that the last term has been created by symmetrising a High Energy approximation. In the extremal $Q\bar{Q}$ vertex, we did not make this symmetrisation and so we should undo this here to find equality. Secondly, we can write the spinor chains in the first two terms as terms involving $\slashed{p}$ which will both shorten the expression and provide a more direct similarity to the extremal $Q\bar{Q}$ amplitude. Thirdly, we will write out in full the $q_1$ and $q_2$ terms, subject to removing terms that will contract to give zero. Doing these steps, we

arrive at

$$j^{\mu\nu}_{uno} = -i\frac{\tilde{C}_1}{s_{2g}}\left(\bar{u}_2\gamma^\nu(\slashed{p}_g+\slashed{p}_2)\gamma^\mu u_b\right) + i\frac{\tilde{C}_2}{s_{bg}}\left(\bar{u}_2\gamma^\mu(\slashed{p}_g-\slashed{p}_b)\gamma^\nu u_b\right)$$
$$+ \frac{\tilde{C}_t}{s_{b2}}\left[\langle 2|\rho|b\rangle\left(2p_g^\rho\eta^{\mu\nu} - 2p_g^\mu\eta^{\nu\rho} - (2p_2-2p_b)^\nu\eta^{\mu\rho}\right) + q_1^2\langle 2|\mu|b\rangle\left(\frac{2p_a^\nu}{s_{ga}}\right)\right].$$
(A.7)

The transformation we now apply is $p_g \to -p_b$ and $p_b \to -p_3$. The first transformation relates to taking the gluon from the outgoing state into the incoming state and the second relates to taking the quark from the incoming state to the outgoing state as an anti-quark. The relabelling of the momenta in this fashion is chosen such that the conventions for incoming and outgoing particles are preserved, which will make the relation to the extremal $Q\bar{Q}$ amplitude much clearer. Applying these rules leads us to

$$j^{\mu\nu}_{uno,crossed} = i\frac{C_2}{s_{2b}}\left(\bar{u}_2\gamma^\nu(-\slashed{p}_b+\slashed{p}_2)\gamma^\mu v_{-3}\right) - i\frac{C_1}{s_{3b}}\left(\bar{u}_2\gamma^\mu(-\slashed{p}_b+\slashed{p}_3)\gamma^\nu v_{-3}\right)$$
$$- \frac{C_t}{s_{23}}\left[\langle 2|\rho|-3\rangle\left(-2p_b^\rho\eta^{\mu\nu} + 2p_b^\mu\eta^{\nu\rho} - (2p_2+2p_3)^\nu\eta^{\mu\rho}\right) - q_1^2\langle 2|\mu|-3\rangle\left(\frac{2p_a^\nu}{s_{ab}}\right)\right],$$
(A.8)

where we have identified that the colour factors after this transformation exactly match the colour factors of the $Q\bar{Q}$ vertex in the specified way. The final step is to notice that we can transform the spinors depending on $-p_3$ to spinors depending on $+p_3$ by pulling out a phase factor. We will define the product of $-i$ with this phase factor as $e^{i\phi}$, such that

$$j^{\mu\nu}_{uno,crossed} = e^{i\phi}\left\{i\frac{C_1}{t_{3b}}\left(\bar{u}_2\gamma^\mu(\slashed{p}_3-\slashed{p}_b)\gamma^\nu v_3\right) - i\frac{C_2}{t_{2b}}\left(\bar{u}_2\gamma^\nu(\slashed{p}_2-\slashed{p}_b)\gamma^\mu v_3\right) -\right.$$
$$\left.+ i\frac{C_t}{s_{23}}\left[\langle 2|\rho|3\rangle\left(2p_b^\rho\eta^{\mu\nu} - 2p_b^\mu\eta^{\nu\rho} + (2p_2+2p_3)^\nu\eta^{\mu\rho}\right) + q_1^2\langle 2|\mu|3\rangle\left(\frac{2p_a^\nu}{s_{ab}}\right)\right]\right\},$$
(A.9)

which is precisely the extremal $q\bar{q}$ vertex up to an overall, irrelevant phase factor.

# Appendix B

# Form of the $H$ Functions in the $gq \to Hgq$ Box Integral

In section 4.2.2, we discussed the $gq \to Hgq$ amplitude with full finite quark masses included. Such an amplitude contained so-called box diagrams, which we parameterised in the following way [28]:

$$M^{\mu\nu\rho} = \frac{g_s^3 m_t^2}{v} 2i f^{bac} J^{\mu\nu\rho}, \tag{B.1}$$

where

$$J^{\mu\nu\rho} = \eta^{\mu\nu}(H_1 k_1^\rho + H_2 k_2^\rho) + \eta^{\mu\rho} H_4 k^\nu + \eta^{\nu\rho} H_5 k^\mu + H_{10} k_2^\rho k^\mu k^\nu + H_{12} k_1^\rho k^\mu k^\nu, \tag{B.2}$$

with the convention that all the $k$s are taken as outgoing. We have that $k_1^\mu$ and $k_2^\mu$ are the momenta corresponding to the on-shell gluons with colour indices $a, b$ and $k = -k_1 - k_2 - k_H$ with colour index $c$. Not all of the $H$ functions above are independent:

$$\begin{aligned} H_2 &= -H_1(k_1 \leftrightarrow k_2) \\ H_5 &= -H_4(k_1 \leftrightarrow k_2) \\ H_{12} &= -H_{10}(k_1 \leftrightarrow k_2). \end{aligned} \tag{B.3}$$

We will therefore only give the forms for $H_1, H_4$ and $H_{10}$. Since there are three distinct orderings for the box diagram, we present this in the form $H_i = E_i +$

$F_i + G_i$, where each term is a function of the scalar integrals

$$B_0(k) = \int \frac{d^4q}{(2\pi)^4} \frac{1}{(q^2 - m_t^2)((q+k)^2 - m_t^2)},$$
$$C_0(p, k) = \int \frac{d^4q}{(2\pi)^4} \frac{1}{(q^2 - m_t^2)((q+p)^2 - m_t^2)((q+p+k)^2 - m_t^2)}, \quad \text{(B.4)}$$
$$D_0(p, k, v) = \int \frac{d^4q}{(2\pi)^4} \left[ \frac{1}{(q^2 - m_t^2)((q+p)^2 - m_t^2)((q+p+k)^2 - m_t^2)} \right.$$
$$\left. \times \frac{1}{((q+p+k+v)^2 - m_t^2)} \right].$$

The $B_0$ integral is divergent in four dimensions but will always appear in combinations such that the divergence is cancelled. We will also define

$$\begin{aligned}
s_{ij} &= (k_i + k_j)^2, \\
S_i &= 2k_i \cdot k, \\
\Delta &= s_{12}s_{34} - S_1 S_2, \\
\Sigma &= 4s_{12}s_{34} - (S_1 + S_2)^2.
\end{aligned} \quad \text{(B.5)}$$

The $E_i$, $F_i$ and $G_i$ are defined as [28]:

$$E_1 = -s_{12}D_0(2, 1, 34) \left[ 1 - \frac{8m_t^2}{s_{12}} + \frac{S_2}{2s_{12}} + \frac{S_2(s_{12} - 8m_t^2)(s_{34} + S_1)}{2s_{12}\Delta} \right.$$
$$\left. + \frac{2(s_{34} + S_1)^2}{\Delta} + \frac{S_2(s_{34} + S_1)^3}{\Delta^2} \right]$$
$$- [(s_{12} + S_2)C_0(2, 134) - s_{12} + (S_1 - S_2)C_0(12, 34) - S_1 C_0(1, 34)]$$
$$\times \left( \frac{S_2(s_{12} - 4m_t^2)}{2s_{12}\Delta} + \frac{2(s_{34} + S_1)}{\Delta} + \frac{S_2(s_{34} + S_1)^2}{\Delta^2} \right)$$
$$+ [C_0(1, 34) - C_0(12, 34)] \left( 1 - \frac{4m_t^2}{s_{12}} \right) - C_0(12, 34) \frac{2s_{34}}{S_1}$$
$$- [B_0(134) - B_0(1234)] \frac{2s_{34}(s_{34} + S_1)}{S_1 \Delta}$$
$$+ [B_0(34) - B_0(1234) + s_{12}C_0(12, 34)] \left( \frac{2s_{34}(s_{34} + S_1)(S_1 - S_2)}{\Delta \Sigma} + \frac{2s_{34}(s_{34} + S_1)}{S_1 \Delta} \right)$$
$$+ [B_0(12) - B_0(1234) - (s_{34} + S_1 + S_2)C_0(12, 34)] \frac{2(s_{34} + S_1)(2s_{12}s_{34} - S_2(S_1 + S_2))}{\Delta \Sigma},$$
$$\text{(B.6a)}$$

$$F_1 = -S_2 D_0(1,2,34) \left[ \frac{1}{2} - \frac{(s_{12} - 8m_t^2)(s_{34} + S_2)}{2\Delta} \right] - \frac{s_{12}(s_{34} + S_2)^3}{\Delta^2}$$

$$+ \left[ (s_{12} + S_1) C_0(1,234) - s_{12} C_0(1,2) - (S_1 - S_2) C_0(12,34) - S_2 C_0(2,34) \right]$$

$$\times \left( \frac{S_2(s_{12} - 4m_t^2)}{2 s_{12} \Delta} + \frac{S_2(s_{34} + S_2)^2}{\Delta^2} \right)$$

$$- [C_0(12,34) - C_0(1,234)] \left( 1 - \frac{4m_t^2}{s_{12}} \right) - C_0(1,234)$$

$$+ [B_0(234) - B_0(1234)] \frac{2(s_{34} + S_2)^2}{(s_{12} + S_1)\Delta}$$

$$- [B_0(34) - B_0(1234) + s_{12} C_0(12,34)] \frac{2 s_{34}(s_{34} + S_2)(S_2 - S_1)}{\Delta \Sigma}$$

$$[B_0(12) - B_0(1234) - (s_{34} + S_1 + S_2) C_0(12,34)] \frac{2(s_{34} + S_2)(2 s_{12} s_{34} - S_2(S_1 + S_2))}{\Delta \Sigma},$$

(B.6b)

$$G_1 = S_2 D_0(1,34,2) \left[ \frac{\Delta}{s_{12}^2} - \frac{4m_t^2}{s_{12}} \right]$$

$$- S_2 \left[ (s_{12} + S_1) C_0(1,234) - S_1 C_0(1,34) \right] \left( \frac{1}{s_{12}^2} - \frac{s_{12} - 4m_t^2}{2 s_{12} \Delta} \right)$$

$$- S_2 \left[ (s_{12} + S_2) C_0(13,2) - S_2 C_0(2,34) \right] \left( \frac{1}{s_{12}^2} + \frac{s_{12} - 4m_t^2}{2 s_{12} \Delta} \right)$$

$$- C_0(1,34) - [C_0(1,234) - C_0(1,34)] \frac{4m_t^2}{s_{12}} + [B_0(134) - B_0(1234)] \frac{2}{s_{12}}$$

$$+ [B_0(134) - B_0(34)] \frac{2 s_{34}}{s_{12} S_1} + [B_0(234) - B_0(1234)] \frac{2(s_{34} + S_2)}{s_{12}(s_{12} + S_1)},$$

(B.6c)

$$E_4 = -s_{12}D_0(2,1,34)\left[\frac{1}{2} - \frac{(S_1 - 8m_t^2)(s_{34} + S_1)}{2\Delta} - \frac{s_{12}(s_{34} + S_1)^3}{\Delta^3}\right]$$
$$+ [(s_{12} + S_2)C_0(2,134) - s_{12}C_0(1,2) + (S_1 - S_2)C_0(12,34) - S_1 C_0(1,34)]$$
$$\times \left(\frac{(S_1 - 4m_t^2)}{2\Delta} + \frac{s_{12}(s_{34} + S_1)^2}{\Delta^2}\right)$$
$$- C_0(12,34) + [B_0(134) - B_0(1234)]\left(\frac{2s_{34}}{\Delta} + \frac{2s_{12}(s_{34} + S_1)}{(s_{12} + S_2)\Delta}\right)$$
$$- [B_0(34) - B_0(1234) + s_{12}C_0(12,34)]\left(\frac{2s_{34}(2s_{12}s_{34} - S_2(S_1 + S_2) + s_{12}(S_1 - S_2))}{\Delta\Sigma}\right)$$
$$+ [B_0(12) - B_0(1234) - (s_{34} + S_1 + S_2)C_0(12,34)]$$
$$\times \left(\frac{2s_{12}(2s_{12}s_{34} - S_1(S_1 + S_2) + s_{34}(S_2 - S_1))}{\Delta\Sigma}\right), \tag{B.6d}$$

$$F_4 = -s_{12}D_0(1,2,34)\left[\frac{1}{2} + \frac{(S_1 - 8m_t^2)(s_{34} + S_2)}{2\Delta} + \frac{s_{12}(s_{34} + S_2)^3}{\Delta^3}\right]$$
$$- [(s_{12} + S_1)C_0(1,234) - s_{12}C_0(1,2) - (S_1 - S_2)C_0(12,34) - S_2 C_0(2,34)]$$
$$\times \left(\frac{S_1 - 4m_t^2}{2\Delta} + \frac{s_{12}(s_{34} + S_2)^2}{\Delta^2}\right)$$
$$- C_0(12,34) - [B_0(234) - B_0(1234)]\frac{2(s_{34} + S_2)}{\Delta}$$
$$+ [B_0(34) - B_0(1234) + s_{12}C_0(12,34)]\frac{2s_{34}(2s_{12}s_{34} - S_1(S_1 + S_2) + s_{12}(S_2 - S_1))}{\Delta\Sigma}$$
$$- [B_0(12) - B_0(1234) - (s_{34} + S_1 + S_2)C_0(12,34)]$$
$$\times \left(\frac{2s_{12}(2s_{12}s_{34} - S_2(S_1 + S_2) + s_{34}(S_1 - S_2))}{\Delta\Sigma}\right), \tag{B.6e}$$

$$G_4 = -D_0(1,34,2)\left[\frac{\Delta}{s_{12}} + \frac{s_{12} + S_1}{2} - 4m_t^2\right]$$
$$+ [(s_{12} + S_1)C_0(1,234) - S_1 C_0(1,34)]\left(\frac{1}{s_{12}} - \frac{S_1 - 4m_t^2}{2\Delta}\right)$$
$$+ [(s_{12} + S_2)C_0(134,2) - S_2 C_0(2,34)]\left(\frac{1}{s_{12}} + \frac{S_1 - 4m_t^2}{2\Delta}\right)$$
$$+ [B_0(1234) - B_0(134)]\frac{2}{s_{12} + S_2}, \tag{B.6f}$$

$$E_{10} = -s_{12}D_0(2,1,34)\left[\frac{s_{34}+S_1}{\Delta} + \frac{12m_t^2 S_1(s_{34}+S_1)}{\Delta^2} - \frac{4s_{12}S_1(s_{34}+S_1)^3}{\Delta^3}\right]$$
$$- [(s_{12}+S_2)C_0(2,134) - s_{12}C_0(1,2) + (S_1-S_2)C_0(12,34) - S_1 C_0(1,34)]$$
$$\times \left(\frac{1}{\Delta} + \frac{4m_t^2 S_1}{\Delta^2} - \frac{4s_{12}S_1(s_{34}+S_1)^2}{\Delta^3}\right)$$
$$+ C_0(12,34)\left(\frac{4s_{12}s_{34}(S_1-S_2)}{\Delta\Sigma} - \frac{4(s_{12}-2m_t^2)(2s_{12}s_{34}-S_1(S_1+S_2))}{\Delta\Sigma}\right)$$
$$+ [B_0(134) - B_0(1234)]\left(\frac{4(s_{34}+S_1)}{(s_{12}+S_2)\Delta} + \frac{8S_1(s_{34}+S_1)}{\Delta^2}\right)$$
$$+ [B_0(34) - B_0(1234) + s_{12}C_0(12,34)]\left(\frac{12s_{34}(2s_{12}+S_1+S_2)(2s_{12}s_{34}-S_1(S_1+S_2))}{\Delta\Sigma^2}\right.$$
$$\left. -\frac{4s_{34}(4s_{!2}+3S_1+S_2)}{\Delta\Sigma} + \frac{8s_{12}s_{34}(s_{34}(s_{12}+S_2)-S_1(s_{34}+S_1))}{\Delta^2\Sigma}\right)$$
$$+ [B_0(12) - B_0(1234) - (s_{34}+S_1+S_2)C_0(12,34)]$$
$$\times \left(\frac{12s_{12}(2s_{34}+S_1+S_2)(2s_{12}S_{34}-S_1(S_1+S_2))}{\Delta\Sigma^2}\right.$$
$$\left. + \frac{8s_{12}S_1(s_{34}(s_{12}+S_2)-S_1(s_{34}+S_1))}{\Delta^2\Sigma}\right)$$
$$+ \frac{i}{4\pi^2}\left(\frac{2s_{12}s_{34}-S_1(S_1+S_2)}{\Delta\Sigma}\right),$$

(B.6g)

$$F_{10} = s_{12}D_0(1,2,34)\left[\frac{s_{34}+S_2}{\Delta} - \frac{4m_t^2}{\Delta} + \frac{12m_t^2 s_{34}(s_{12}+S_1)}{\Delta^2}\right.$$
$$\left. - \frac{4s_{12}(s_{34}+S_2)^2}{\Delta^2} - \frac{4s_{12}S_1(s_{34}+S_2)^3}{\Delta^3}\right]$$
$$+ [(s_{12}+S_1)C_0(1,234) - s_{12}C_0(1,2) - (S_1-S_2)C_0(12,34) - S_2 C_0(2,34)]$$
$$\times \left(\frac{1}{\Delta} + \frac{4m_t^2 S_1}{\Delta^2} - \frac{4s_{12}(s_{34}+S_2)}{\Delta^2} - \frac{4s_{12}S_1(s_{34}+S_1)^2}{\Delta^3}\right)$$
$$- C_0(12,34)\left(\frac{4s_{12}s_{34}}{S_2\Delta} + \frac{4s_{12}s_{34}(S_2-S_1)}{\Delta\Sigma} + \frac{4(s_{12}-2m_t^2)(2s_{12}s_{34}-S_1(S_2+S_2))}{\Delta\Sigma}\right)$$
$$- [B_0(234) - B_0(1234)]\left(\frac{4s_{34}}{S_2\Delta} + \frac{8s_{34}(s_{12}+S_1)}{\Delta^2}\right)$$
$$- [B_0(34) - B_0(1234) + s_{12}C_0(12,34)]\left(-\frac{12s_{34}(2s_{12}+S_1+S_2)(2s_{12}s_{34}-S_1(S_1+S_2))}{\Delta\Sigma^2}\right.$$
$$\left. - \frac{4s_{12}s_{34}^2}{S_2\Delta^2} + \frac{4s_{34}S_1}{\Delta\Sigma} - \frac{4s_{34}(s_{12}s_{34}(2s_{12}+S_2) - S_1^2(2s_{12}+S_1))}{\Delta^2\Sigma}\right)$$
$$- [B_0(12) - B_0(1234) - (s_{34}+S_1+S_2)C_0(12,34)]$$
$$\left(-\frac{12s_{12}(2s_{34}+S_1+S_2)(2s_{12}s_{34}-S_1(S_1+S_2))}{\Delta\Sigma^2} + \frac{8s_{12}(2s_{34}+S_1)}{\Delta\Sigma}\right.$$
$$\left. - \frac{8s_{12}s_{34}(2s_{12}s_{34}-S_1(S_1+S_2)+s_{12}(S_2-S_1))}{\Delta^2\Sigma}\right)$$
$$+ \left(\frac{i}{4\pi^2}\right)\left(\frac{2s_{12}s_{34}-S_1(S_1+S_2)}{\Delta\Sigma}\right),$$
(B.6h)

$$G_{10} = -D_0(1,34,2)\left(1 + \frac{4S_1 m_t^2}{\Delta}\right)$$
$$+ [(s_{12}+S_1)C_0(1,234) - S_1 C_0(1,34)]\left(\frac{1}{\Delta} + \frac{4S_1 m_t^2}{\Delta^2}\right)$$
$$- [(s_{12}+S_2)C_0(134,2) - S_2 C_0(2,34)]\left(\frac{1}{\Delta} + \frac{4S_1 m_t^2}{\Delta^2}\right)$$
$$+ [B_0(1234) - B_0(134)]\frac{4(s_{34}+S_1)}{\Delta(s_{12}+S_2)} + [B_0(34) - B_0(234)]\frac{4s_{34}}{\Delta S_2},$$
(B.6i)

where the arguments of the scalar integrals are written with the number of the parton momentum for notational brevity. A combined number means that the momenta are added: for example, $C_0(12,34) \equiv C_0(k_1+k_2, k_3+k_4)$.